\documentclass[fleqn,usenatbib]{mnras}
\usepackage[authoryear]{natbib}
\usepackage{graphicx}
\usepackage{amssymb}
\usepackage{longtable}
\usepackage{xtab}
\usepackage{threeparttable}
\usepackage{enumerate}
\usepackage{enumitem}

\newcommand{\MC}{\multicolumn}
\newcommand{\kms}{km~s$^{-1}$}

\newcommand{\HI}{H{\sc i}}
\newcommand{\HII}{H{\sc ii}}
\newcommand{\sunn}{$_{\odot}$}
\newcounter{qub}
\newcounter{qub1}
\newcounter{qub2}
\newcommand{\qq}{\addtocounter{qub}{1}\arabic{qub}}
\newcommand{\cc}{\addtocounter{qub1}{1}\arabic{qub1}}
\newcommand{\qqq}{\addtocounter{qub2}{1}\arabic{qub2}}
\DeclareRobustCommand{\ion}[2]{%
\relax\ifmmode
\ifx\testbx\f
{\mathrm{#1\,\textsc{#2}}}\else
{\mathrm{#1\,\mathsc{#2}}}\fi
\else\textup{#1\,{\mdseries\textsc{#2}}}%
\fi}

\title[Galaxies in the Eridanus void. Sample and O/H ]
{Study of galaxies in the Eridanus void. Sample and oxygen
abundances\footnotemark[0]\thanks{Based on observations obtained with
the Southern African Large Telescope (SALT), programmes
\mbox{2012-1-RSA-001}, \mbox{2012-2-RSA\_OTH-004} and
\mbox{2013-1-RSA\_OTH-008} and with the Special Astrophysical Observatory
of RAS 6-m telescope (BTA).}}
\author[A.Y.~Kniazev, E.S.~Egorova, S.A.~Pustilnik]
{A.Y. Kniazev,$^{1,2,3,4}$\thanks{E-mail: akniazev@saao.ac.za} E.S.~Egorova,$^{3}$ S.A. Pustilnik$^{4}$ \\
\rule{-4pt}{20pt}
$^1$ South African Astronomical Observatory, PO Box 9, 7935 Observatory,
   Cape Town, South Africa\\
$^2$ Southern African Large Telescope Foundation, PO Box 9, 7935 Observatory,
   Cape Town, South Africa \\
$^{3}$ Sternberg Astronomical Institute, Lomonosov Moscow State University,
Moscow, Russia \\
$^4$ Special Astrophysical Observatory of RAS, Nizhnij Arkhyz,
  Karachai-Circassia 369167, Russi
}

\begin{document}

\label{firstpage}

\date{Accepted on 2018 March 15. Received on 2018 February 10}

\pagerange{\pageref{firstpage}--\pageref{lastpage}} \pubyear{2018}

\maketitle

\begin{abstract}

We present a sample of 66 galaxies belonging to the equatorial part 
(Dec.= --7\degr, +7\degr) of the large so called Eridanus void
\citep[after][]{Fairall98}. The void galaxies are selected as to be separated
from the luminous galaxies {\bf ($M_{\rm B} < M_{\rm B}^{*} +1$)},
delineating the void, by more than
2 Mpc. Our main goal is to study systematically the evolutionary parameters
of the void sample (metallicity and gas content) and to compare the void
galaxy properties with their counterparts residing in denser environments.
Besides the general galaxy parameters, compiled mainly from the literature,
we present the results of dedicated observations to measure the oxygen
abundance O/H in  HII-regions of 23 void galaxies obtained with the 11-m
SALT telescope (SAAO) and the 6-m BTA telescope (SAO),
as well as the O/H estimates derived from the analysis of the SDSS DR12
spectra for 3 objects. We compiled all available data on O/H in 36 these
void galaxies, including those for 11 galaxies available in the literature
(for one object both SDSS and SALT spectra were used), and
analyze this data in relation to galaxy luminosity ($\log$(O/H) versus
$M_{\rm B}$). Comparing them with the control sample of similar type
galaxies from the Local Volume, we find clear evidence for a substantially
lower average metallicity of the Eridanus void galaxies. This result
matches well the conclusions of our recent similar study for galaxies in the
Lynx-Cancer void.

\end{abstract}


\section[]{INTRODUCTION}
\label{sec:intro}

The modern cosmological $\Lambda$CDM models of the large-scale structure
and galaxy formation, including the state-of-art N-body simulations, predict
that galaxy properties and evolution can significantly depend on their global
environment
\citep[e.g.,][and references therein]{Peebles01, Gottlober03, Hoeft06,
Hoeft10, Hahn07, Hahn09, Kreckel11b}. However, the role of the most rarefied
environment (typical of voids) in galaxy formation and evolution is not well
studied both theoretically and observationally.

In particular, most of the mass studies of galaxies in voids based on large
samples from the SDSS spectral database, are limited by rather distant giant
voids (at $D \sim 100 - 200$ Mpc), and hence, probe only the upper part of
the void galaxy luminosity function. Namely, those with
$M_{\rm B} \lesssim -16^m$ to $\lesssim -17^m$,
\citet{Rojas04,Rojas05,Patiri06,Hoyle05,Hoyle12}.
Besides, these
works used for their analysis only the data available in the SDSS database
that precluded the direct study of evolutionary status of void galaxies.
While some differences of void galaxy global properties with respect of
wall population were found, they were rather modest.
Thus one could conclude that for the upper
mass/luminosity range the effect of void environment is rather subtle.

The authors of the more recent Void Galaxy Survey (VGS)
\citep{Kreckel11a,Kreckel12,Kreckel15,Beygu12}, while also dealing mostly
with more distant voids (typical $D \sim 80$ Mpc), partly approached closer
voids that allowed them to include some part of void dwarf galaxies with
absolute magnitudes down to $M_{\rm r} \lesssim -14^m$. They also undertook
a detailed multi-wavelength study of their galaxy sample, including \HI\
mapping, gas metallicity measurements and photometry. This allowed them to
find several unusual cases probably indicating the role of filaments in voids.
 
The nearby voids have the important advantage for the study of galaxies in
voids since they allow to probe the least massive void dwarfs which are
expected to be the most susceptible to external perturbations.

In the course of the systematic study of dwarf galaxies in the nearby
Lynx-Cancer void, we have already discovered a half-dozen unusual objects,
including DDO~68 \citep*{DDO68,IT07}, J0926+3343 \citep{J0926} and other
very metal-poor LSBDs
\citep[][hereafter Paper~III]{void_LSBD}. The recent Giant Meterwave Radio
Telescope (GMRT) \HI\ mapping of a subsample of this void galaxies revealed
three extremely gas-rich blue unevolved dwarfs \citep{CP2013,CPE2017}.
The analysis and statistical study
of photometric parameters \citep{Pap4} and of integrated \HI\ data
\citep{Pap6}, as well as of the significantly extended data on metallicities
of this void galaxies \citep{Pap7} allows one to conclude that void galaxies
evolve slower then their analogs in denser environments.

Moreover, there exists a sizable fraction of extremely gas-rich void galaxies
(mostly the least massive LSBDs) with the lowest gas metallicities
($Z \lesssim Z$\sunn/20, or 12+$\log$(O/H) $\lesssim$7.38) and atypically
blue colours of stellar population in the outer parts. The straightforward
interpretation implies that the main star formation episode in these objects
took place rather recently - between one and a few Gyr ago - and transformed
only a few percent of the original gas proto-galaxy to stars.

In this paper we address properties of galaxies in another very low density
region known as the Eridanus void. The paper is organized as follows.
In Sec.~\ref{sec:void} we present the void description.
In Sec.~\ref{sec:sample} we describe the selection criteria and summarize
the Eridanus void galaxy sample. In Sec.~\ref{sec:obs} we briefly describe the
observations and reduction of the obtained data. Sec.~\ref{sec:results}
presents the results of observations and their analysis.
In Sec.~\ref{sec:dis} we discuss the results and their implications in a
broader context and summarize our conclusions. The value of the Hubble
constant is adopted as $H_{\mathrm 0} =$ 73~\kms~Mpc$^{-1}$.

\section[]{Eridanus void}
\label{sec:void}

\citet{Guseva09} presented, among other new metal-poor galaxies,
the second most metal-poor (at that time) galaxy SDSS J0015+0104,
with 12+$\log$(O/H)=7.07.
In a paper by \citet{Guseva17} its parameter 12+$\log$(O/H) was corrected
to 7.17, equal to that of IZw18.
Our subsequent study of this galaxy \citep{Eridan_XMD} revealed
atypical blue colours in the outer parts and a very high mass fraction of gas.
This LSBD appears to reside in a large region devoid
of luminous galaxies, known as the Eridanus void after \citet{Fairall98}.

In fact, this name does not relate to the sky region of the respective
constellation, but goes back to the Greek word for river. As  \citet{KF1991}
have commented, the name relates to the extended tubular form of this void
and not to the constellation. More specifically, the region is situated
partly in Aquarius (19.5~h -- 0.5~h, mainly
to the South of the equator) and in  Pisces (23~h--02~h) and Cetus
(00~h -- 3.5~h).

To study the evolutionary status of galaxies in this region in a similar way
to our project for the Lynx-Cancer void \citep{PaperI}, we define below the
borders of the adopted  equatorial region of this void. Then we compile the
sample of galaxies residing in this void based on NED, HyperLeda\footnote{http://leda.univ-lyon1.fr/} \citep{Makarov2014}, and other
sources.

\begin{figure*}
 \centering
 \includegraphics[angle=-90,width=0.50\linewidth]{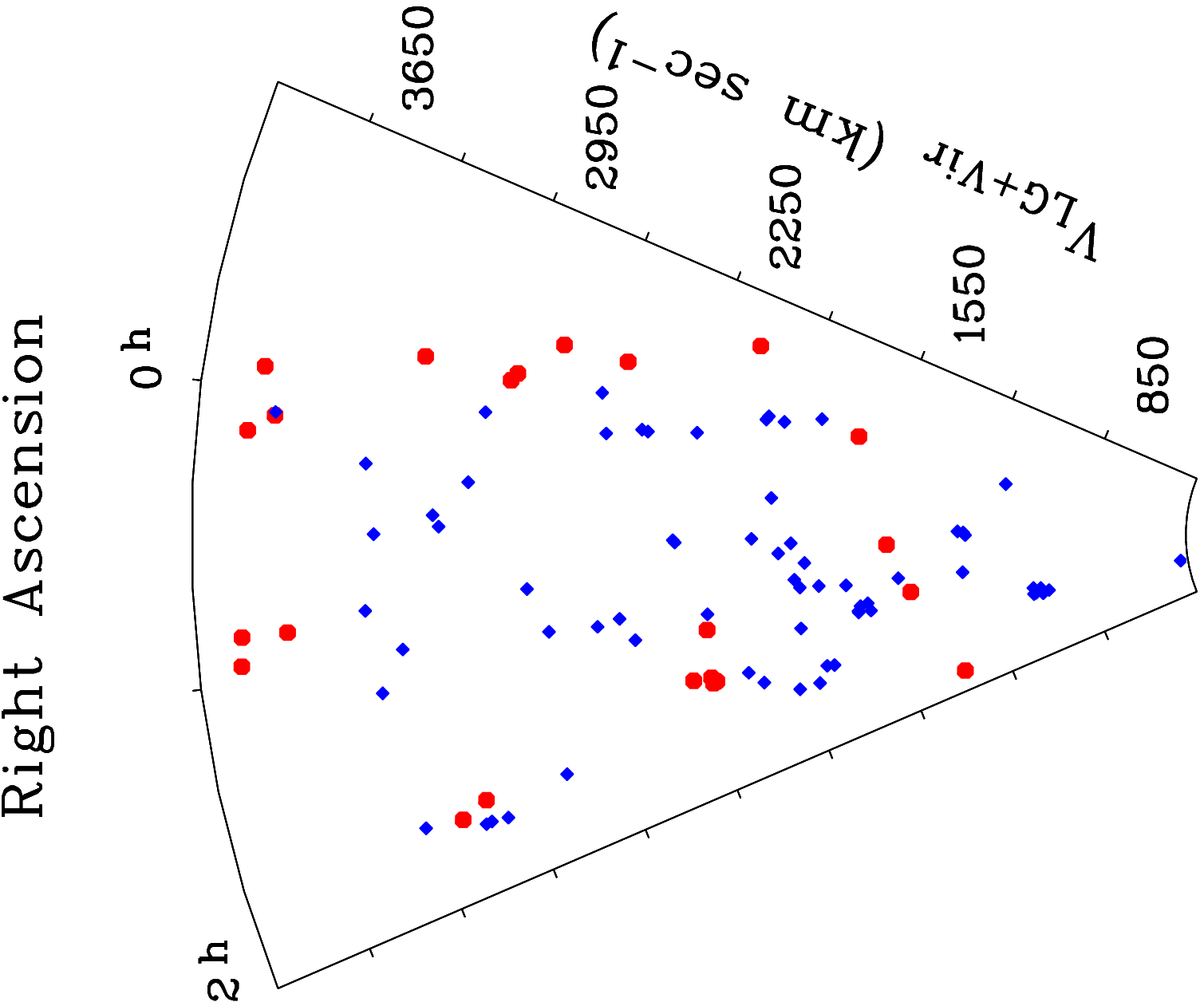}
\includegraphics[angle=-90,width=0.27\linewidth]{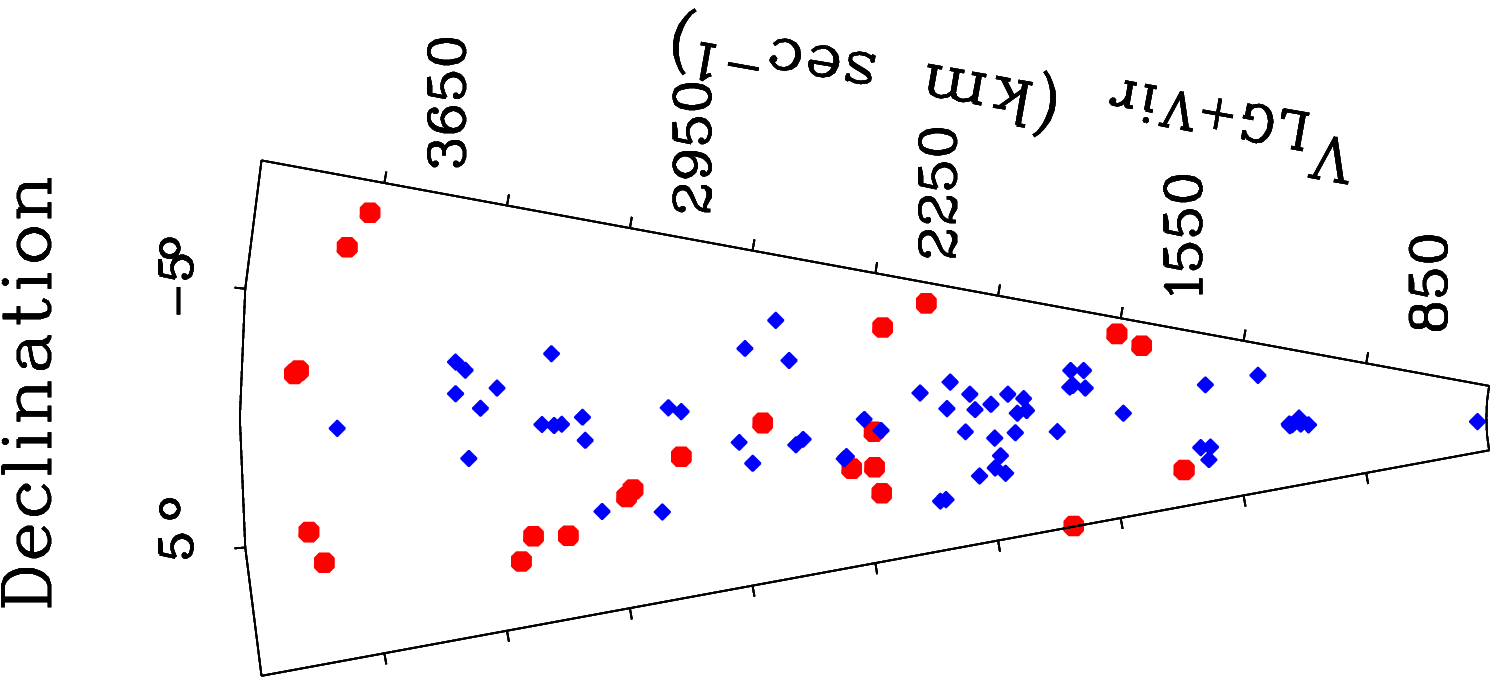}
\caption{\label{fig:Wedge}
Red circles mark bright galaxies with $M_{\rm B}^{*}<-19.5$, while the
blue diamonds mark galaxies from our sample.
{\bf Left panel:} The wedge diagram for Dec.= --10\degr, +10\degr.
{\bf Right panel:} The wedge diagram for the range of $RA = 23^{h}, 2^{h}$
}
\end{figure*}

To illustrate the void galaxy distribution, we show in Fig.~\ref{fig:Wedge}
the  wedge diagram of this sky section. In some cases points that mark the
positions of our sample galaxies are overlapped with the points that mark
the positions of bright galaxies because of the projection effect.

For a preliminary selection of this void sample, we use the void boundaries
outlined by \citet{Fairall98}. As luminosity to delineate `luminous' galaxies
 we adopt ($M_{\rm B} < M_{\rm B}^{*} +1.0 = -19.5^{m}$), where for
$L_{\rm B}^{*}$
(the characteristic 'knee' value of the galaxy luminosity function) we adopt
$M_{\rm B}^{*} = -20.5^{m}$ \citep{Hill2010}.
Then we look in the wedge diagrams with the narrow
ranges in declination for candidate void objects residing within the `empty'
volume.  As the last step we check whether a candidate void galaxy is
sufficiently distant from a `luminous' border galaxy. The more advanced
procedure of void separation will be applied in a forthcoming paper
(Pustilnik et al. 2018, in prep.).

\section[]{Eridanus void galaxy sample - the current version}
\label{sec:sample}

Similar to the selection criteria used for the Lynx-Cancer void galaxy
sample, we picked up galaxies within the predefined void region which
are situated at distances of more than 2.0 Mpc from the bordering luminous
galaxies. While the latter threshold leaves a fraction of real void galaxies
residing in the transitional void zone outside of the formed sample, this
allows to keep the sample rather clean, since such limit eliminates the
cases of interlopers belonging to typical groups around bordering luminous
galaxies. This also allows to skip cases of near-border galaxies of unclear
origin, which could have migrated from the wall regions during the last
several Gyrs.

\begin{figure*}
 \centering
 \includegraphics[angle=-90,width=0.49\linewidth]{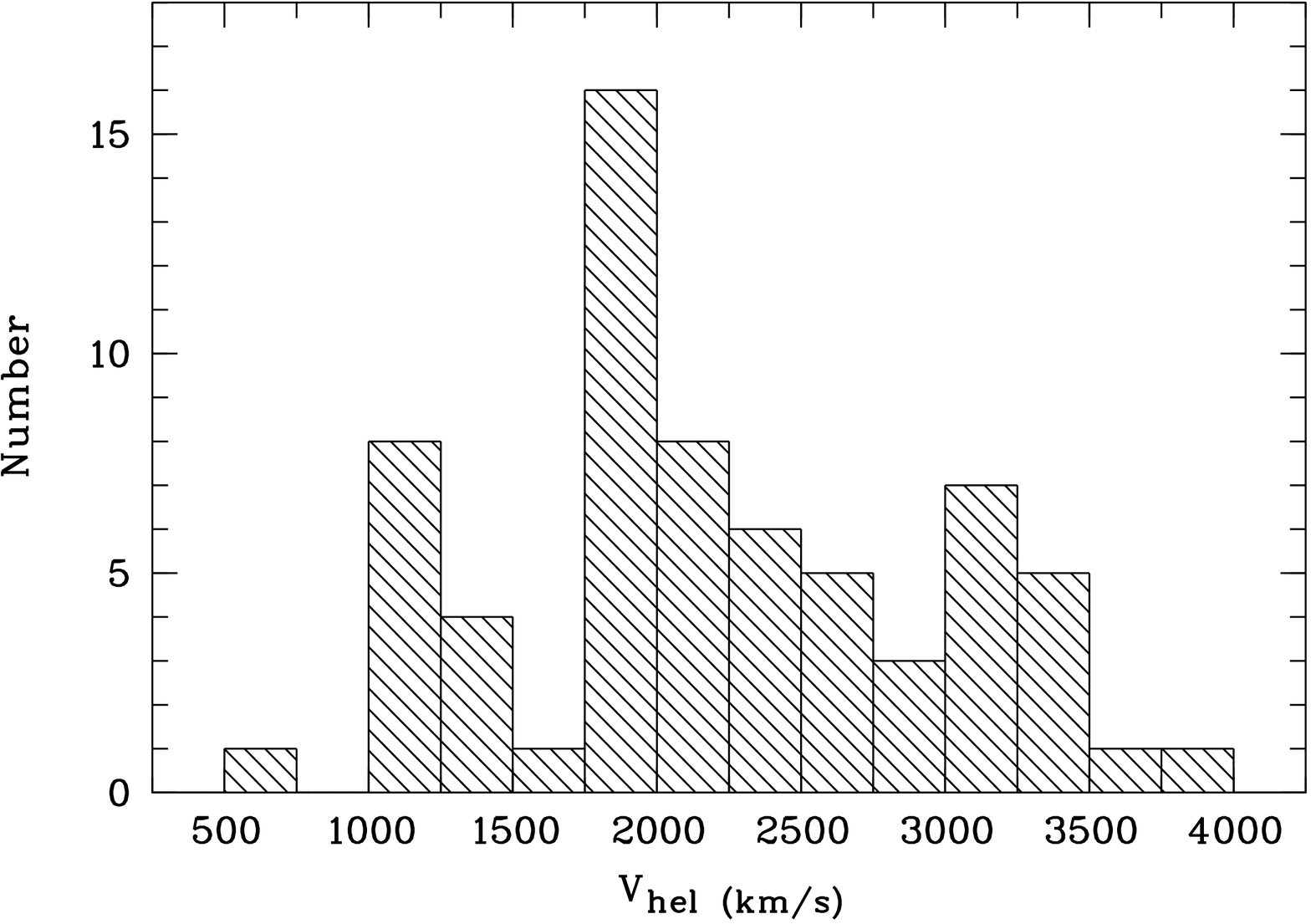}
 \includegraphics[angle=-90,width=0.49\linewidth]{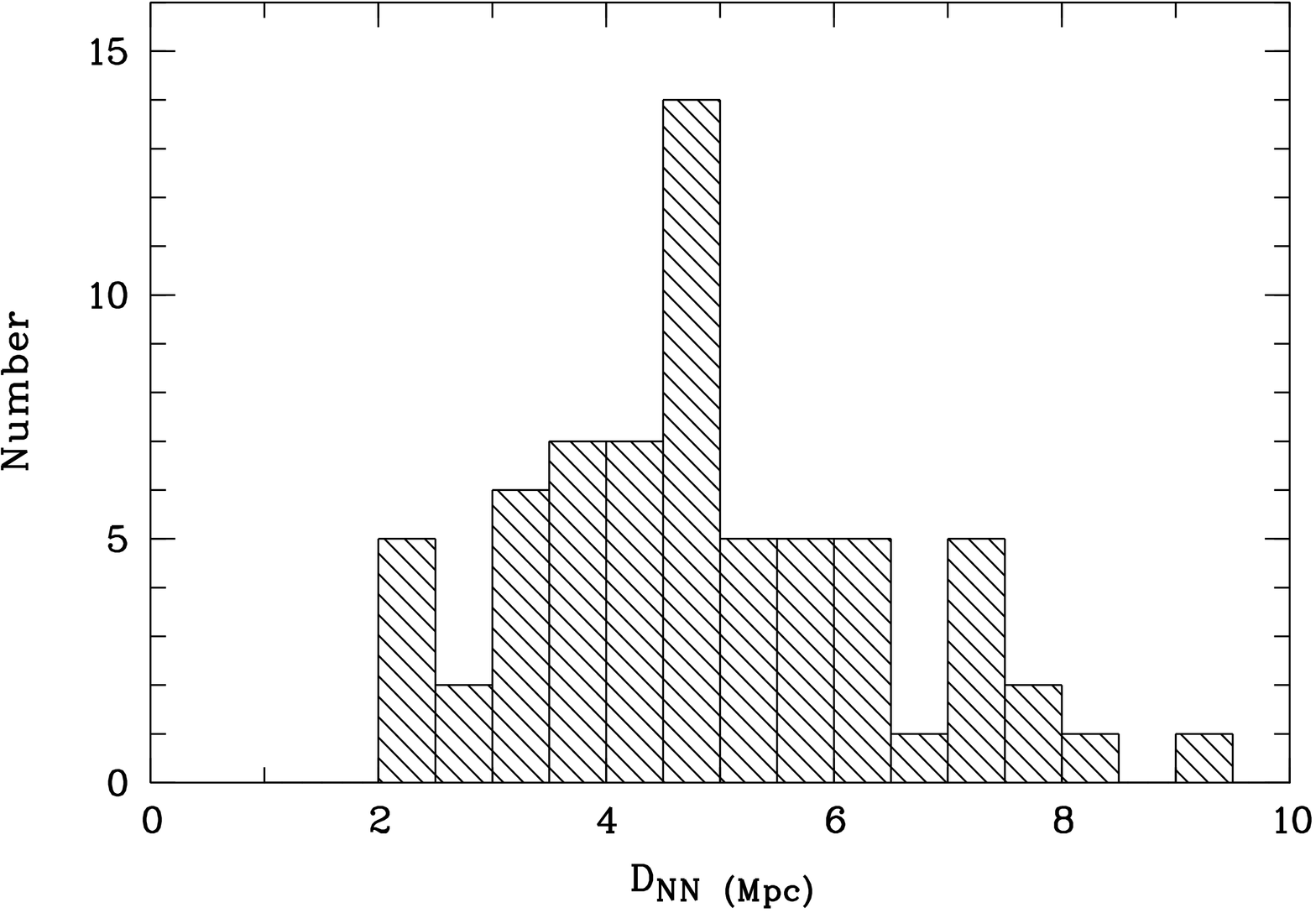}
  \caption{\label{fig:Vel}
{\bf Left panel:} The distribution of velocities $V_{\rm hel}$ for the void
galaxy sample;
{\bf Right panel:} The distribution of distances $D_{\rm NN}$ to the nearest
luminous galaxy for the void galaxy sample.
}
\end{figure*}

\begin{figure*}
 \centering
 \includegraphics[angle=-90,width=0.49\linewidth]{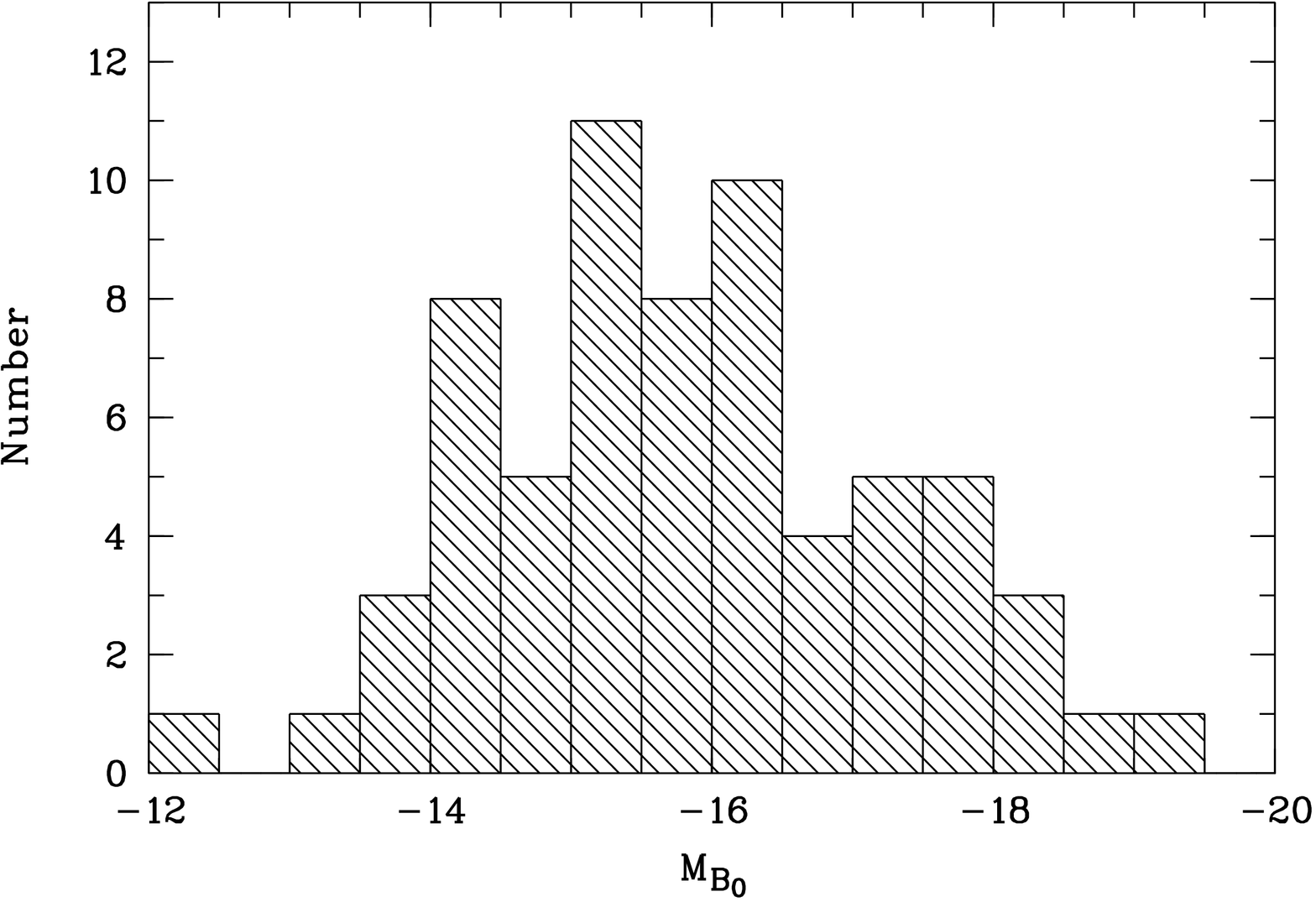}
  \caption{\label{fig:MbOH}
The distribution of absolute blue magnitudes $M_{\rm B}$ for the Eridanus
void galaxies.
}
\end{figure*}

In Table~\ref{tab:sample} we present the current version of the Eridanus
void sample with coordinates and global parameters available from the
literature.
\begin{description}
\item[$\bullet$]Column 1. Common name or SDSS prefix.

\item[$\bullet$]Columns 2 and 3. Epoch J2000 R.A. and Declination.

\item[$\bullet$]Column 4. $V_{\rm hel}$ (almost all are either from NED
or SDSS).

\item[$\bullet$]Column 5. Distance D, in Mpc, adopted as $V_{\rm LG}/73$.
Here $V_{\rm LG}$ is the recession velocity in the Local Group coordinate
system.

\item[$\bullet$]Column 6. The apparent total $B$-band magnitude $B_{\rm tot}$
 (from NED or from the literature). The sources of $B$-mag are given by a
letter in the superscript for the respective values.

\item[$\bullet$]Column 7. Absolute $B$-band magnitudes $M_{\rm B}$ calculated
 using $B_{\rm tot}$ corrected for the Galactic foreground extinction
$A_{\rm B}$ (from NED, following \citet{Schlafly11}).

\item[$\bullet$]Column 8. Type, mainly from NED and HyperLeda (where
available).

\item[$\bullet$]Column 9. The distance in Mpc to the nearest luminous
galaxy, $D_{\rm NN}$.
\end{description}

\subsection{The sample overview}

The Eridanus void is extended and has a complicated form. For our current
project we restricted the studied volume to the equatorial zone with
Dec.= (--7\degr, +7\degr), RA = ($23^{h}40^{m}$ to $01^{h}45^{m}$),
and $V_{\rm hel} < 3500-3800$~\kms\ (the latter is near the distant
void border). It is worth to note that the presented sample of galaxies
in the considered volume is not complete,  in particular for the lower
edge of the luminosity range. This is natural in view of the above mentioned
effects of observational selection.

The resulting sample of the Eridanus void equatorial region is characterized
by the following ranges of the main parameters.  Blue luminosities range from
 $M_{\rm B} = -12^m$ to --19$^m$ with a median value of --15.8$^m$. The whole
 range of radial velocities $V_{\rm hel} = 600 - 3800$~\kms, with a median
value of $\sim$2030~\kms. The distances to the nearest luminous neighbour
$D_{\rm NN}$ range from 2 to 10~Mpc with a median of $\sim$4.7~Mpc.

Our sample consists mostly of disk and irregular galaxies. There are several
blue compact objects of round or elliptical shape at distances of $D\geq$30
Mpc with spectra typical of HII regions. We assume that they represent BCGs
(blue compact galaxies). Also, $\sim$10\% (6 objects) of our sample galaxies
have morphological types E or S0/S0-a in NED or HyperLeda. Though for 2 such
galaxies, deeper images from SDSS Stripe82 or
DECaLS\footnote{\label{decals}http://legacysurvey.org/decamls/}
(The Dark Energy Camera Legacy Survey, \citet{Blum2016}) reveal low surface
brightness structures around the bright central part (that is, discs or signs
of interaction). In fact, several sample objects reveal such low surface
brightness features on deeper images.

With respect to the galaxy sample in the Lynx-Cancer void, this sample
contains a higher fraction of more luminous objects and is almost devoid
of the lowest luminosity representatives. This is a natural result of the
observational selection bias, since the Eridanus void is roughly 1.7 times
more distant than the Lynx-Cancer void. In view of these differences it
will be interesting to examine how the conclusions obtained for the
Lynx-Cancer void
 sample galaxies will look in comparison to the current sample analysis.

\begin{figure*}
 \centering
 \includegraphics[]{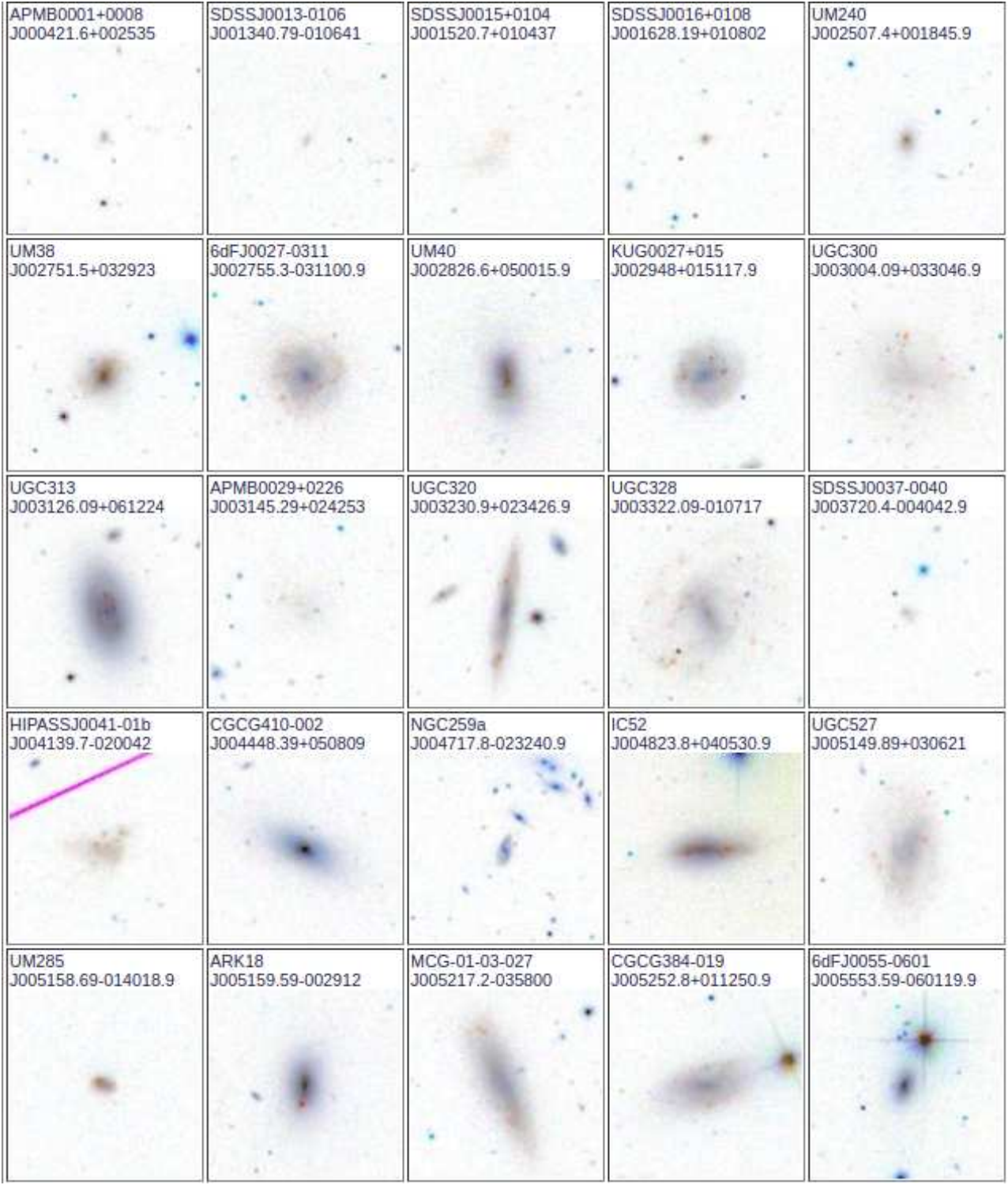}
  \caption{\label{fig:SDSSim1}
The finding charts of 25 galaxies from the Eridanus void sample
(see Table~\ref{tab:sample}). They were prepared with the SDSS DR12
Navigate Tool with inverted colours and are shown in the order of
their RA. The sides of each square measure $\approx$$100''$.
}
\end{figure*}

\begin{figure*}
 \centering
 \includegraphics[]{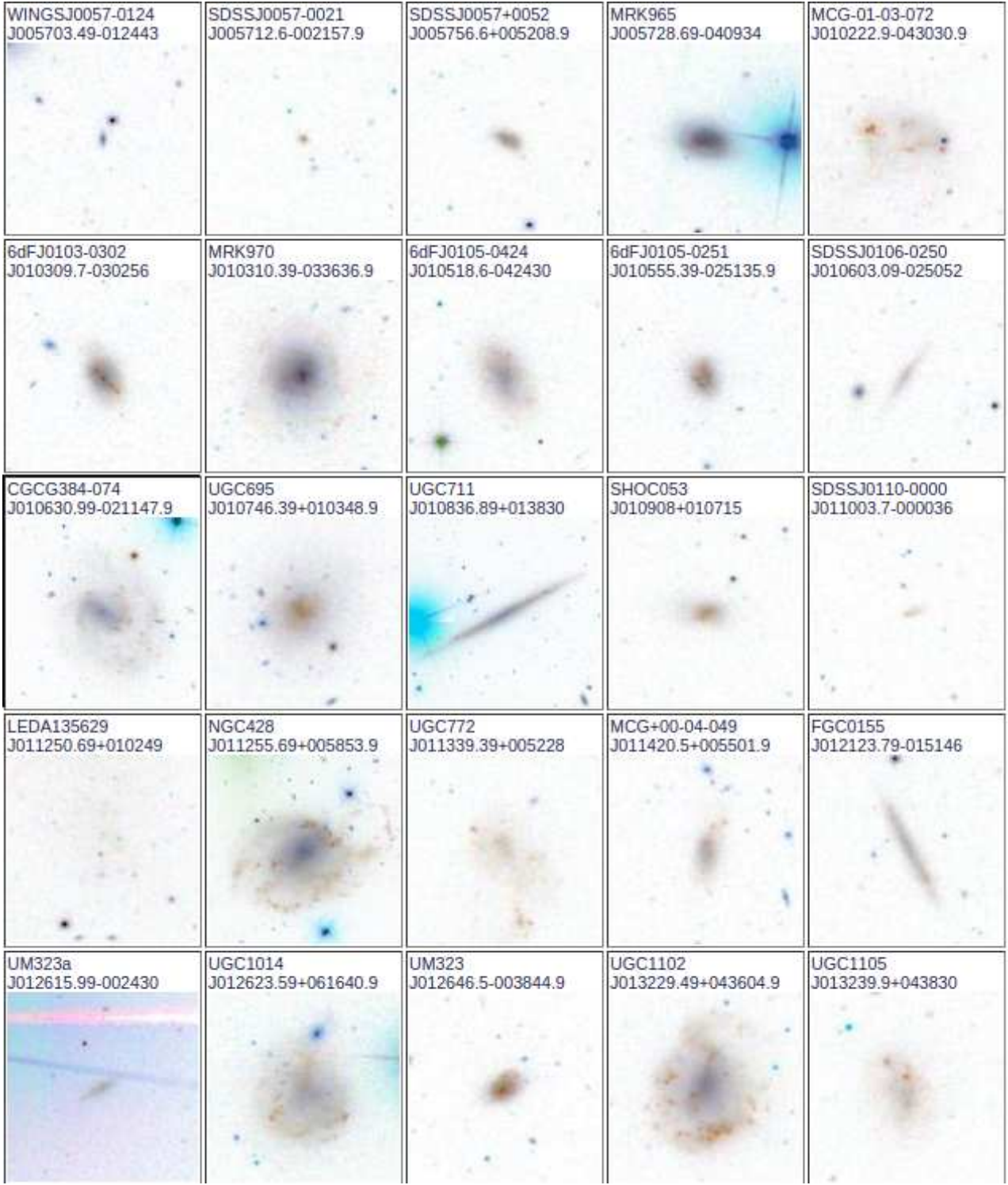}
  \caption{\label{fig:SDSSim2}
The same as for Fig.~\ref{fig:SDSSim1} for the next 25 galaxies from
the Eridanus void sample. The finding chart size for larger galaxies
are: $\approx$$150''$ for CGCG384-074, $\approx$$250''$ for UGC711,
$\approx$$300''$ for NGC428.
}
\end{figure*}

\begin{figure*}
 \centering
 \includegraphics[]{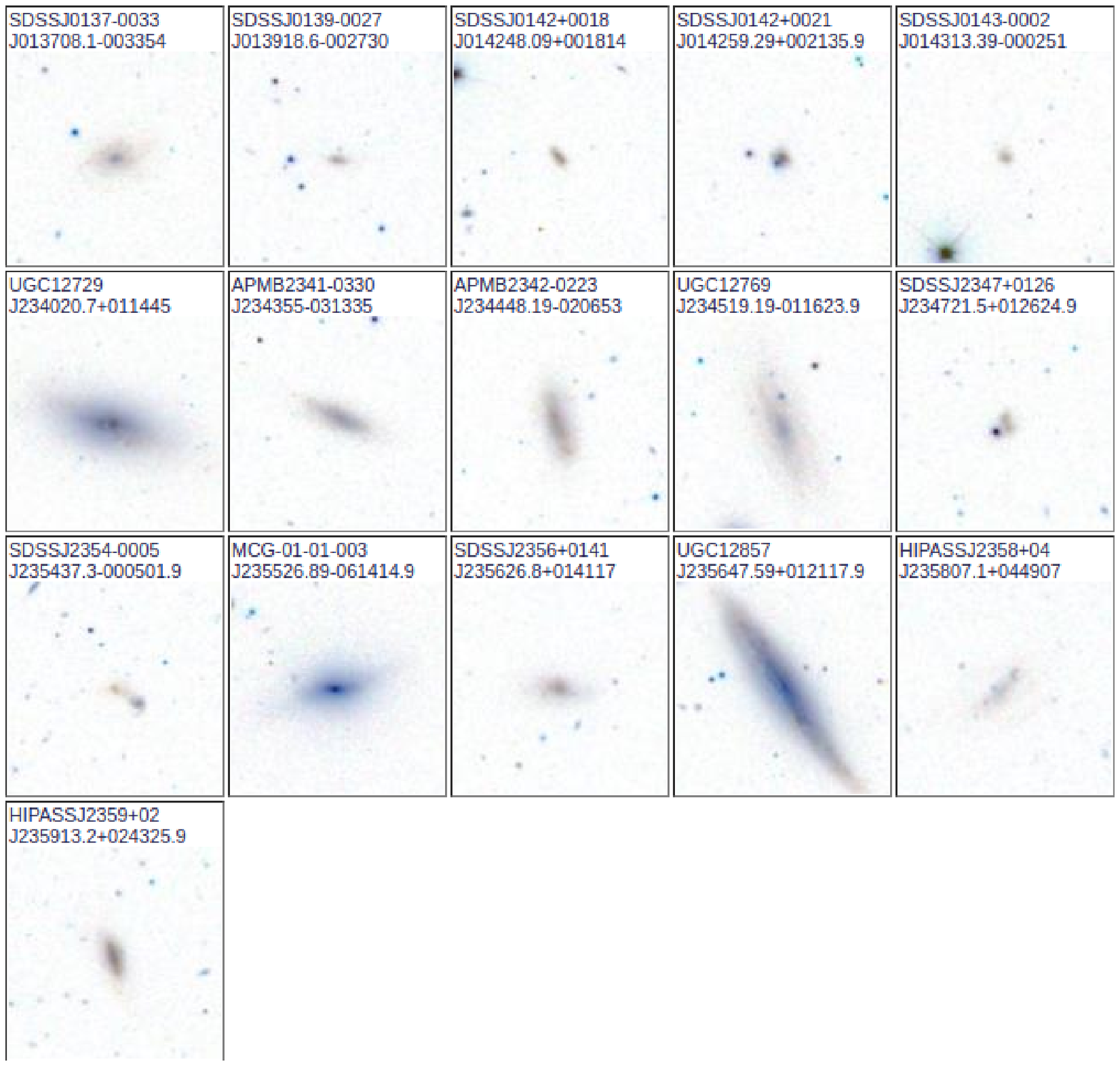}
  \caption{\label{fig:SDSSim3}
The same as for Fig.~\ref{fig:SDSSim1} for the remaining 16 galaxies
from the Eridanus void sample.
}
\end{figure*}

In Fig.~\ref{fig:Vel}--\ref{fig:MbOH} we show the distributions of the
sample galaxies in blue absolute magnitude $M_{\rm B}$, radial velocity
$V_{\rm hel}$ and the distance to the nearest luminous neighbour
$D_{\rm NN}$.
In Fig.~\ref{fig:SDSSim1}--\ref{fig:SDSSim3} we present a mosaic of
66 Eridanus void galaxy images prepared with the SDSS Navigate Tool.

\section[]{SPECTRAL OBSERVATIONS AND DATA REDUCTION}
\label{sec:obs}

\subsection{SALT and BTA data}
\label{sec:salt}

For 21 Eridanus void galaxies, the observations  were conducted with the
the Robert Stobie Spectrograph \citep[RSS;][]{Burgh03,Kobul03} installed at
the Southern African Large Telescope \citep[SALT;][]{Buck06,Dono06}.
Additionally, four other void galaxies were observed with multi-mode
instrument SCORPIO \citep{SCORPIO} installed at the SAO 6-m Telescope (BTA).
All observations with both telescopes were taken with the long-slit mode.
Details of SALT and BTA observations are presented in Table~\ref{tab:Obs}.

In the course of the spectral observations for this project we formed
a back-up list of galaxies with no redshifts as potential Eridanus void
members.
We obtained their radial velocities during weather conditions not suitable
for the main task.
Nine of them were observed with SALT in the same mode as the main program.
Besides,
3 more galaxies were observed at BTA with grism VPHV1200R, getting spectra
in the range of 5500 to 7500~\AA.
Information on the results of these observations is presented
in Table~\ref{tab:vel}.
Most of the objects appear to be background object, and only two galaxies lie
in the equatorial part of the Eridanus void. For several more galaxies
observations were performed, but the $H\alpha$ line was undetectable (though
continuum was seen).
So it was not possible to estimate the velocities for
galaxies SDSS J000224.09-013147.8, J000634.13+153037.5,  J003933.37+010652.3.

The VPH grating GR900 was used for observations with  RSS spectrograph at SALT
to cover the total range from 3600~\AA\ to 6700~\AA, with a slit width of
1.5 arcsec and the typical spectral resolution of FWHM $\sim$ 5~\AA.
For BTA observations we used grism VPHG550G covering the range from 3500 to
7500~\AA. The slit width 1.0 arcsec resulted in a spectral resolution of
$\sim$12~\AA.

The primary data reduction for SALT data was done with the SALT science
pipeline \citep{Cra2010}. The following long-slit data reduction for the
BTA and SALT observations and the emission-line measurements was performed as described in \citet{DDO68}.
Finally, all BTA spectra were transformed to absolute fluxes.
For SALT, spectrophotometric standard stars were used for the relative flux
calibration only,
since the absolute flux calibration is not feasible with SALT because the
unfilled entrance pupil of the telescope moves during the observations.
At the end, the individual 1D spectra of the target \HII-regions were
extracted by summing-up, without weighting, several rows along the slit.
For \HII\ regions, where the principal faint line [O{\sc iii}]$\lambda$4363
was well above the noise, we selected the pixel range where the line was
visible.
For the majority of spectra, where [O{\sc iii}]$\lambda$4363 was absent, the 1D spectra were obtained via extraction of signal in the pixel
range where the H$\beta$ line was visible.

To measure the emission-line intensities with their errors in the individual
1D spectra, we used the way described in detail in \citet{Kniazev04b}.
In case of the good quality of 1D spectra with clear signatures of Balmer
absorptions, we have used the model spectra of stellar populations
constructed with the use of the ULySS
\citep[http://ulyss.univ-lyon1.fr,][]{Koleva2009}
to better subtract the underlying continuum.

\begin{figure*}
 \centering \includegraphics[angle=-90,width=0.4\linewidth]{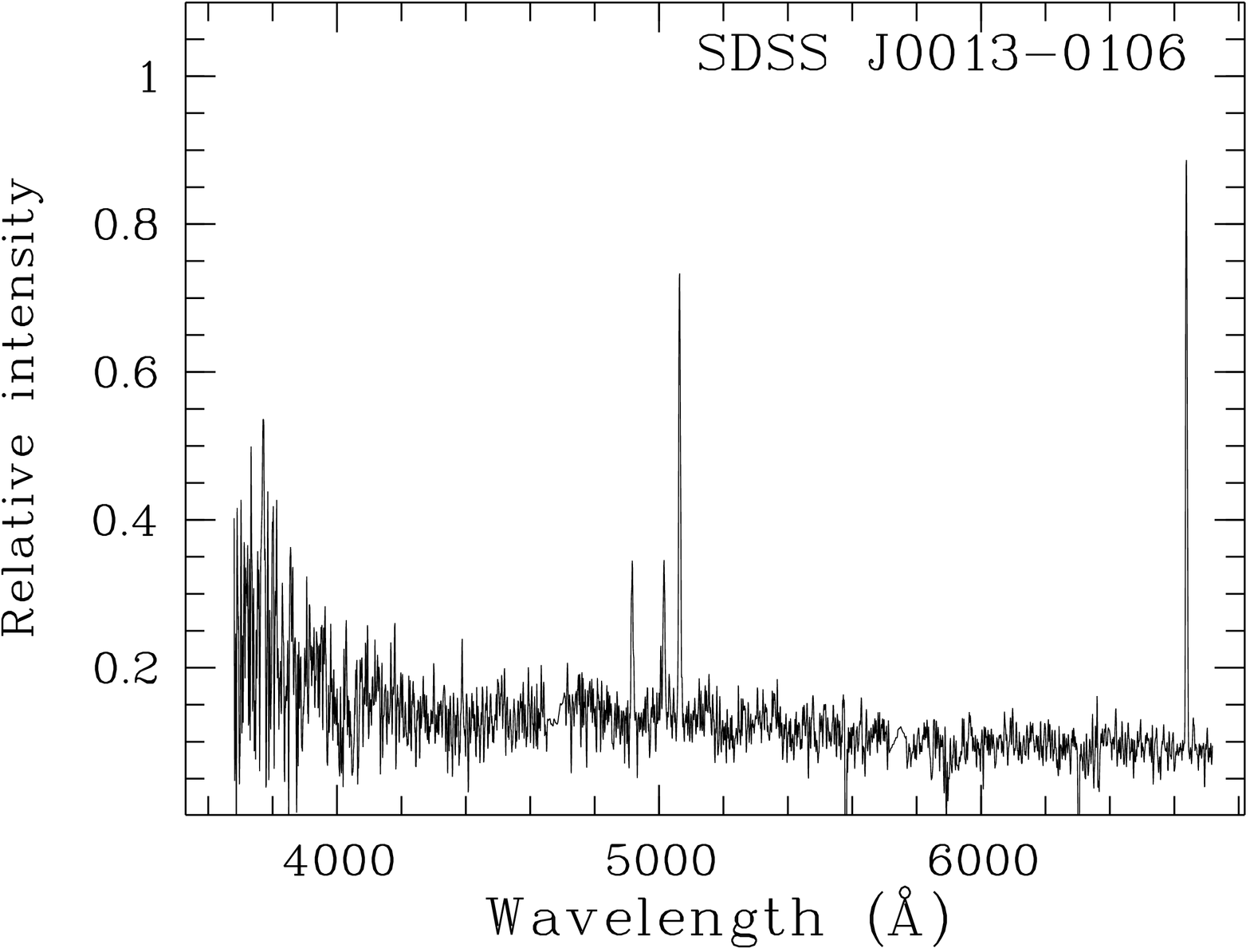}
\includegraphics[angle=-90,width=0.4\linewidth]{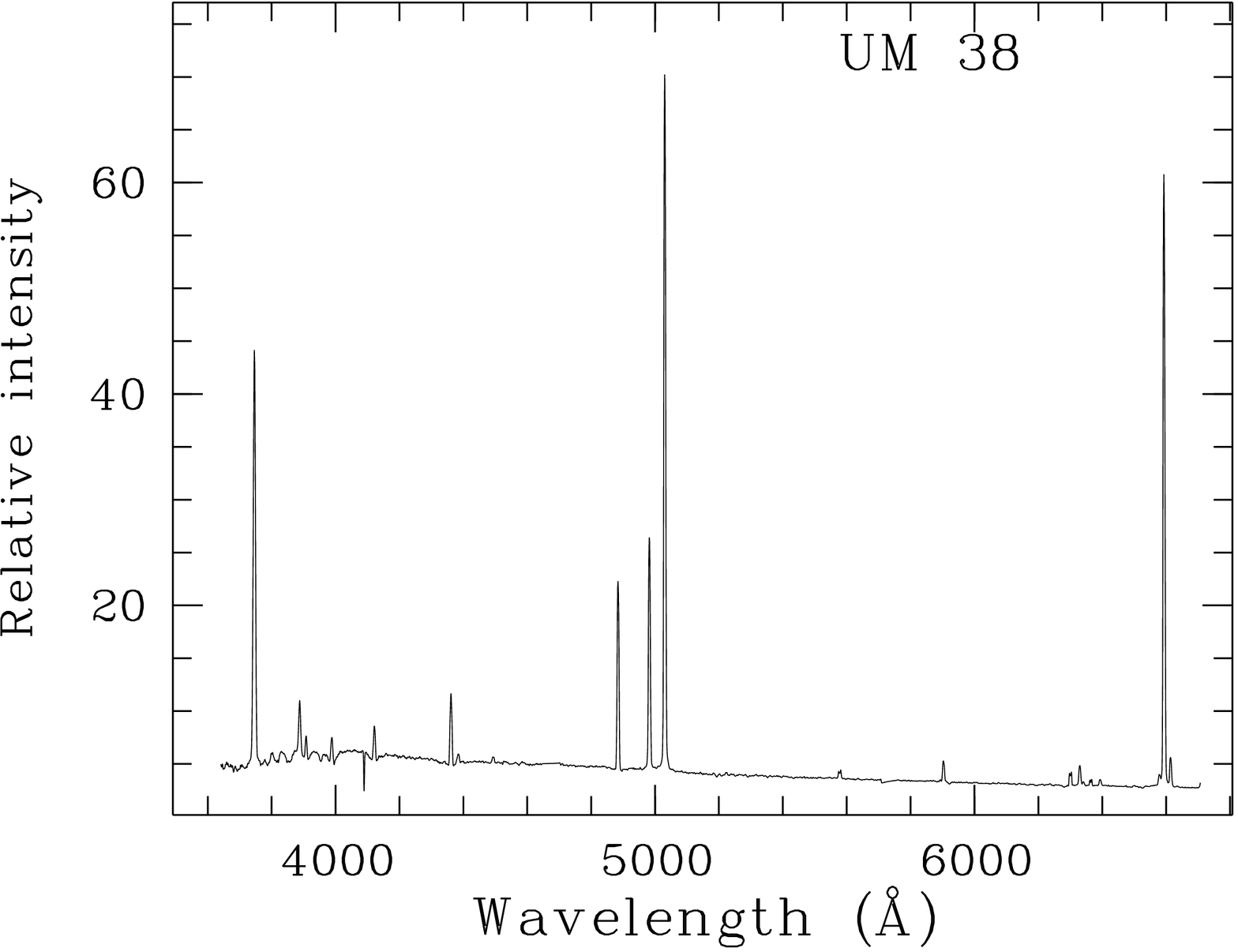}
\includegraphics[angle=-90,width=0.4\linewidth]{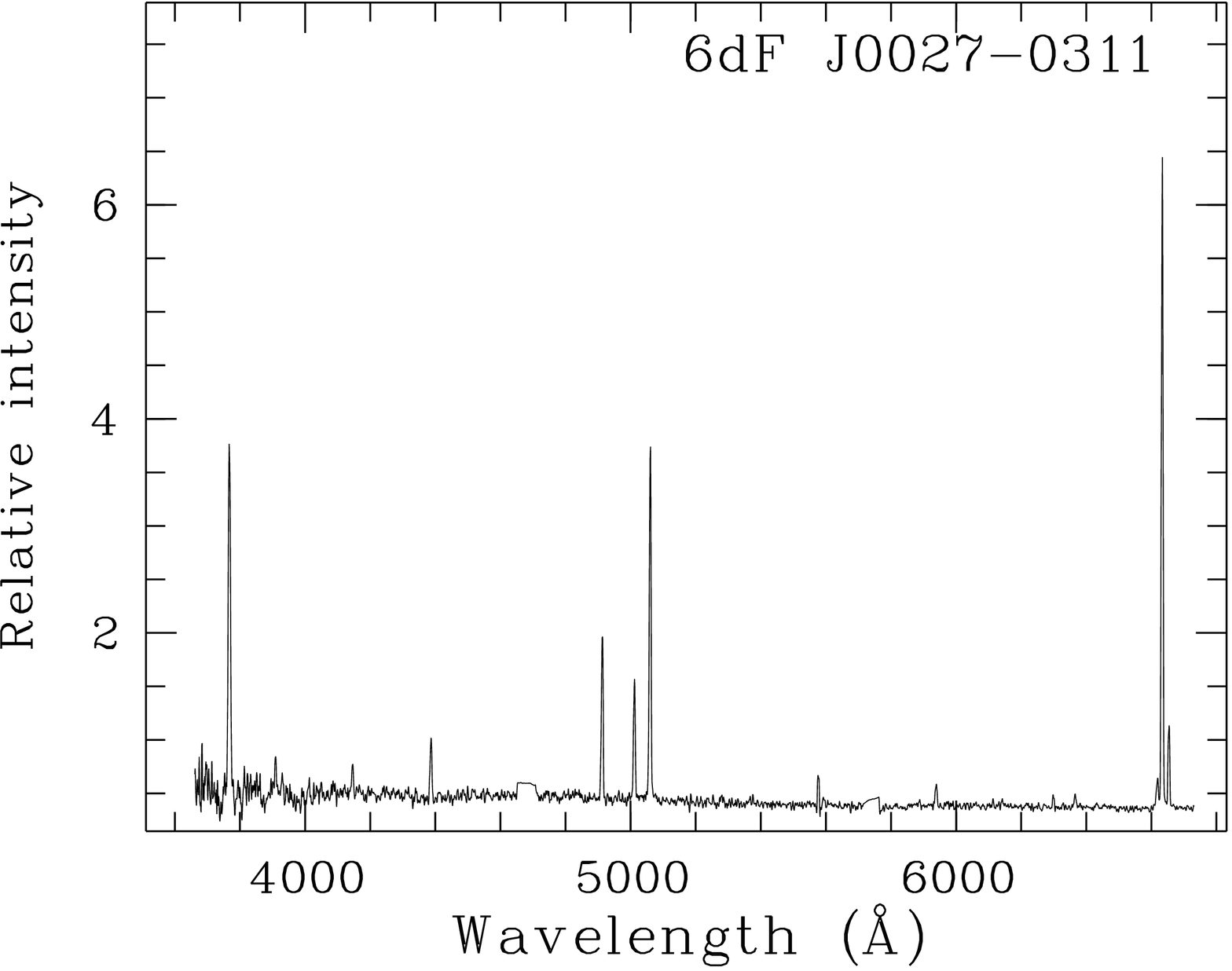}
\includegraphics[angle=-90,width=0.4\linewidth]{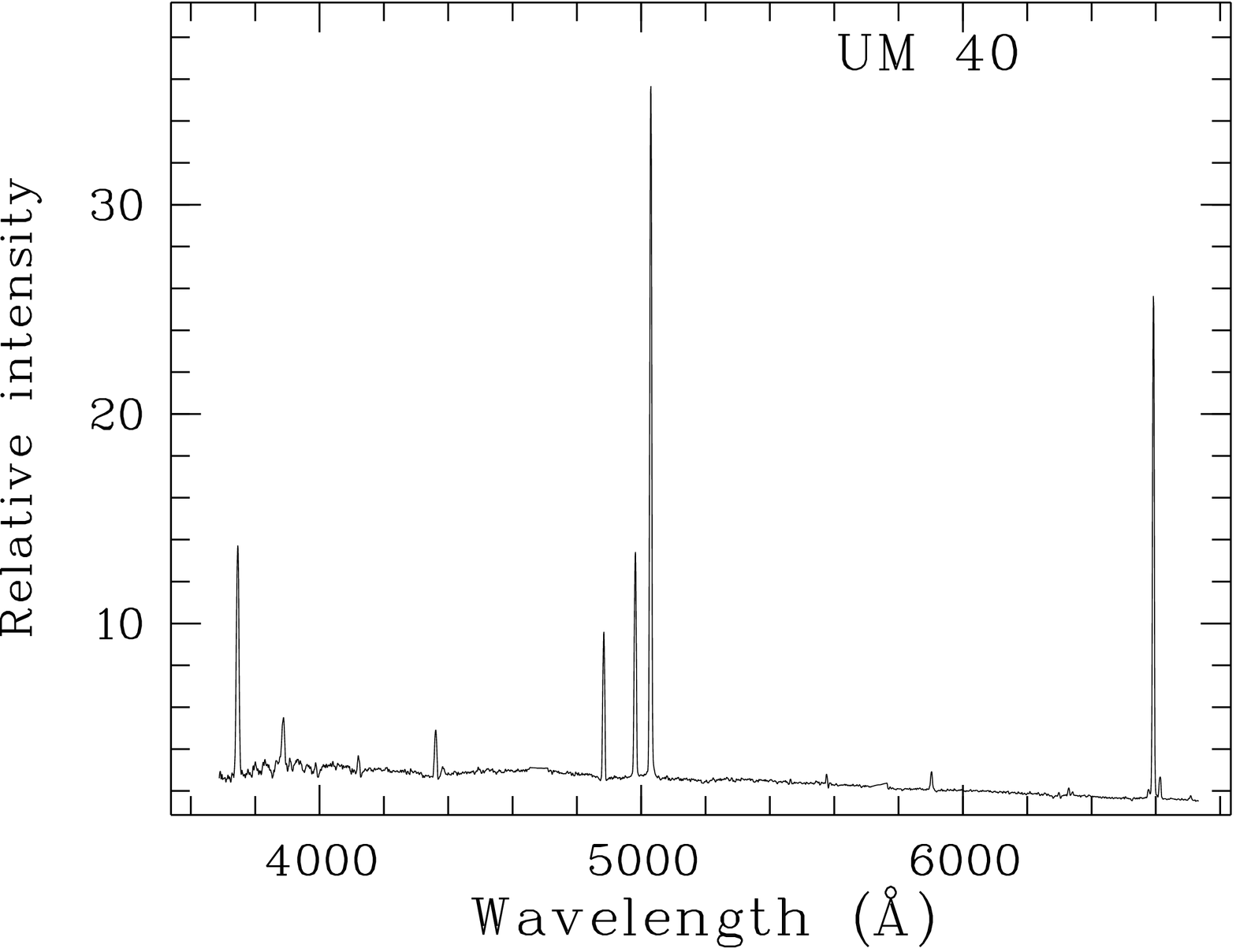}
\includegraphics[angle=-90,width=0.4\linewidth]{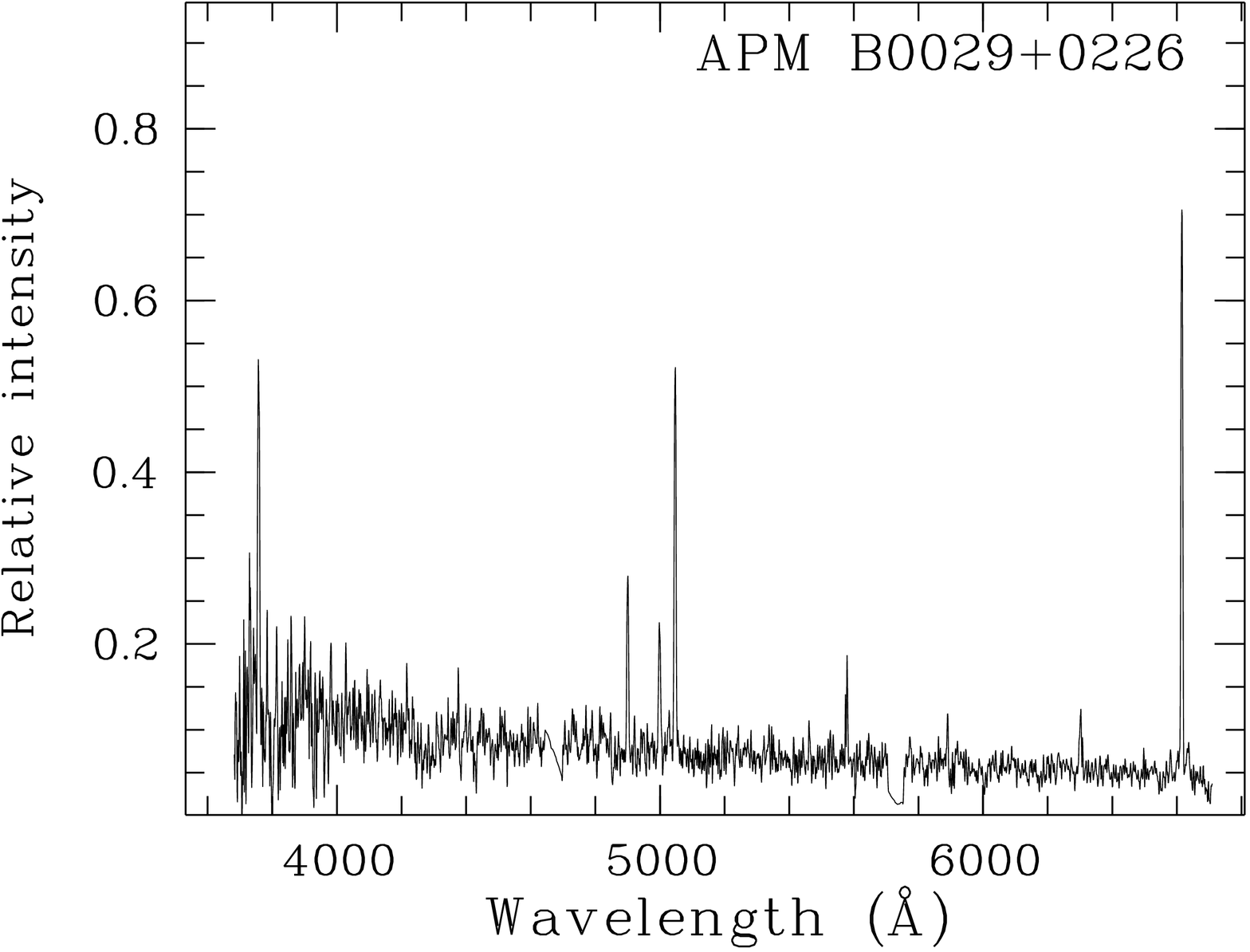}
\includegraphics[angle=-90,width=0.4\linewidth]{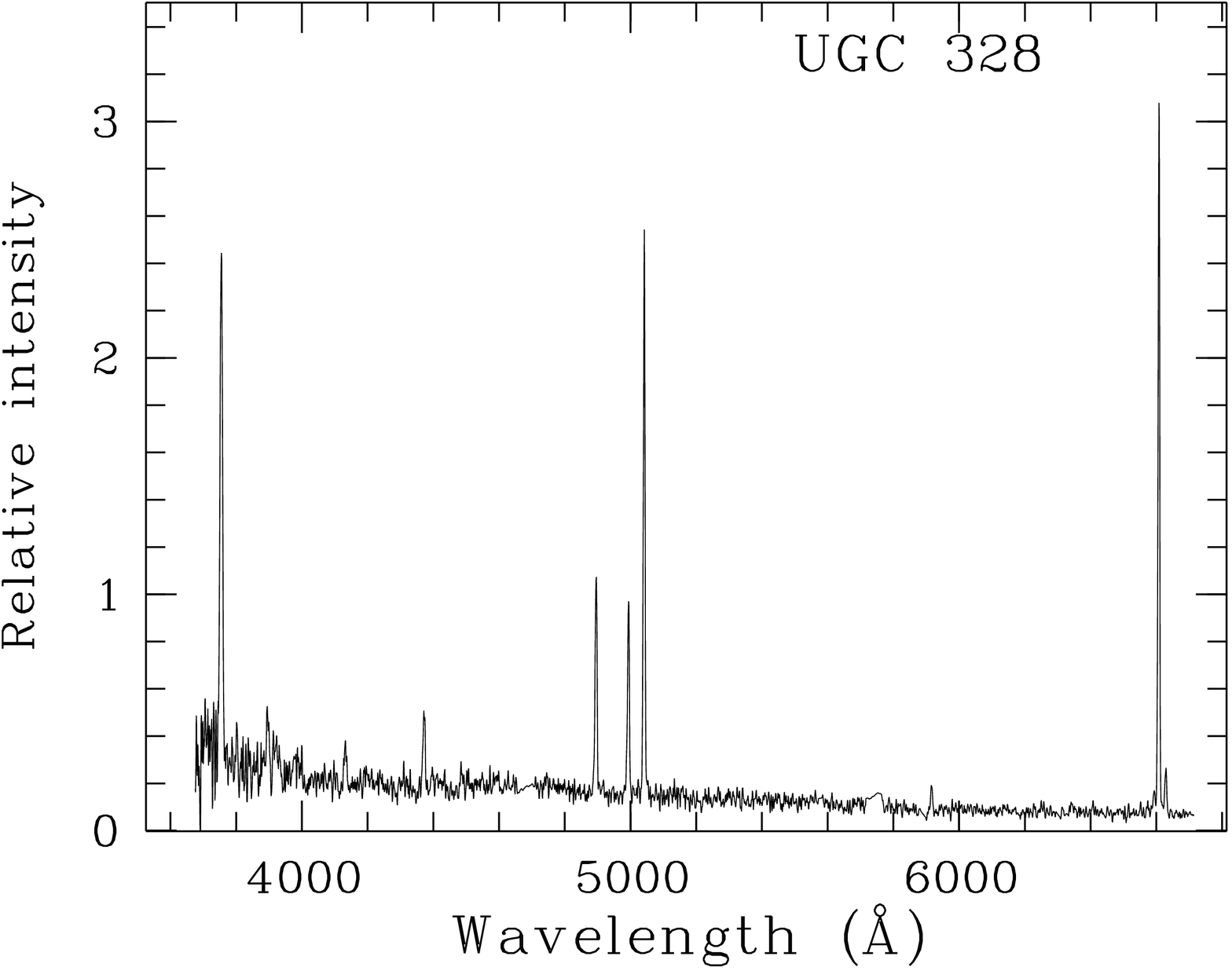}
\includegraphics[angle=-90,width=0.4\linewidth]{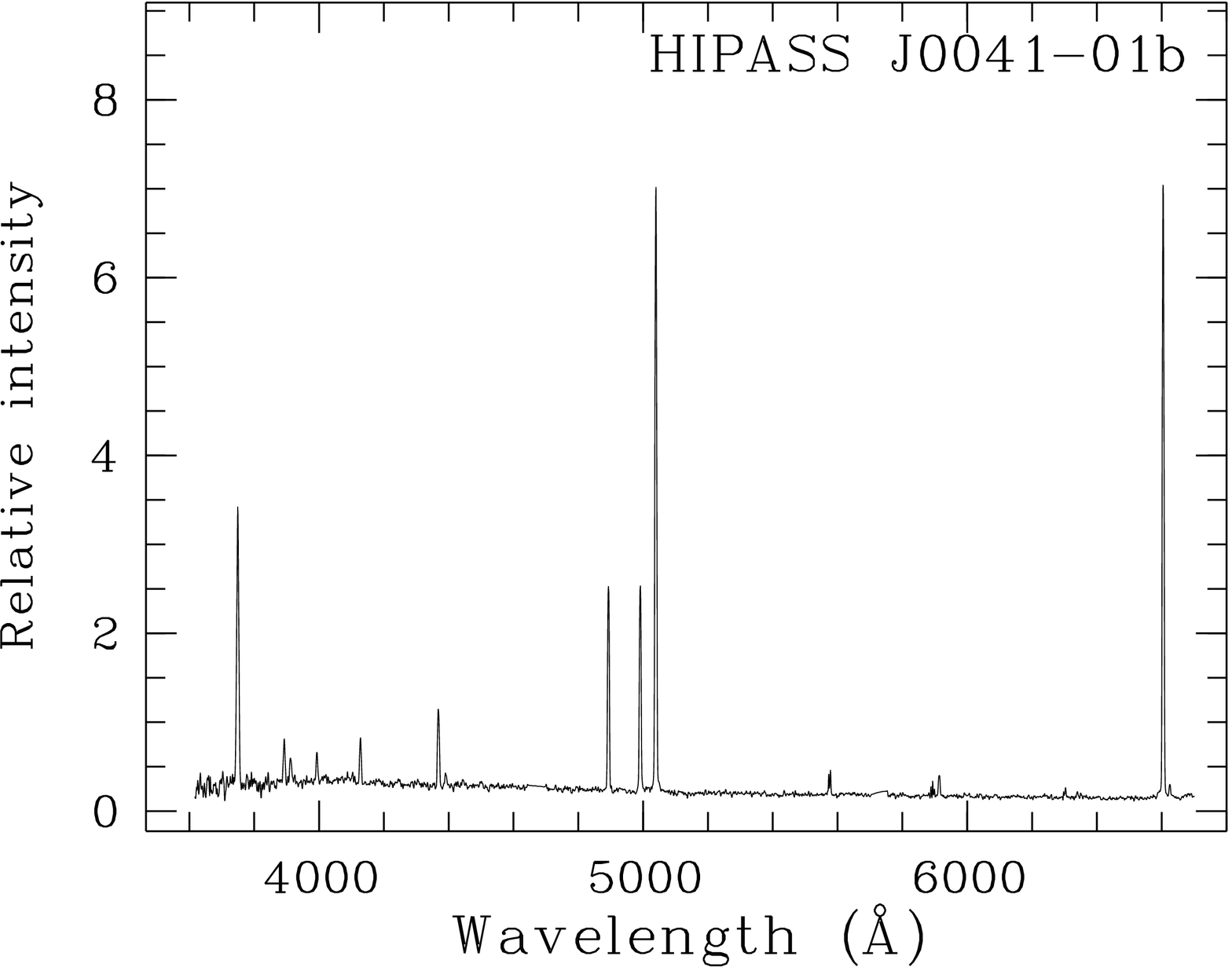}
\includegraphics[angle=-90,width=0.4\linewidth]{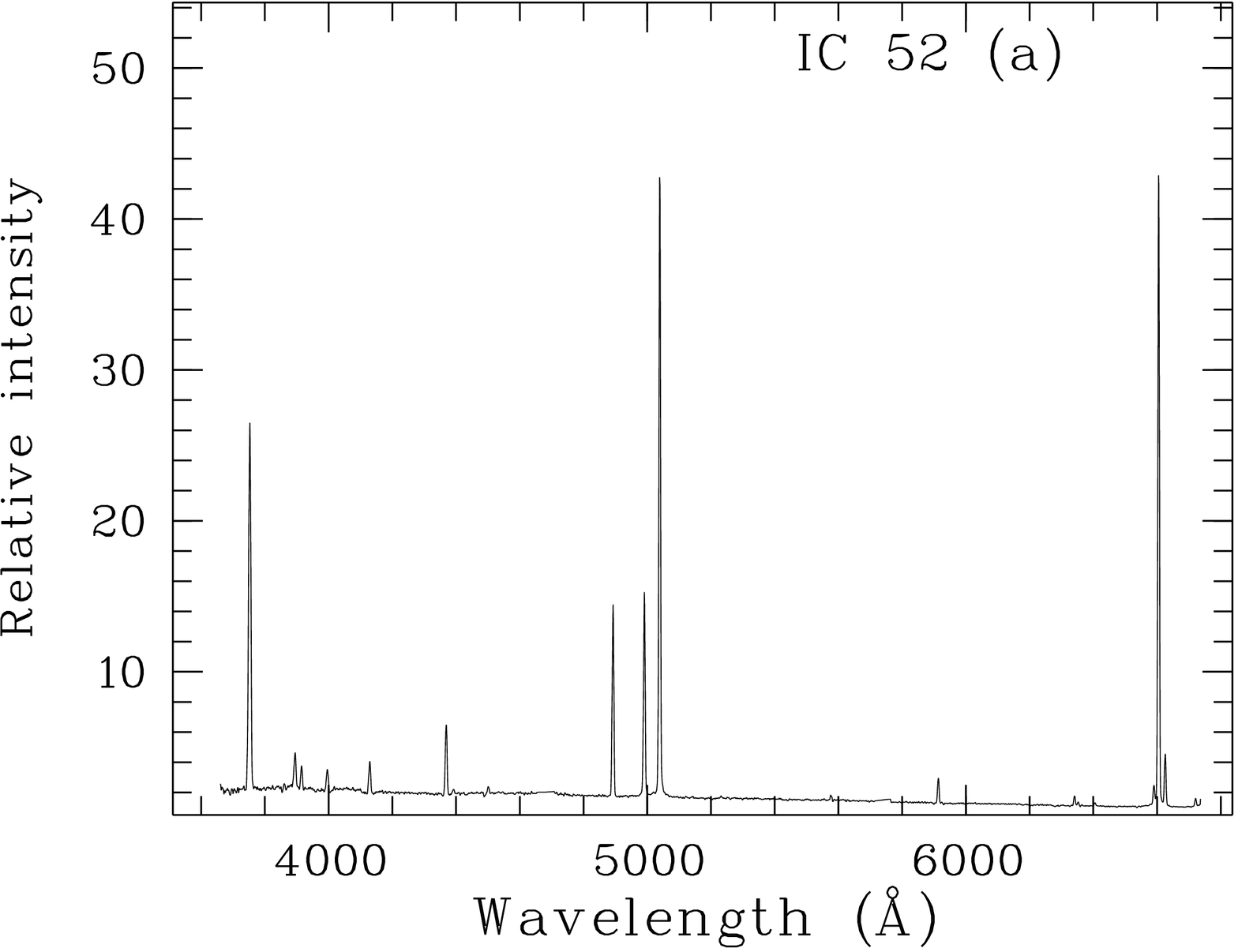}

\caption{\label{fig:SALTspecs1}
Spectra of 8 HII regions in 8 Eridanus void galaxies obtained with SALT.
}
\end{figure*}

\begin{figure*}
 \centering 
\includegraphics[angle=-90,width=0.4\linewidth]{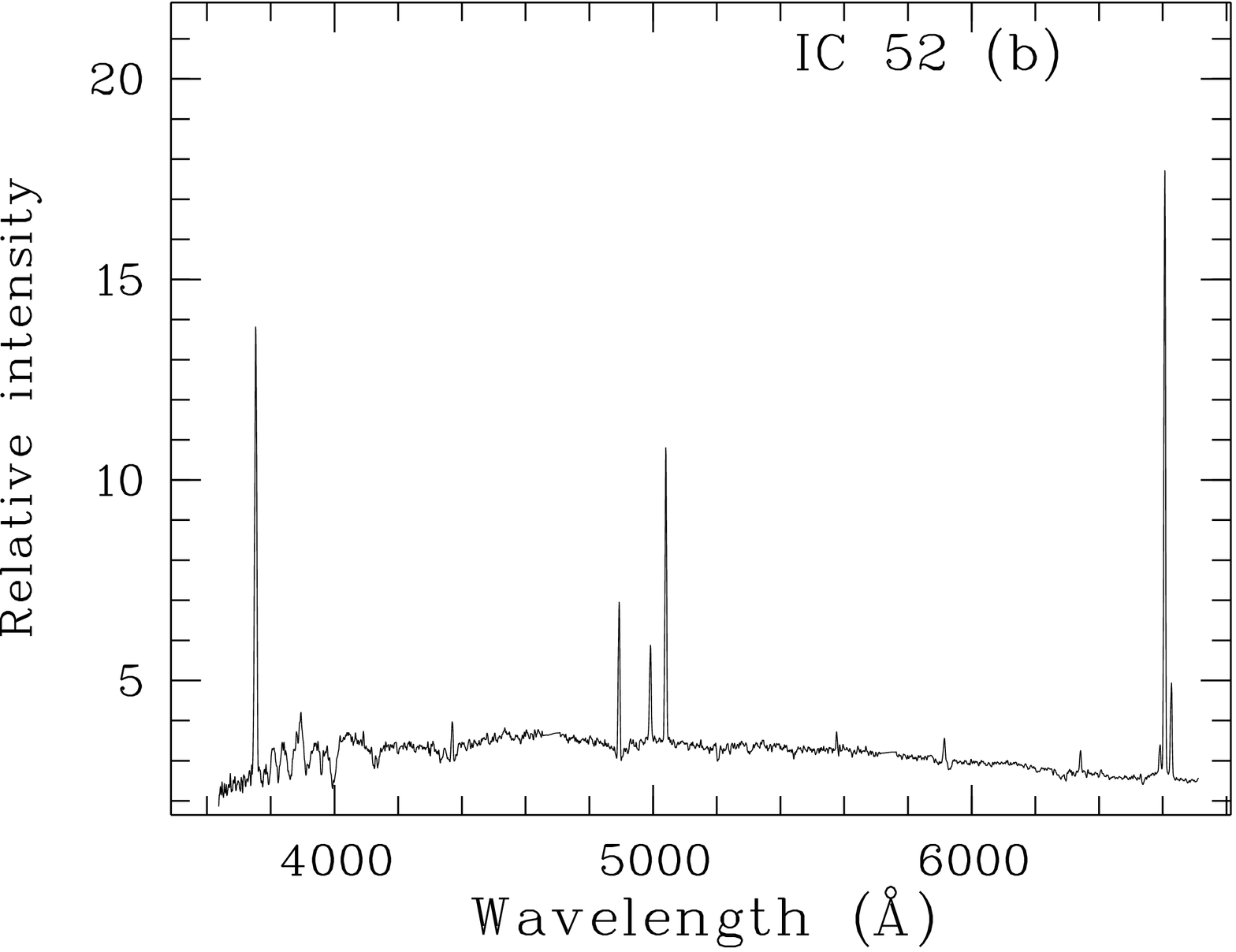}
\includegraphics[angle=-90,width=0.4\linewidth]{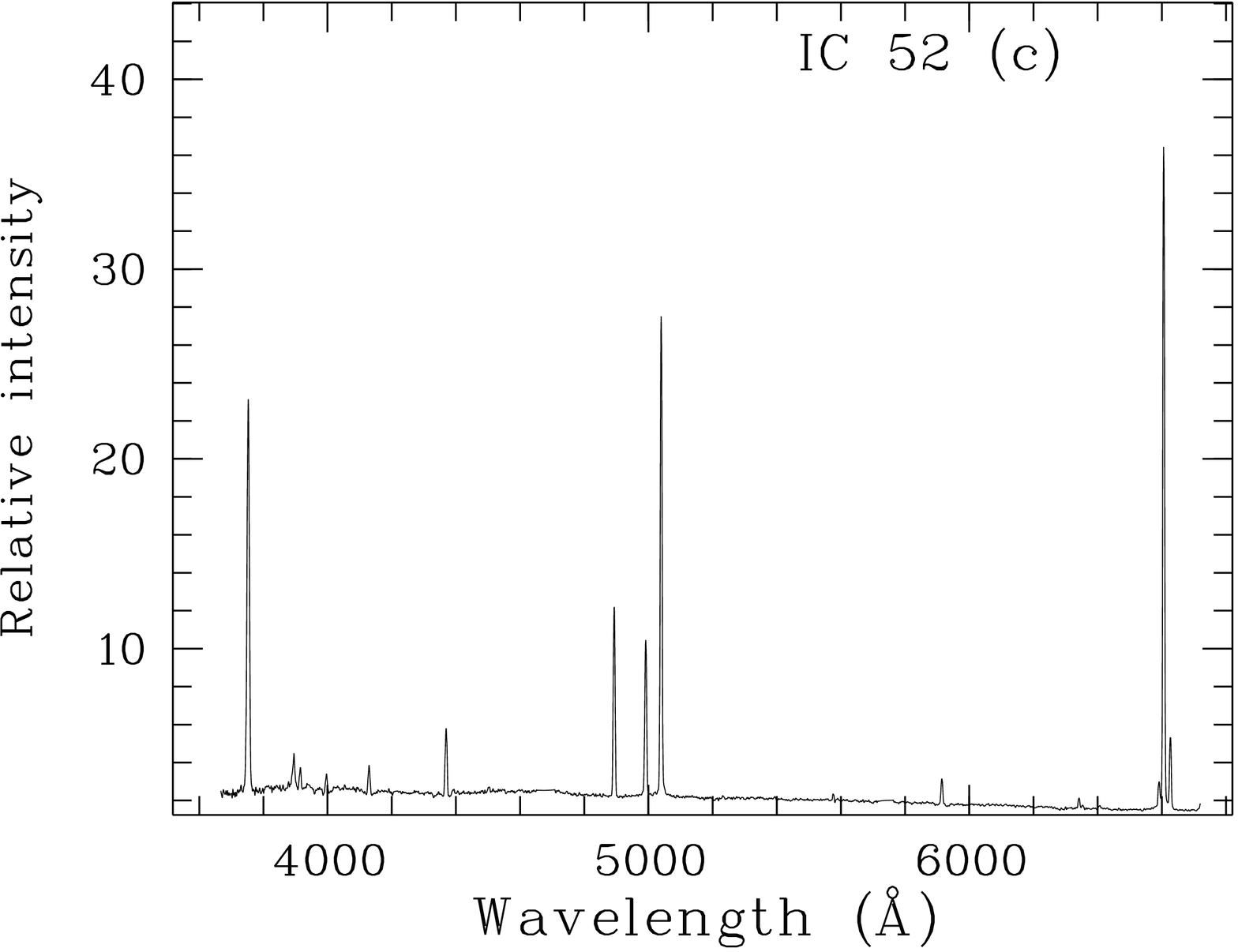}
\includegraphics[angle=-90,width=0.4\linewidth]{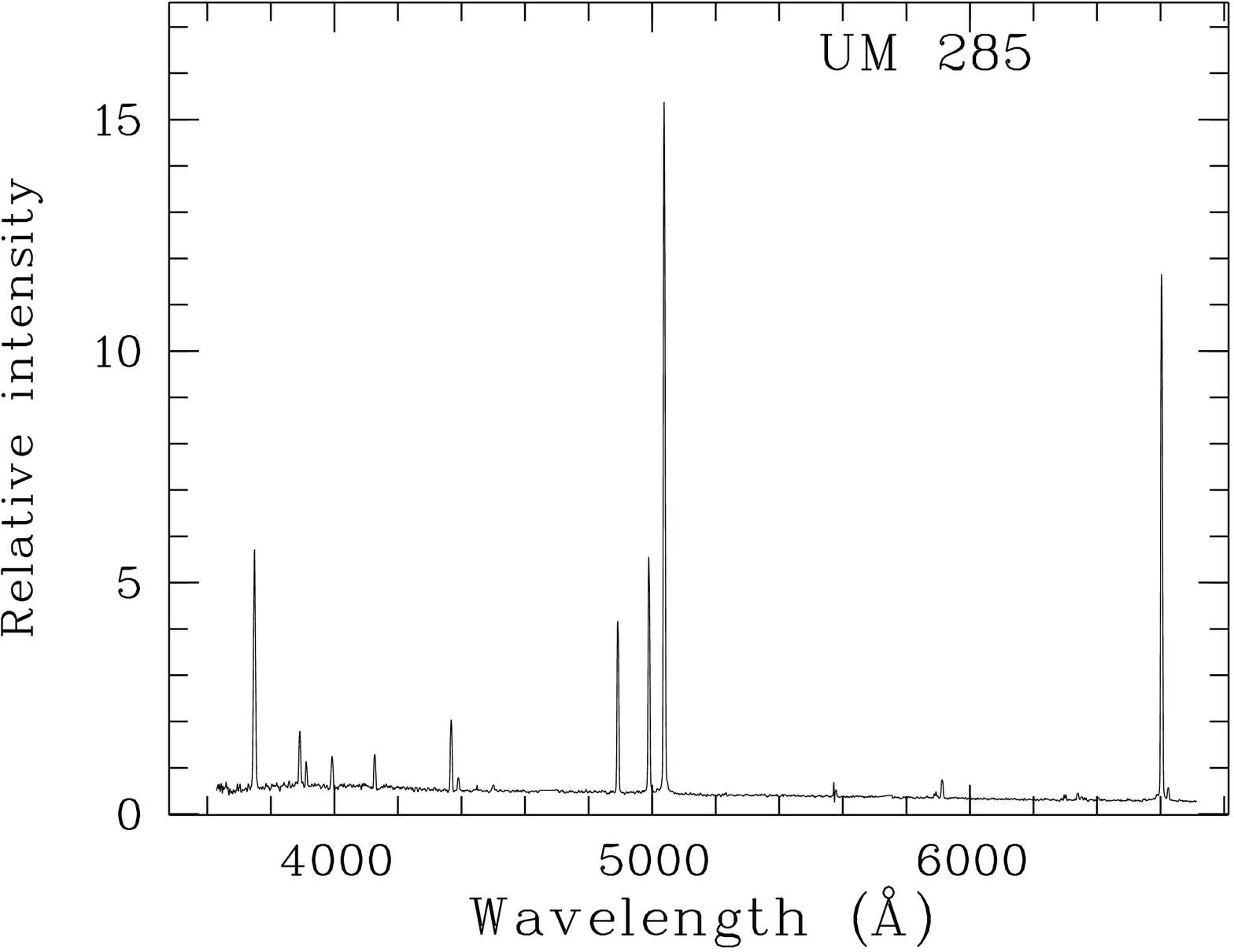}
\includegraphics[angle=-90,width=0.4\linewidth]{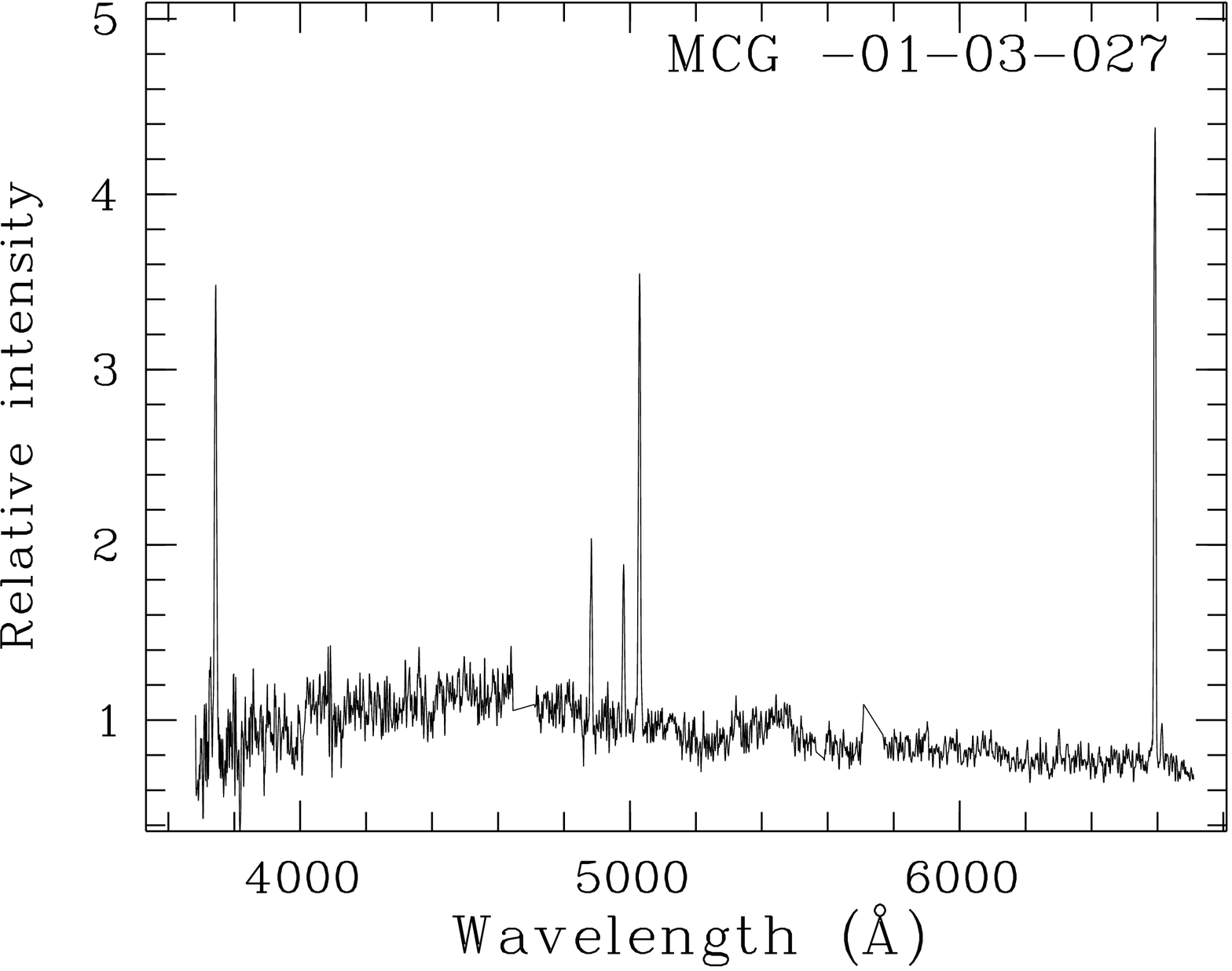}
\includegraphics[angle=-90,width=0.4\linewidth]{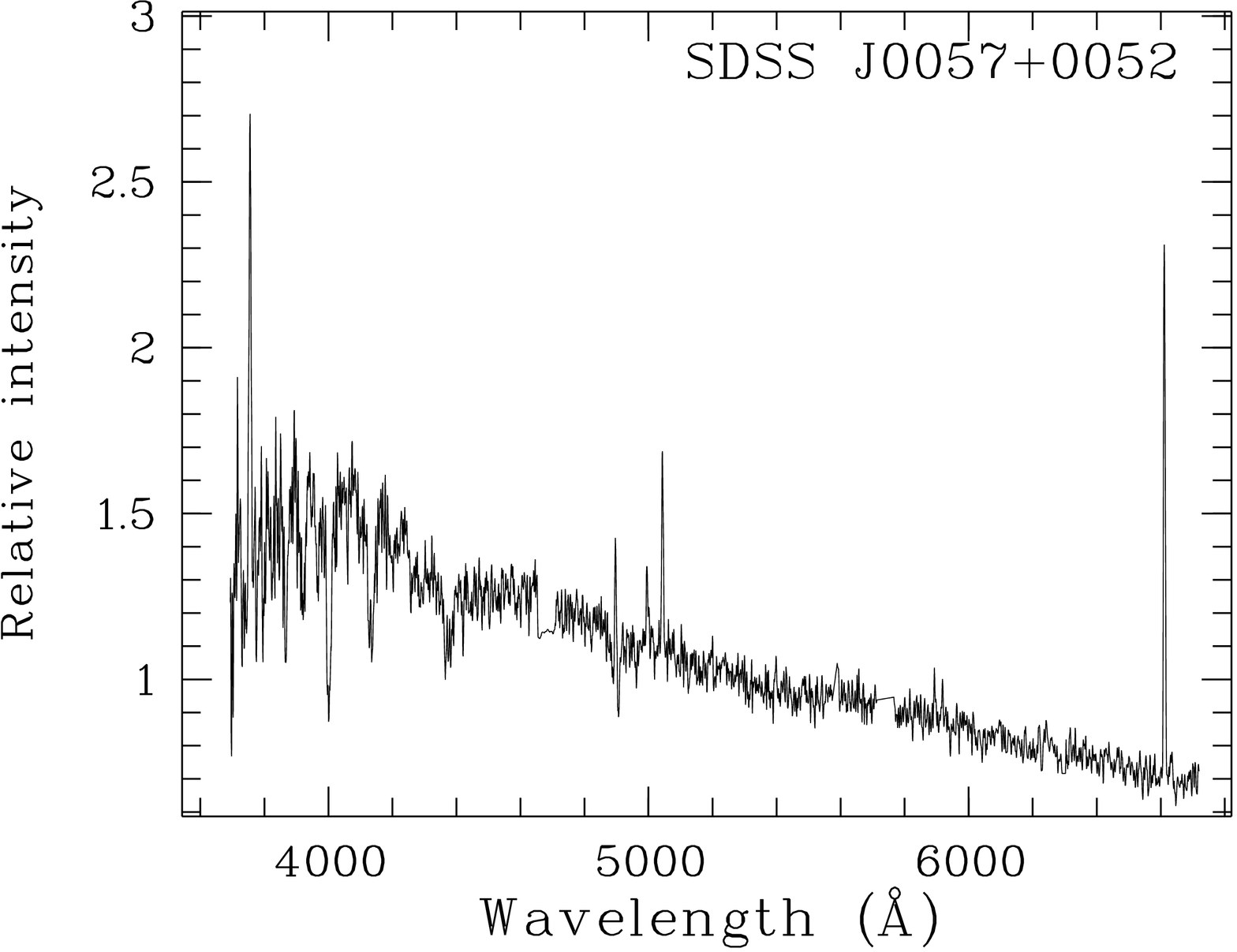}
\includegraphics[angle=-90,width=0.4\linewidth]{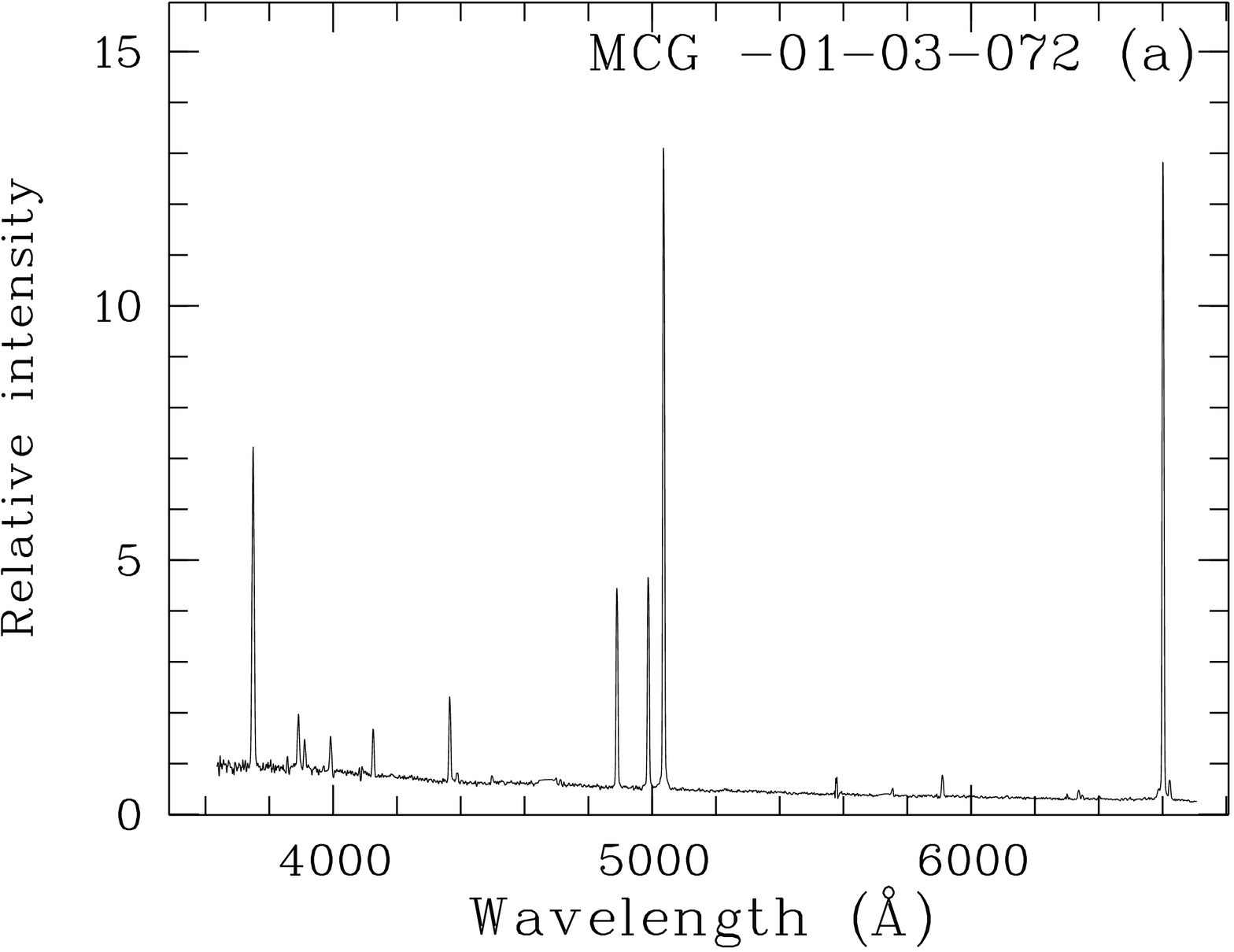}
\includegraphics[angle=-90,width=0.4\linewidth]{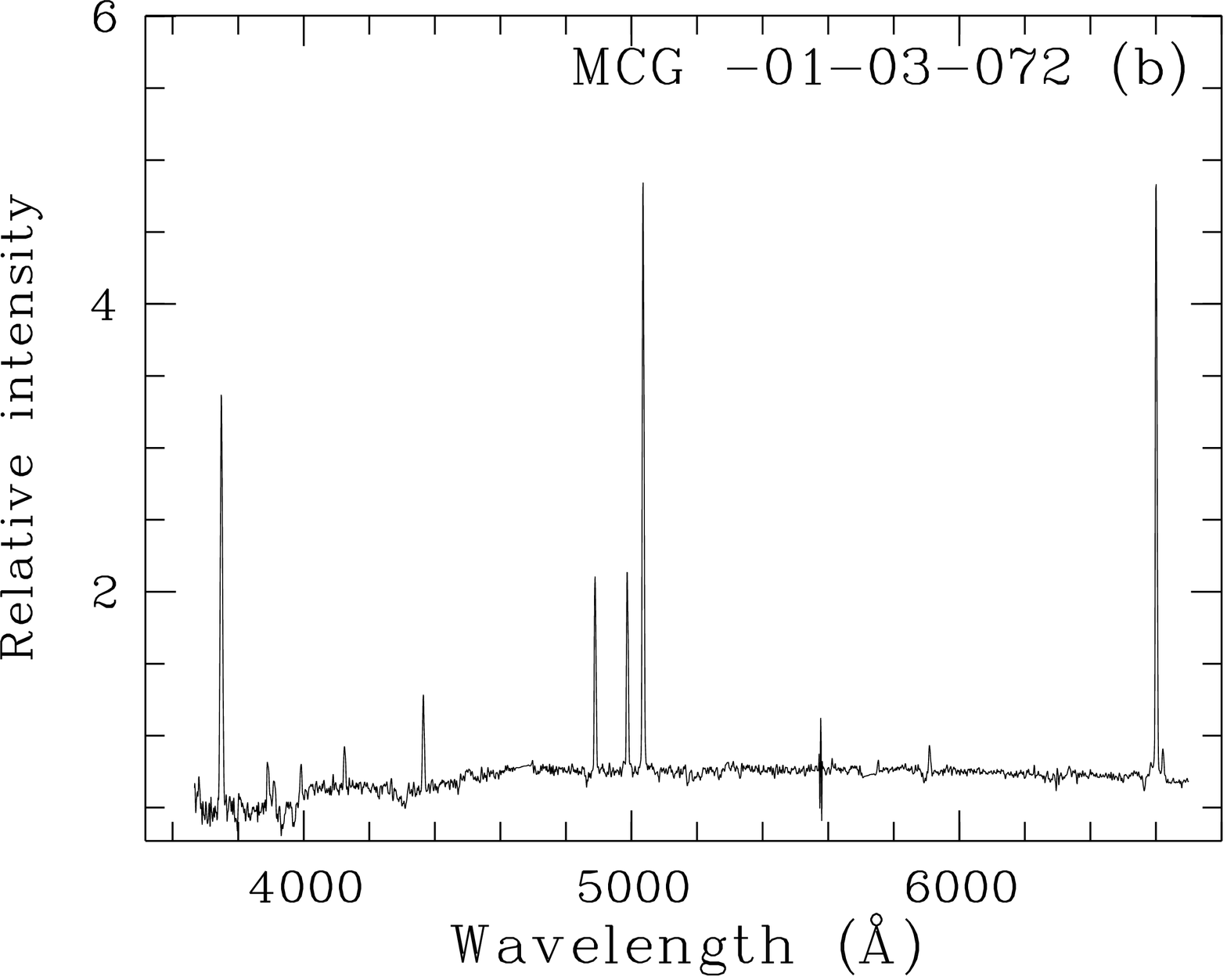}
\includegraphics[angle=-90,width=0.4\linewidth]{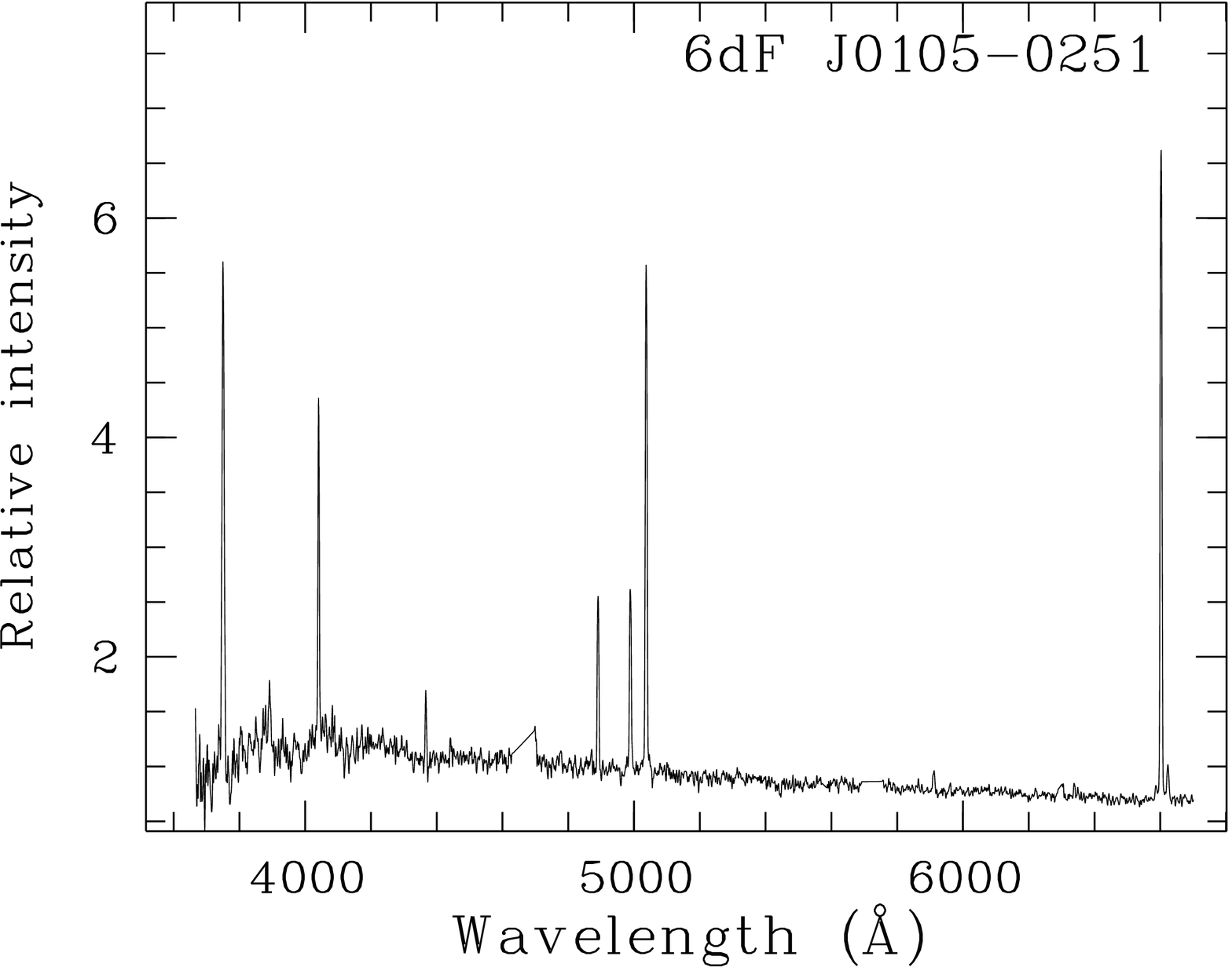}

\caption{\label{fig:SALTspecs2}
Spectra of 8 HII regions in 8 Eridanus void galaxies obtained with SALT.
}
\end{figure*}

\begin{figure*}
 \centering 
\includegraphics[angle=-90,width=0.4\linewidth]{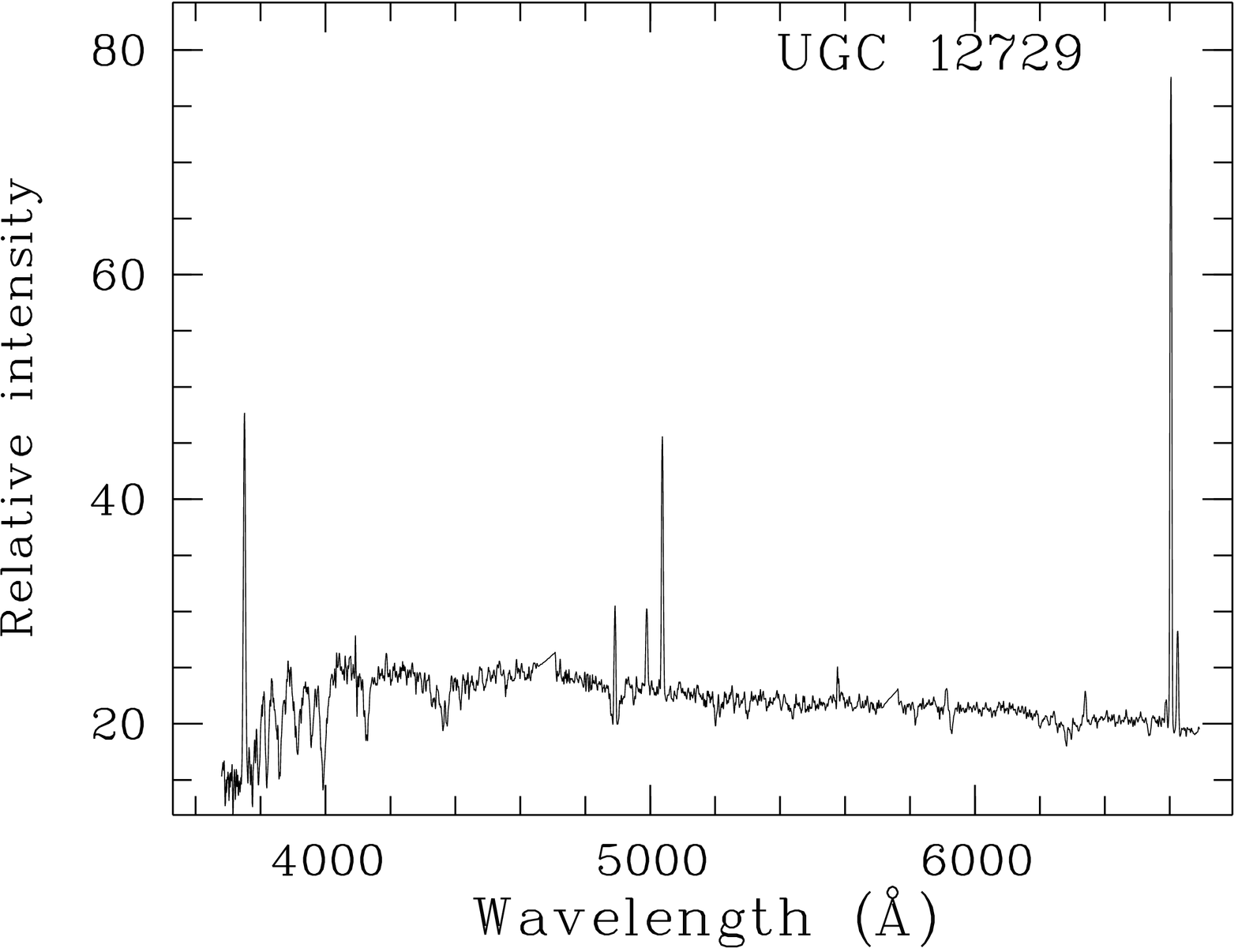}
\includegraphics[angle=-90,width=0.4\linewidth]{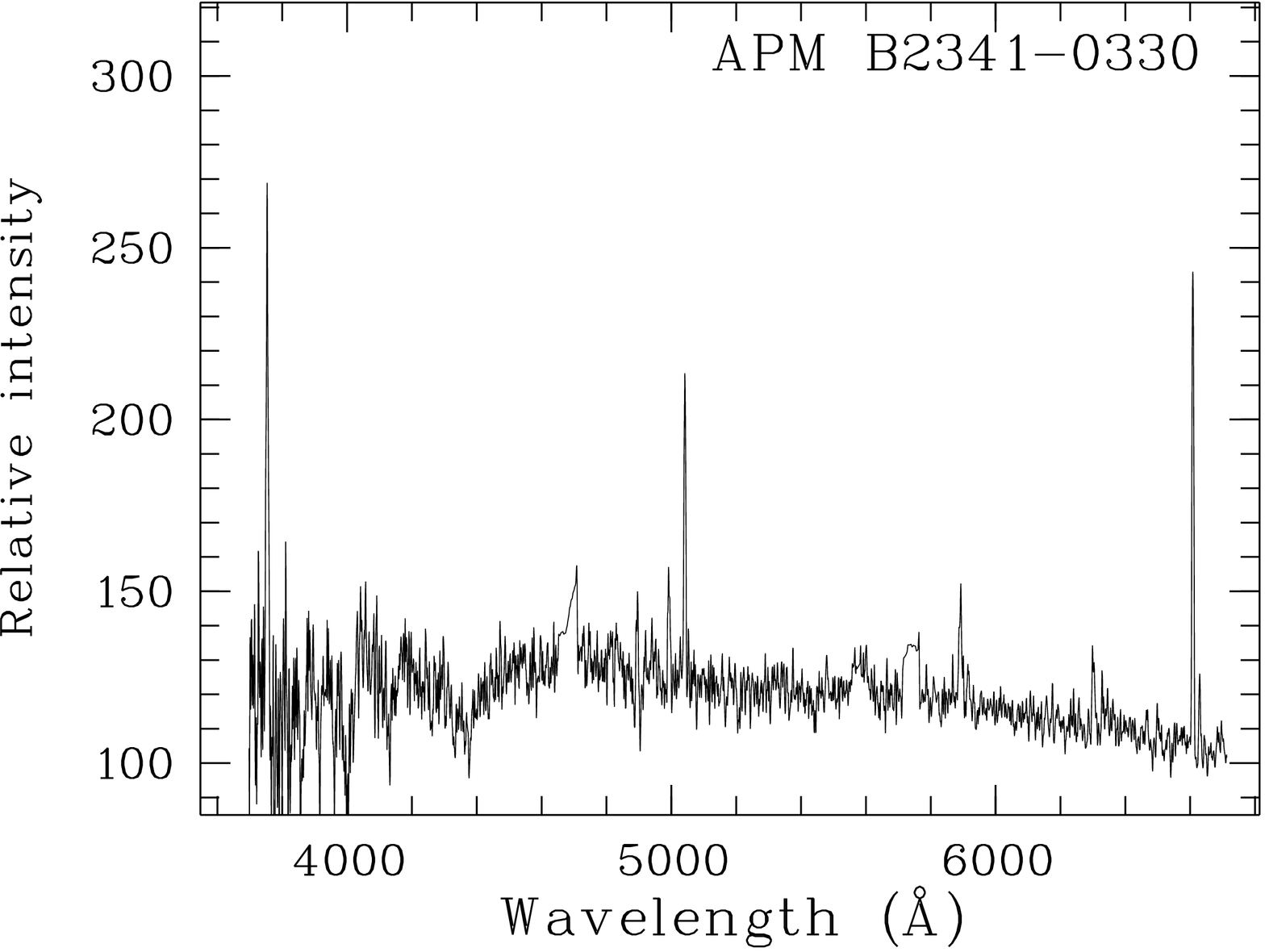}
\includegraphics[angle=-90,width=0.4\linewidth]{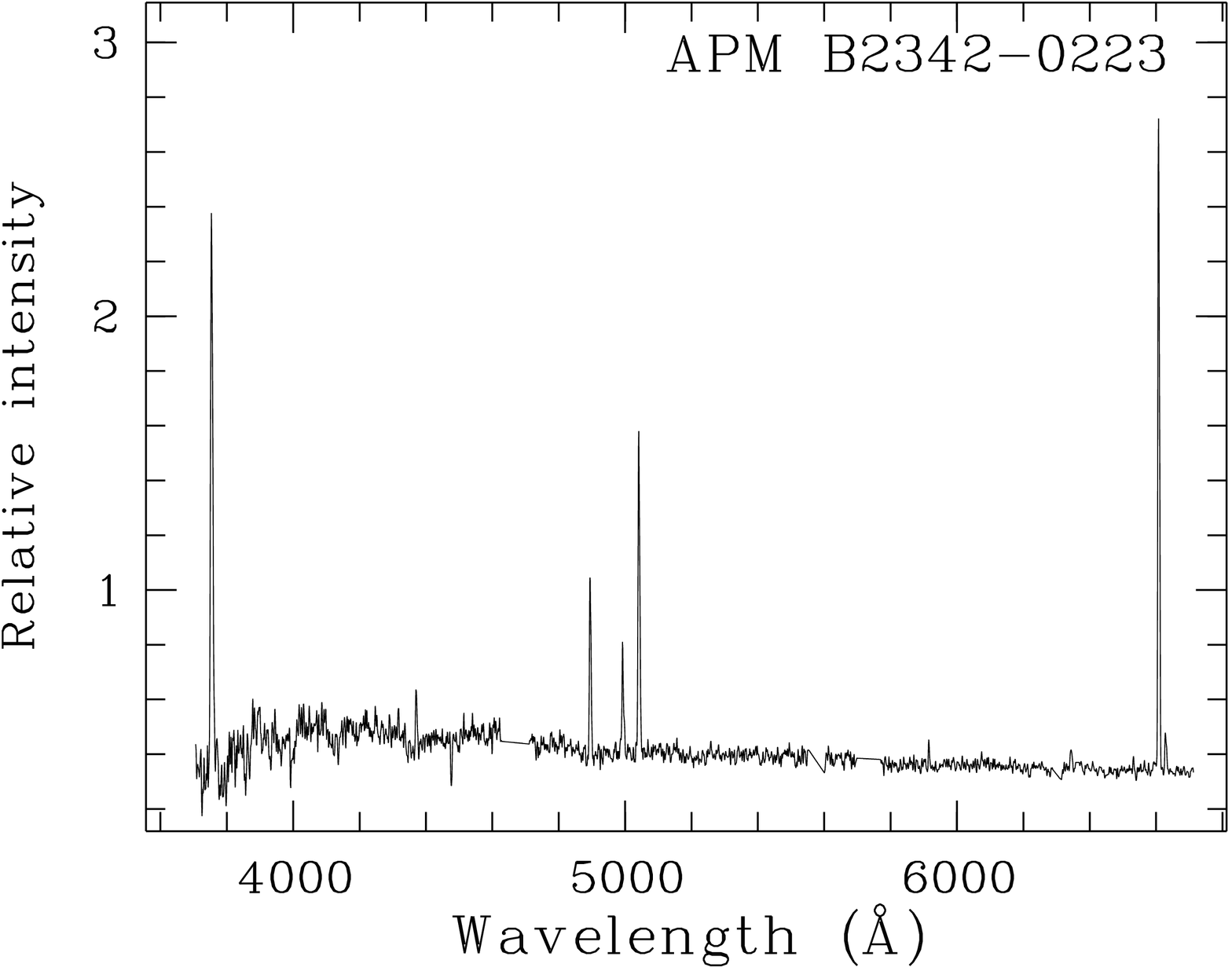}
\includegraphics[angle=-90,width=0.4\linewidth]{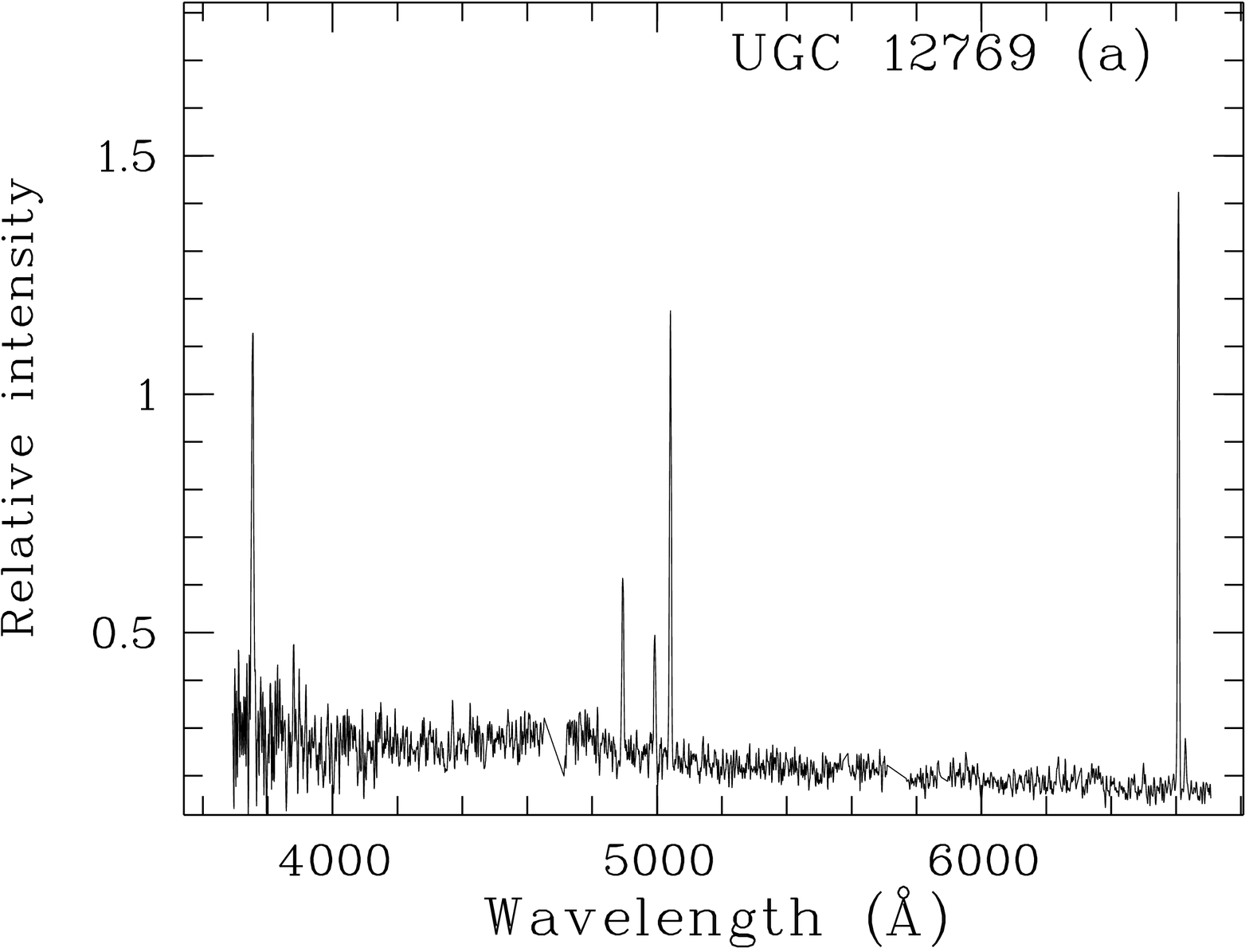}
\includegraphics[angle=-90,width=0.4\linewidth]{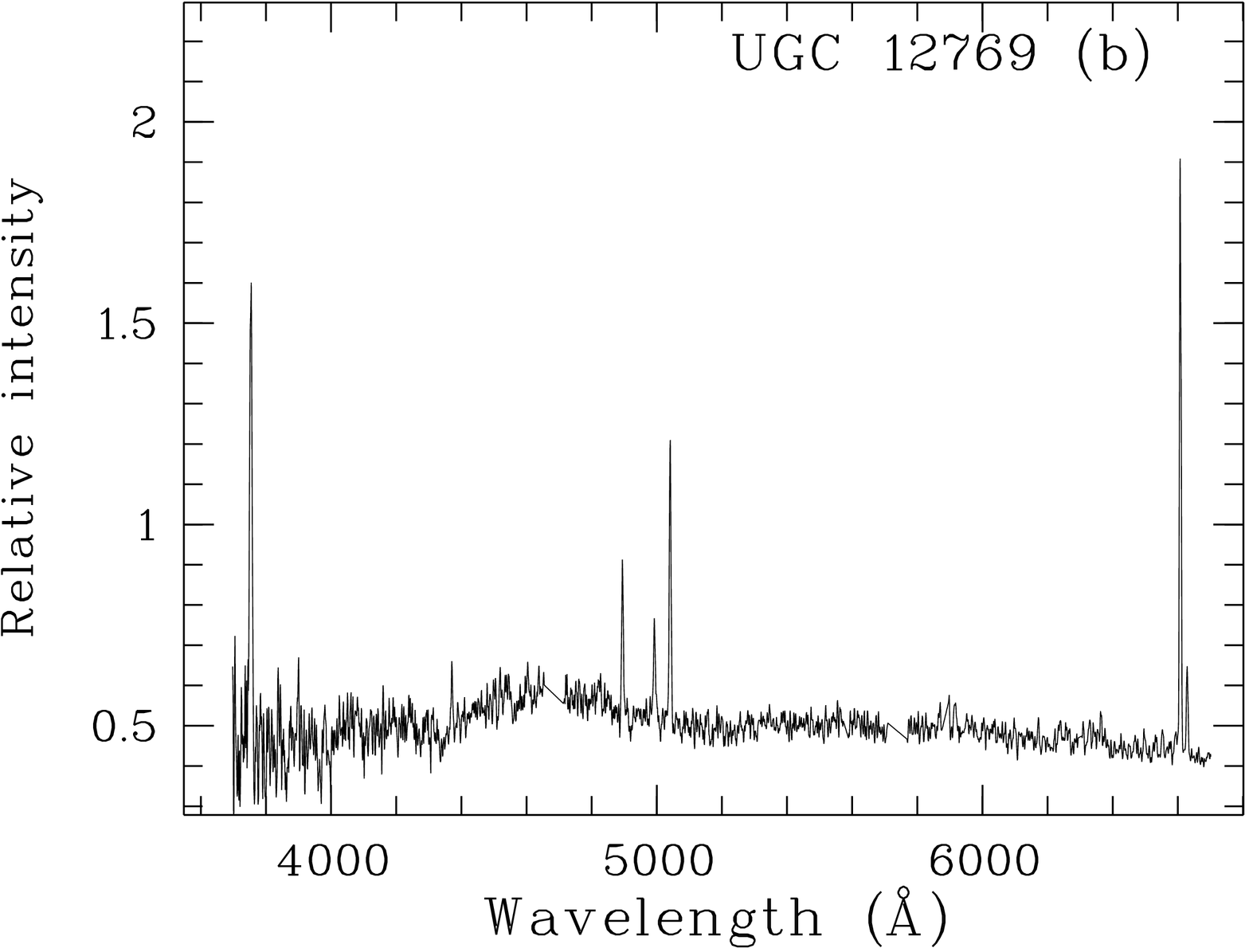}
\includegraphics[angle=-90,width=0.4\linewidth]{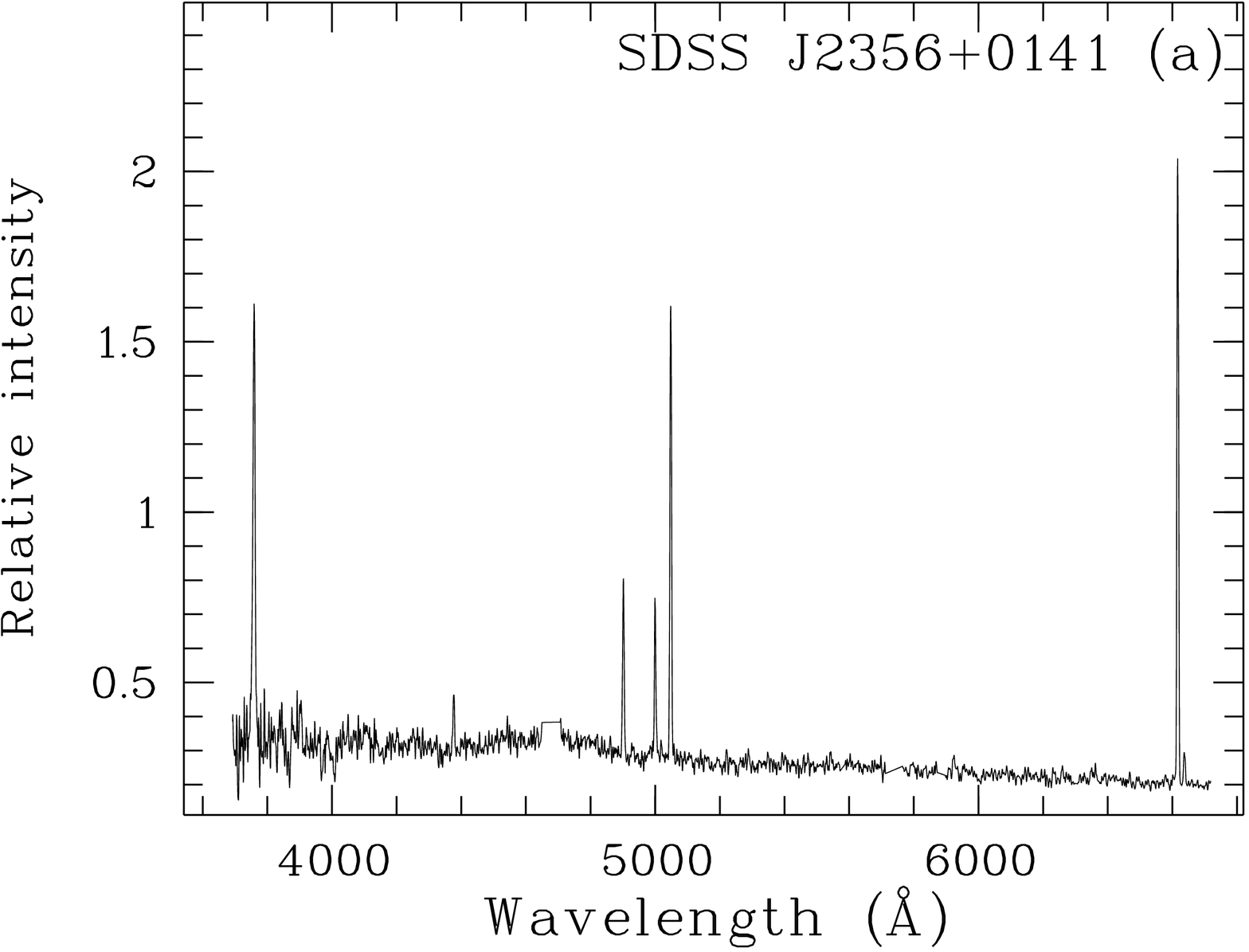}
\includegraphics[angle=-90,width=0.4\linewidth]{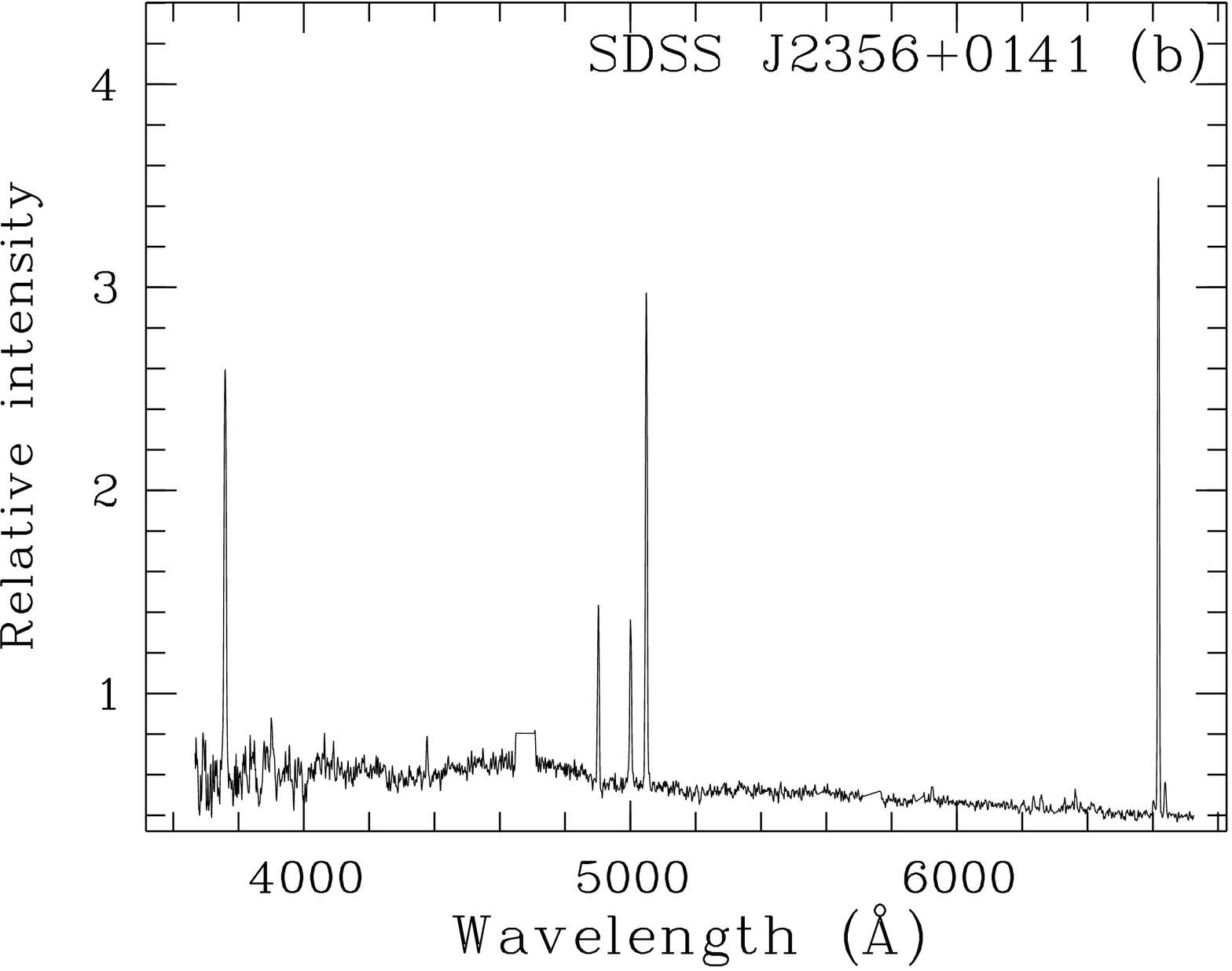}
\includegraphics[angle=-90,width=0.4\linewidth]{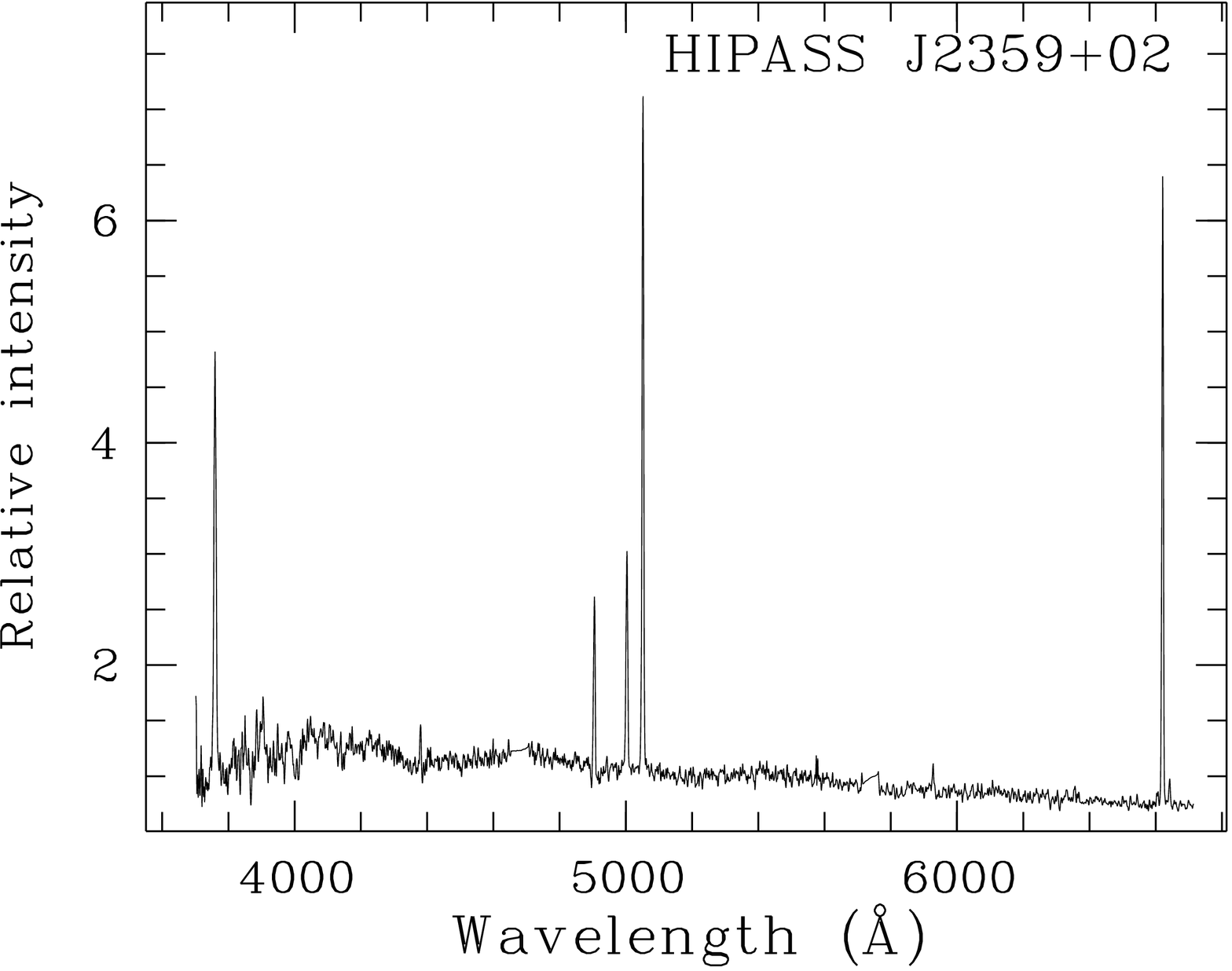}

\caption{\label{fig:SALTspecs3}
Spectra of 8 HII regions in 8 Eridanus void galaxies obtained with SALT.
}
\end{figure*}

\begin{figure*}
 \centering 
\includegraphics[angle=-90,width=0.4\linewidth]{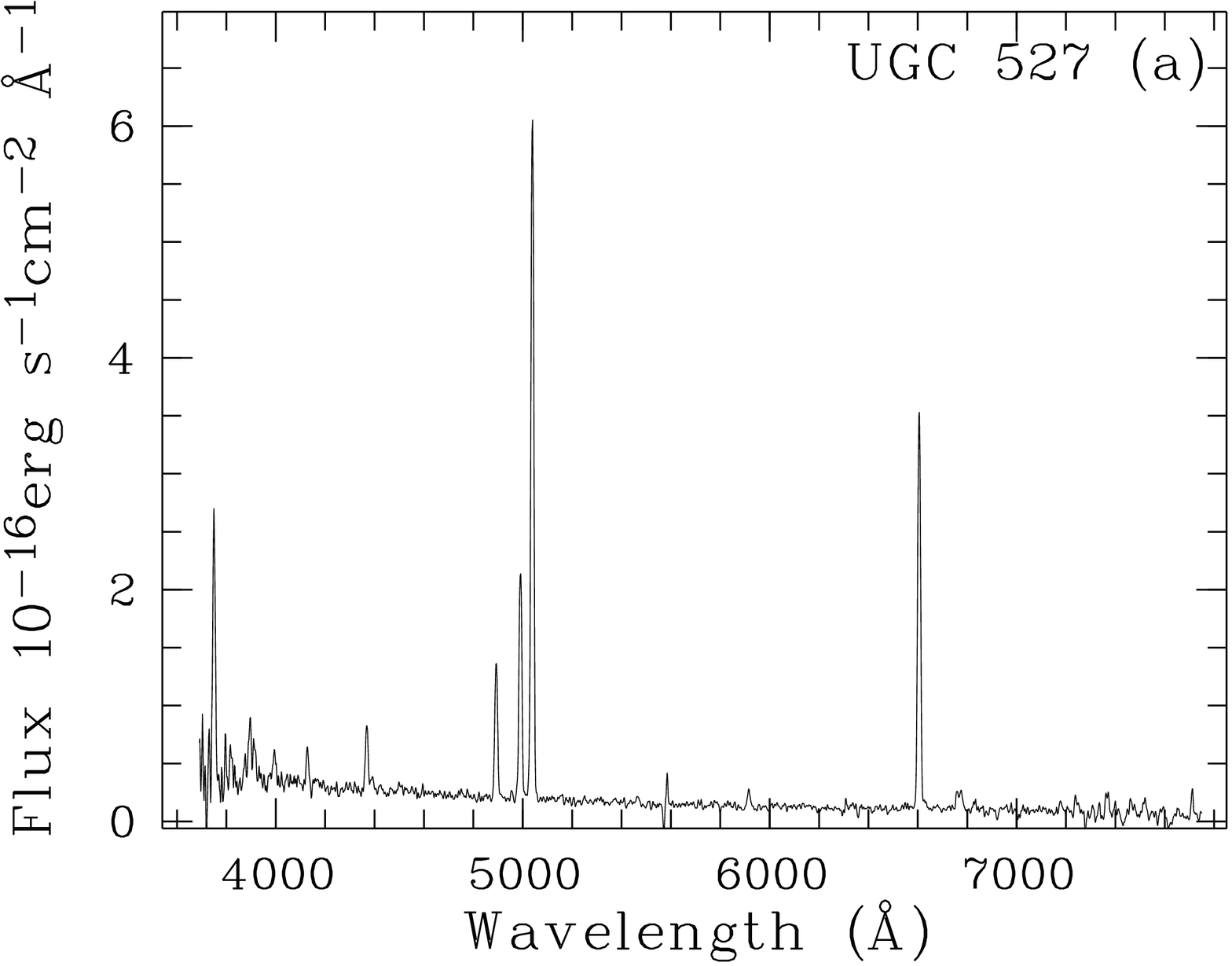}
\includegraphics[angle=-90,width=0.4\linewidth]{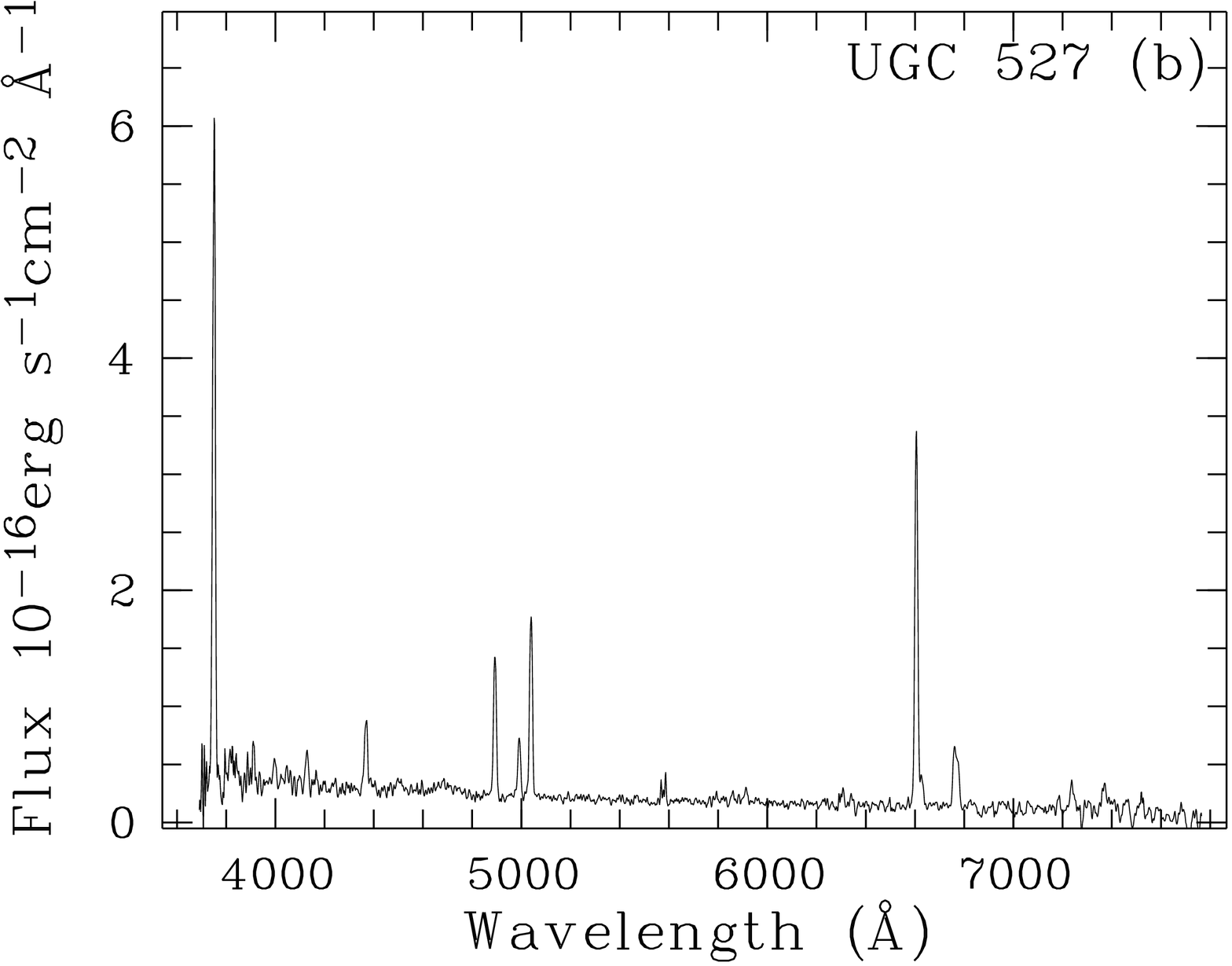}
\includegraphics[angle=-90,width=0.4\linewidth]{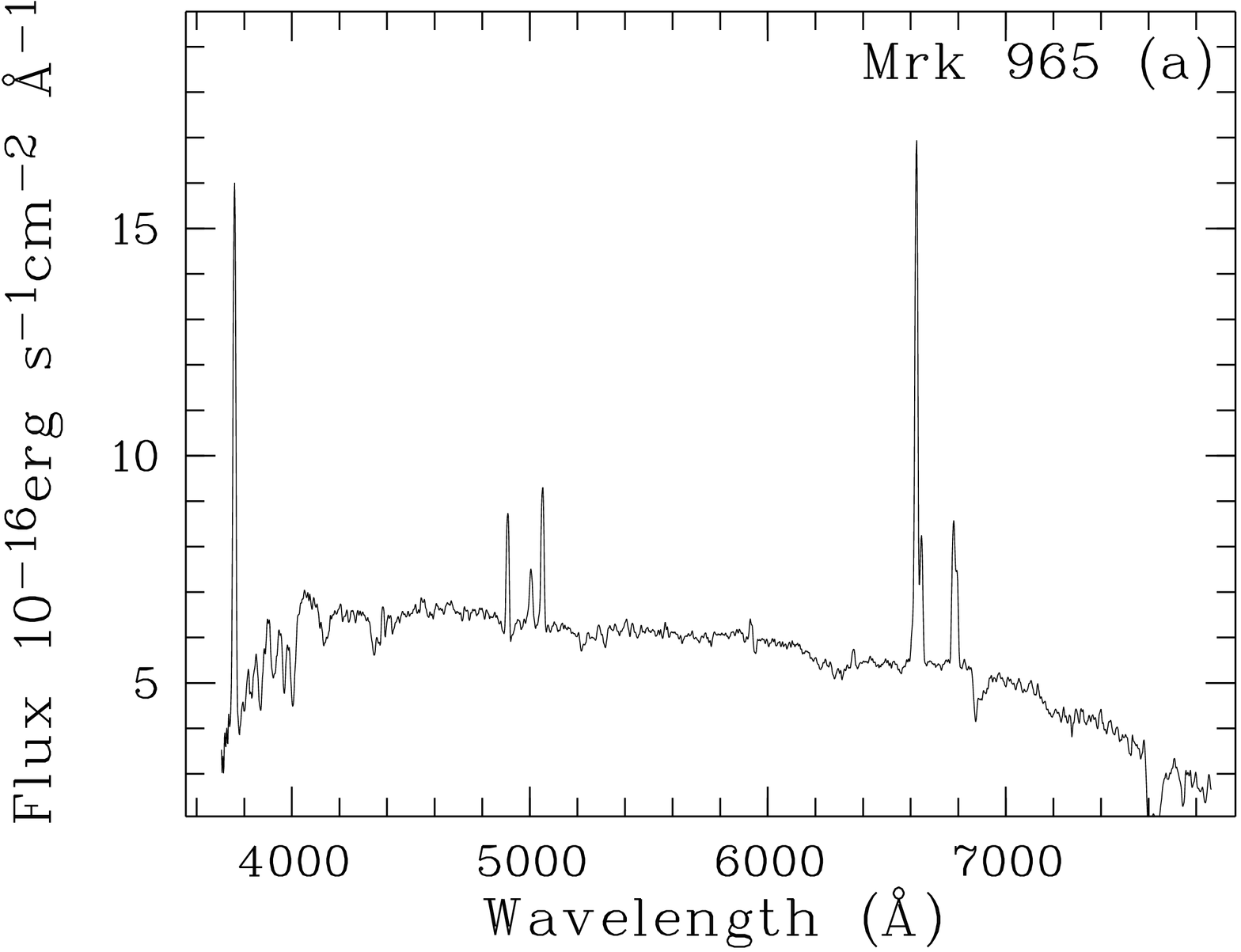}
\includegraphics[angle=-90,width=0.4\linewidth]{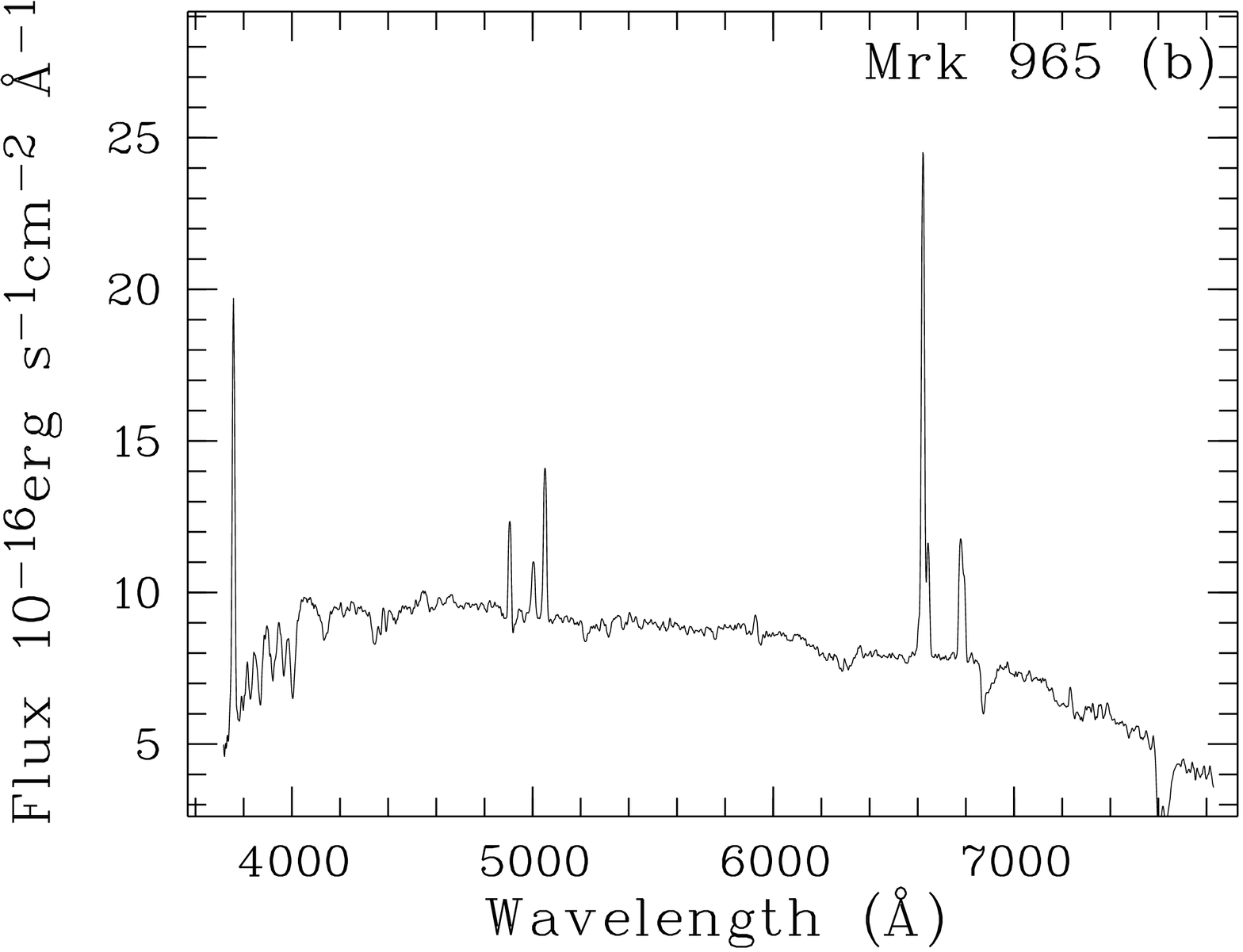}
\includegraphics[angle=-90,width=0.4\linewidth]{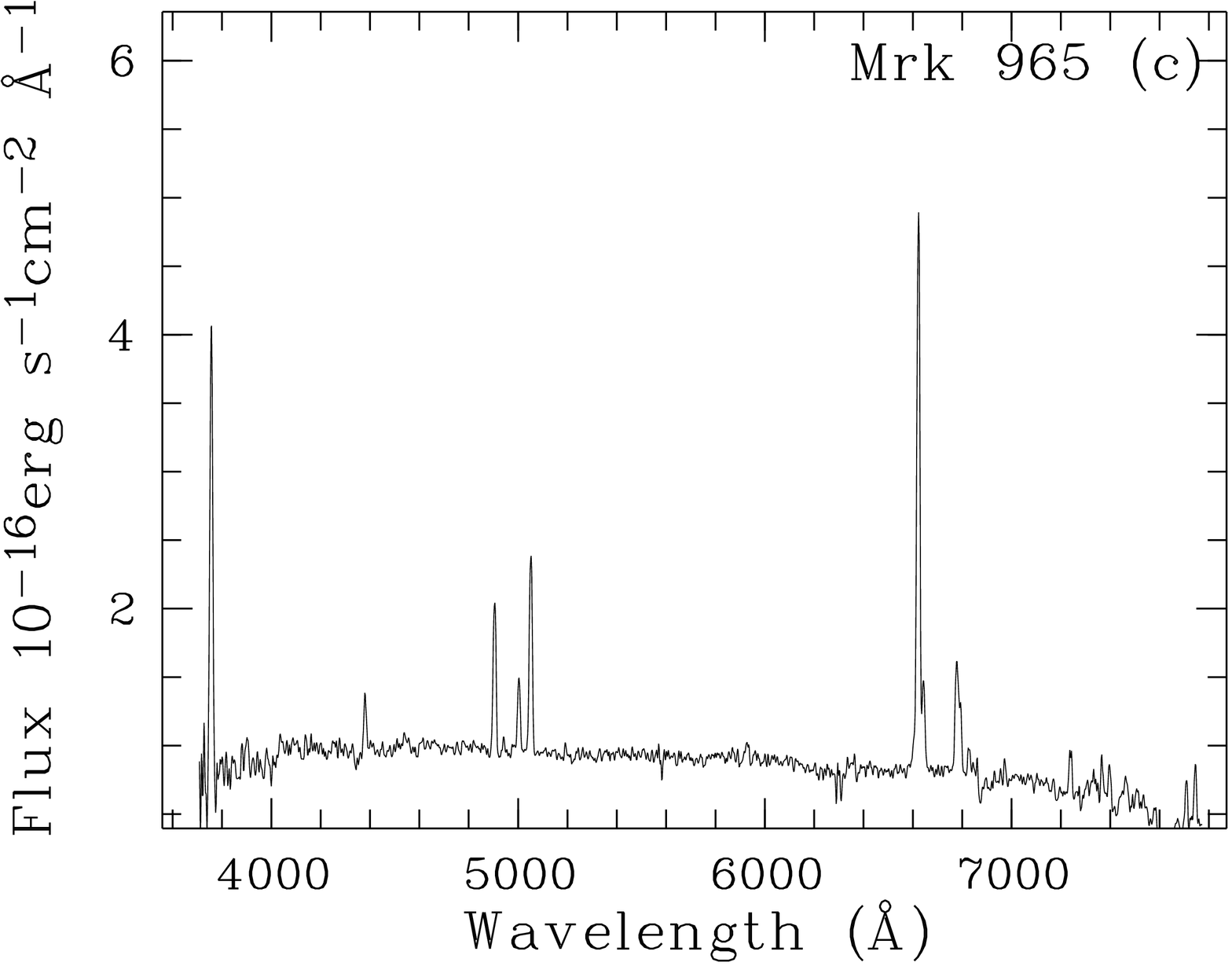}
\includegraphics[angle=-90,width=0.4\linewidth]{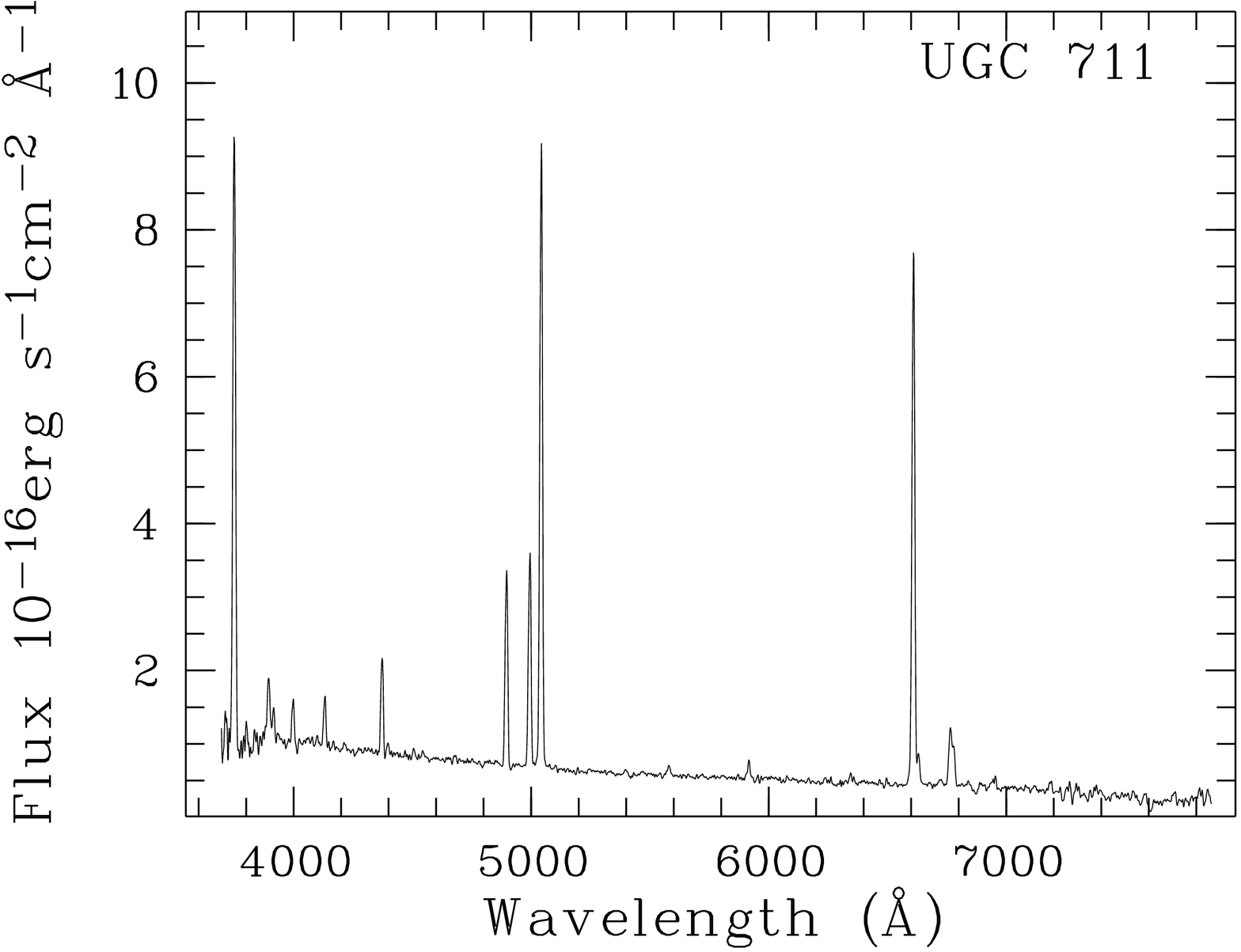}
\includegraphics[angle=-90,width=0.4\linewidth]{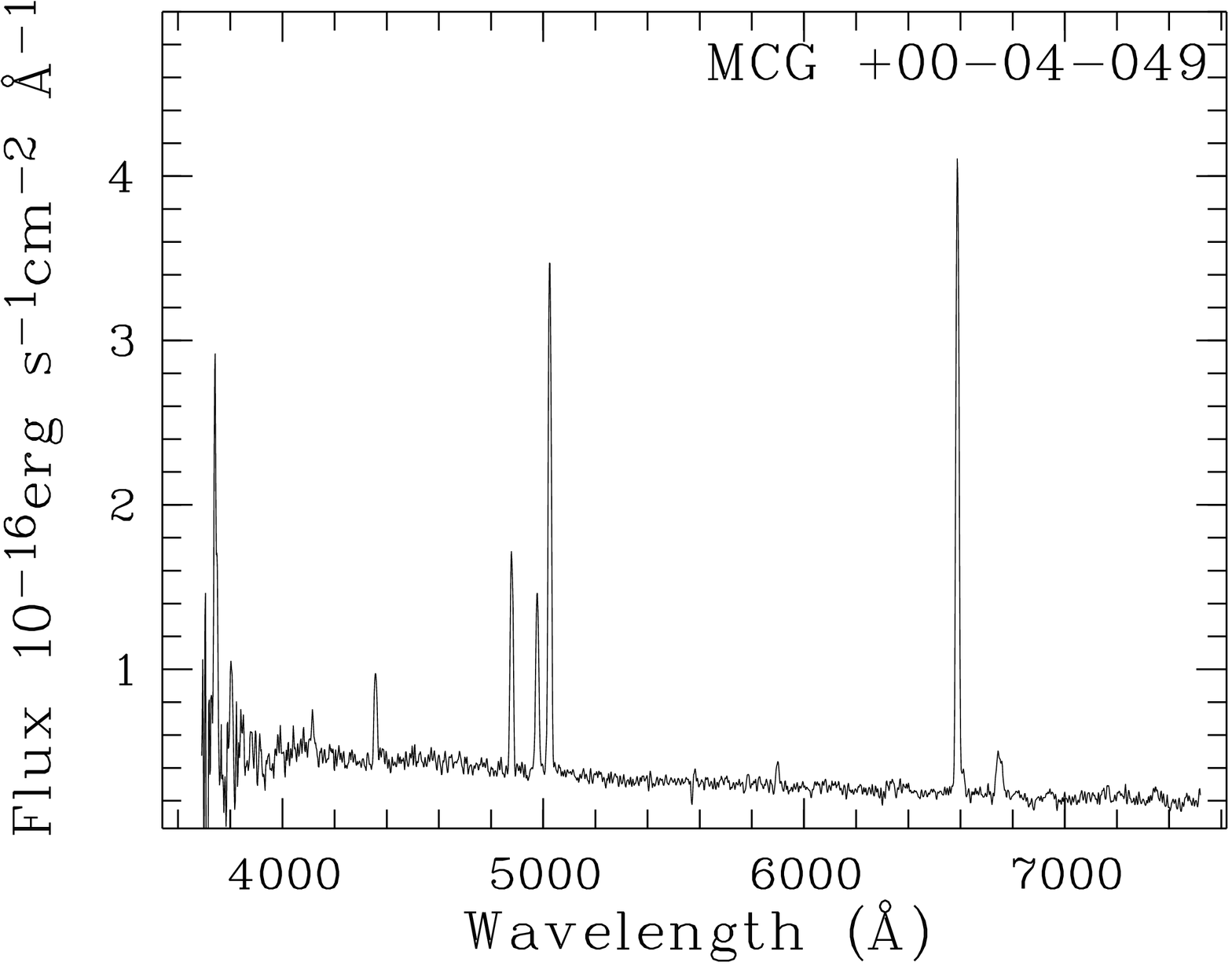}

\caption{\label{fig:BTAspecs}
Spectra of 7 HII regions in 4 Eridanus void galaxies obtained with BTA.
}
\end{figure*}

The measured emission-line intensities with their errors for Eridanus
void galaxies
are presented in Tables \ref{t:Intens_SALT1}-\ref{t:Intens_MCG} for
SALT spectra and
in Tables~\ref{t:Intens_BTA1}-\ref{t:Intens_BTA3} for BTA spectra.
The measured line flux values in the $H\beta$ line are given
in units of $10^{-16}$~erg~s$^{-1}$~cm$^{-2}$.
The 1D spectra of all objects studied are presented in
Figs~\ref{fig:SALTspecs1}-\ref{fig:SALTspecs3} (SALT spectra)
and Fig~\ref{fig:BTAspecs} (BTA spectra).

\subsection{SDSS data}

We use spectra from the SDSS DR12 data base for two void galaxies (NGC 428 and Ark 18). For the detailed description of SDSS observations and their
principles see \citet{Gunn98}, \citet{York2000} and \citet{DR7}.
Three more galaxies (SDSS J0142+0018, SDSS J0142+0021, SDSS J0143-0002)
have available SDSS spectra of poor quality, so reliable estimates of
the oxygen abundances are not possible.

\begin{figure*}
 \centering 
 \includegraphics[angle=-90,width=0.4\linewidth]{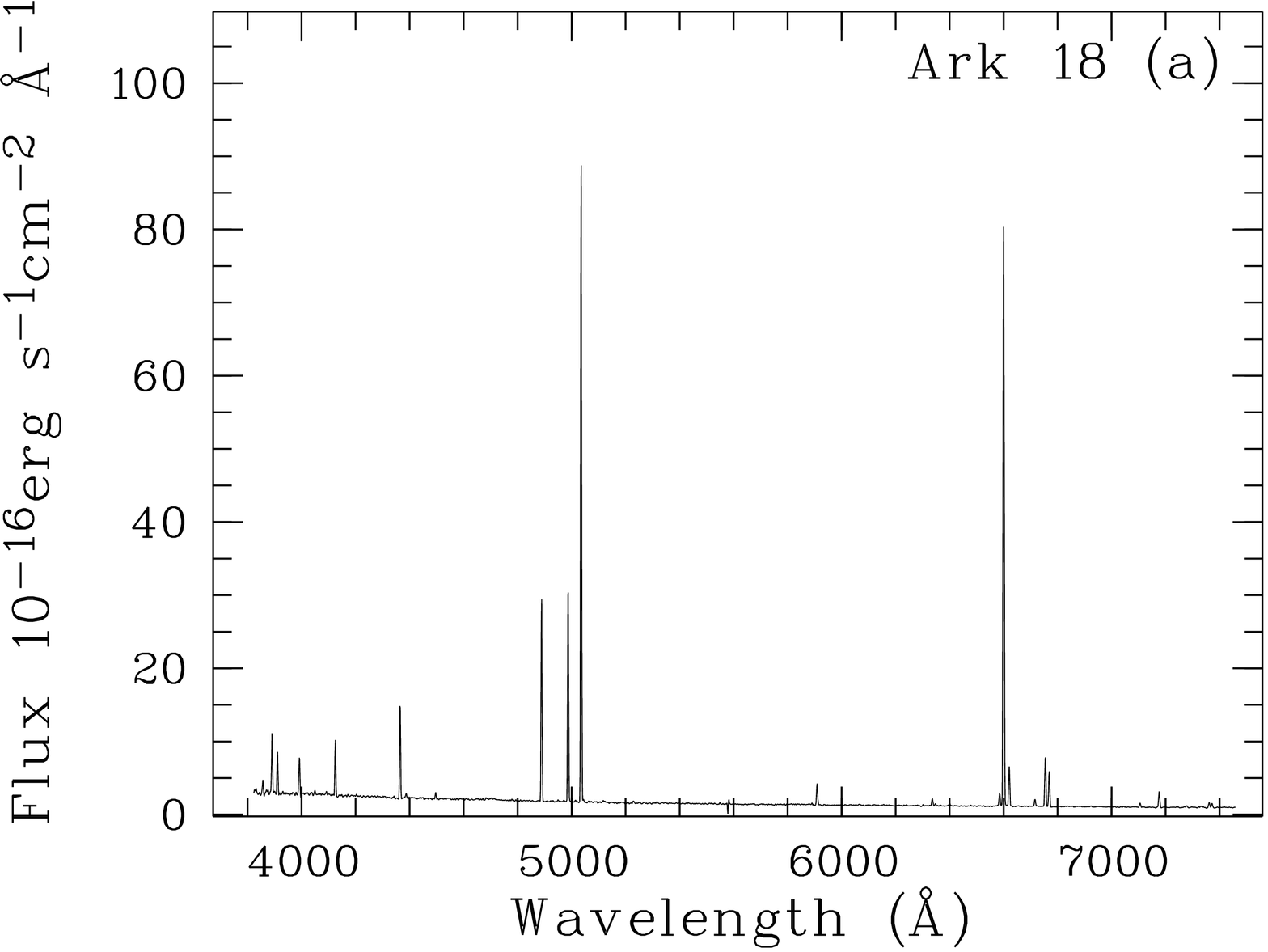}
 \includegraphics[angle=-90,width=0.4\linewidth]{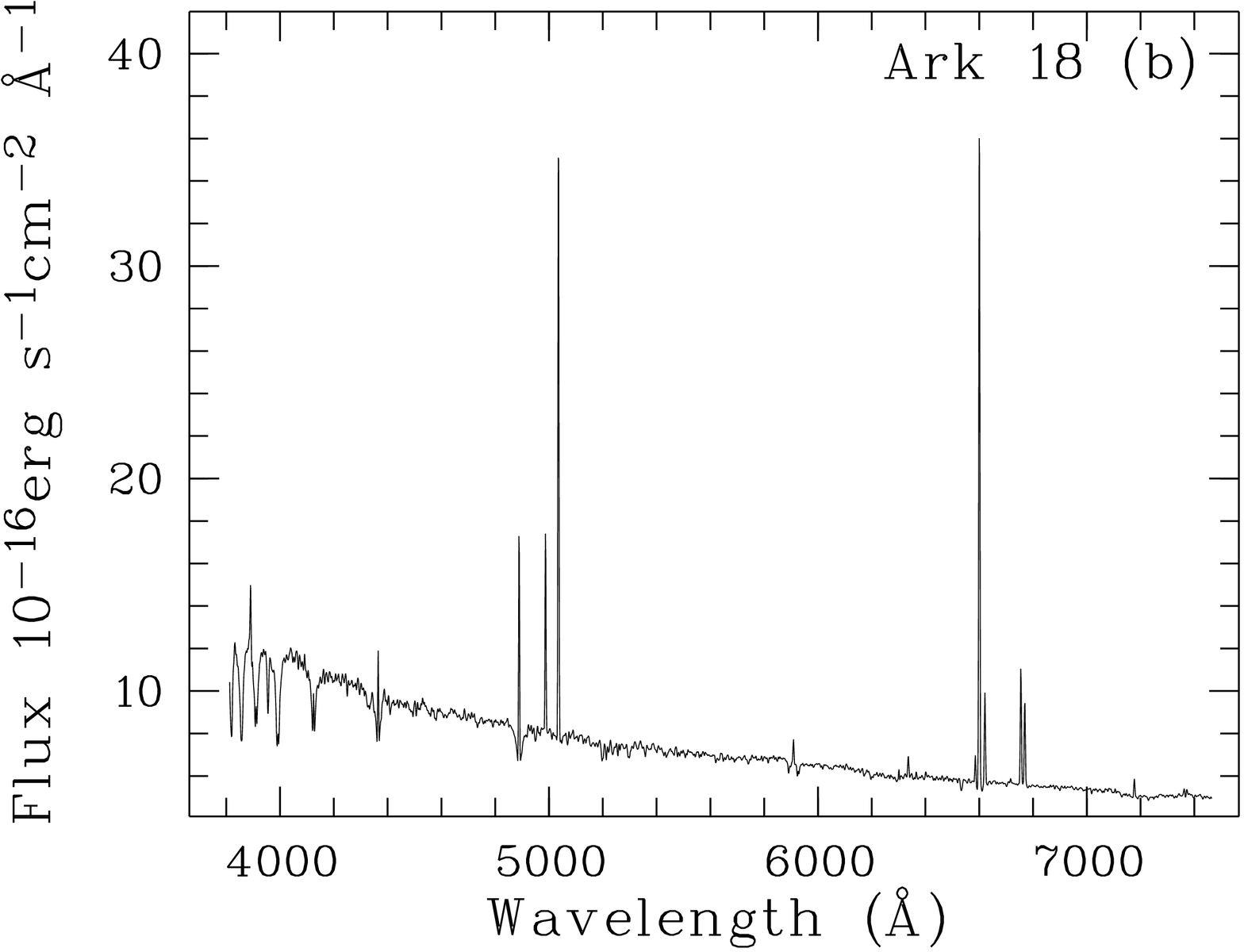}
 \includegraphics[angle=-90,width=0.4\linewidth]{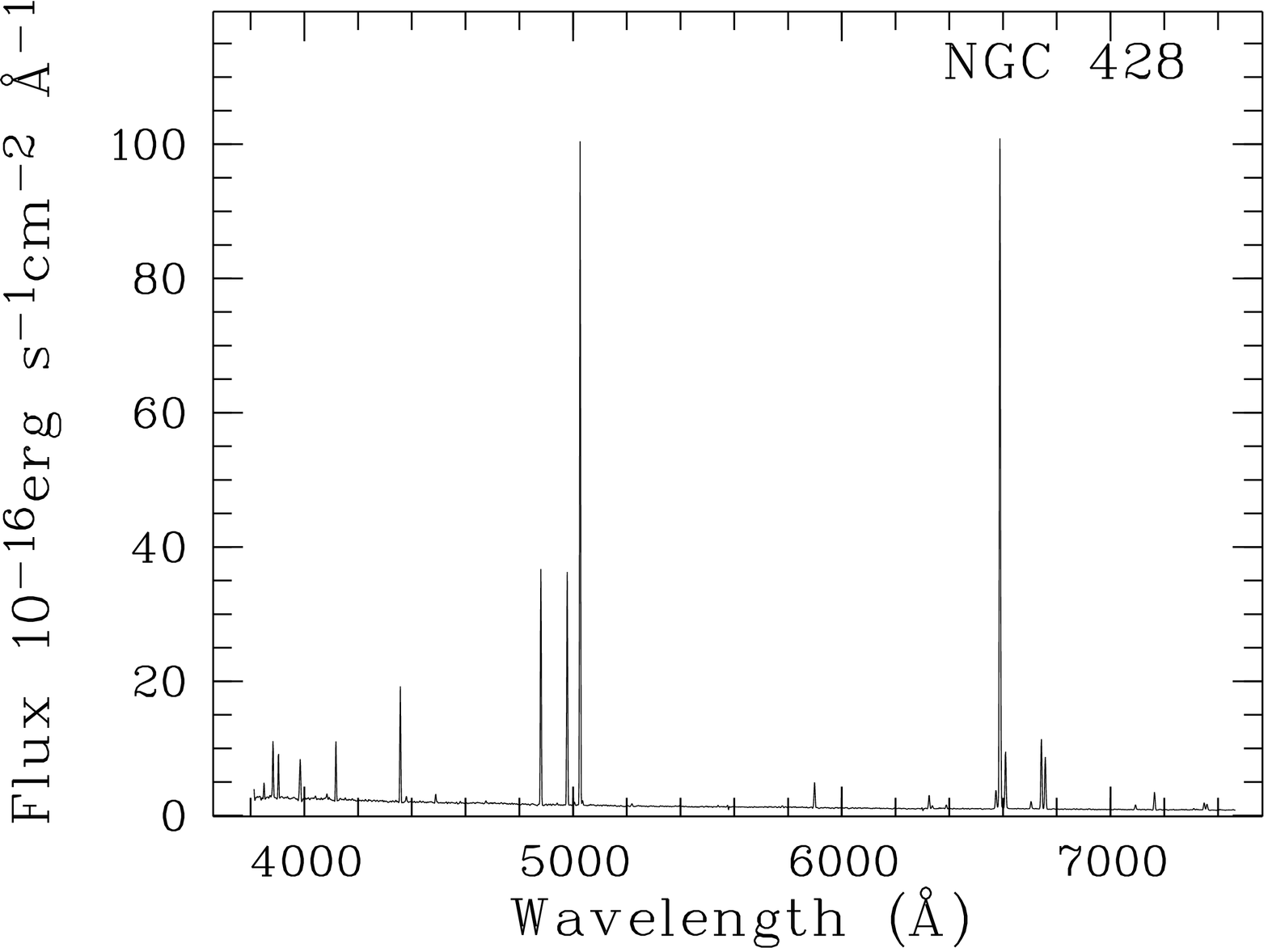}
 \includegraphics[angle=-90,width=0.4\linewidth]{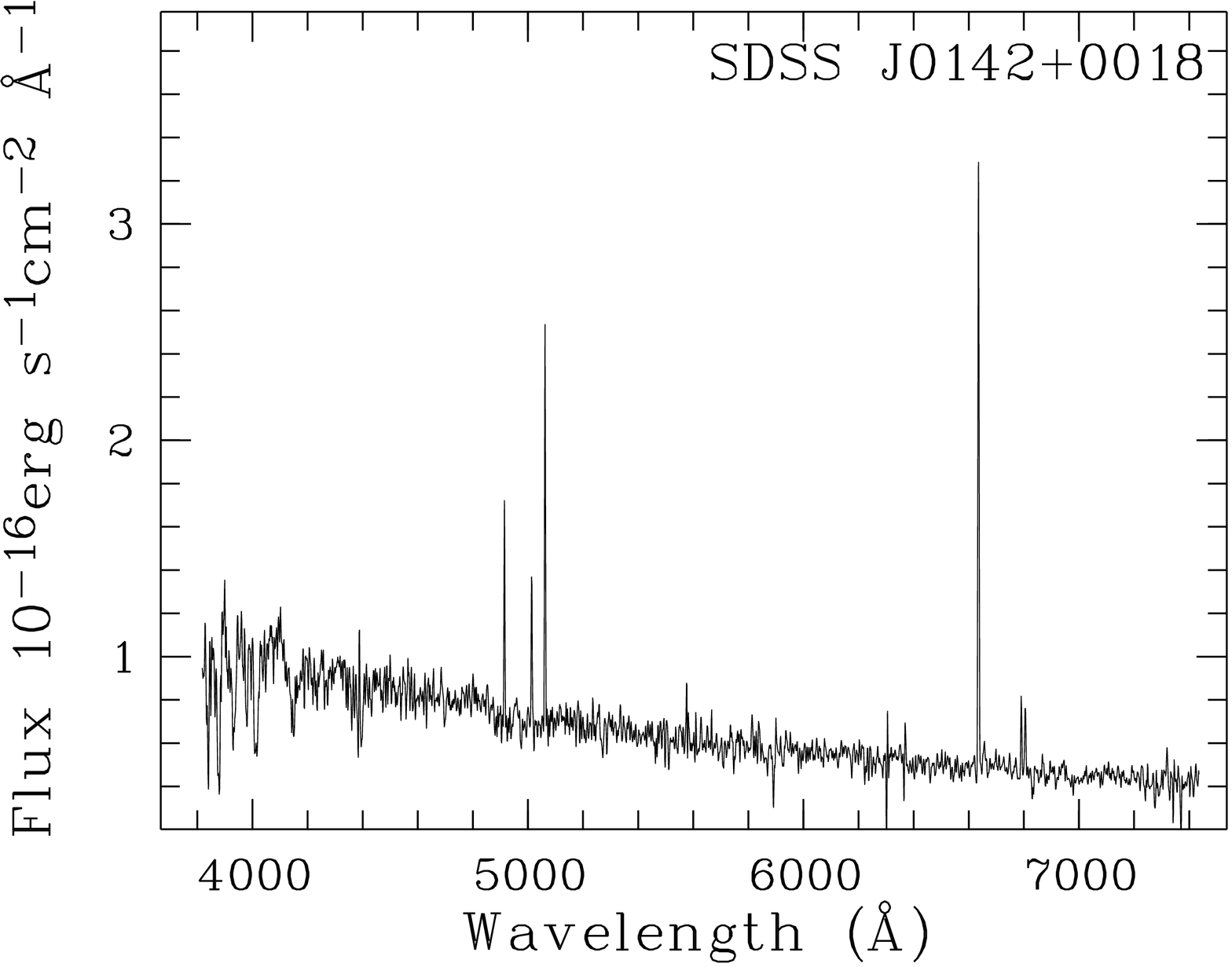}
 \includegraphics[angle=-90,width=0.4\linewidth]{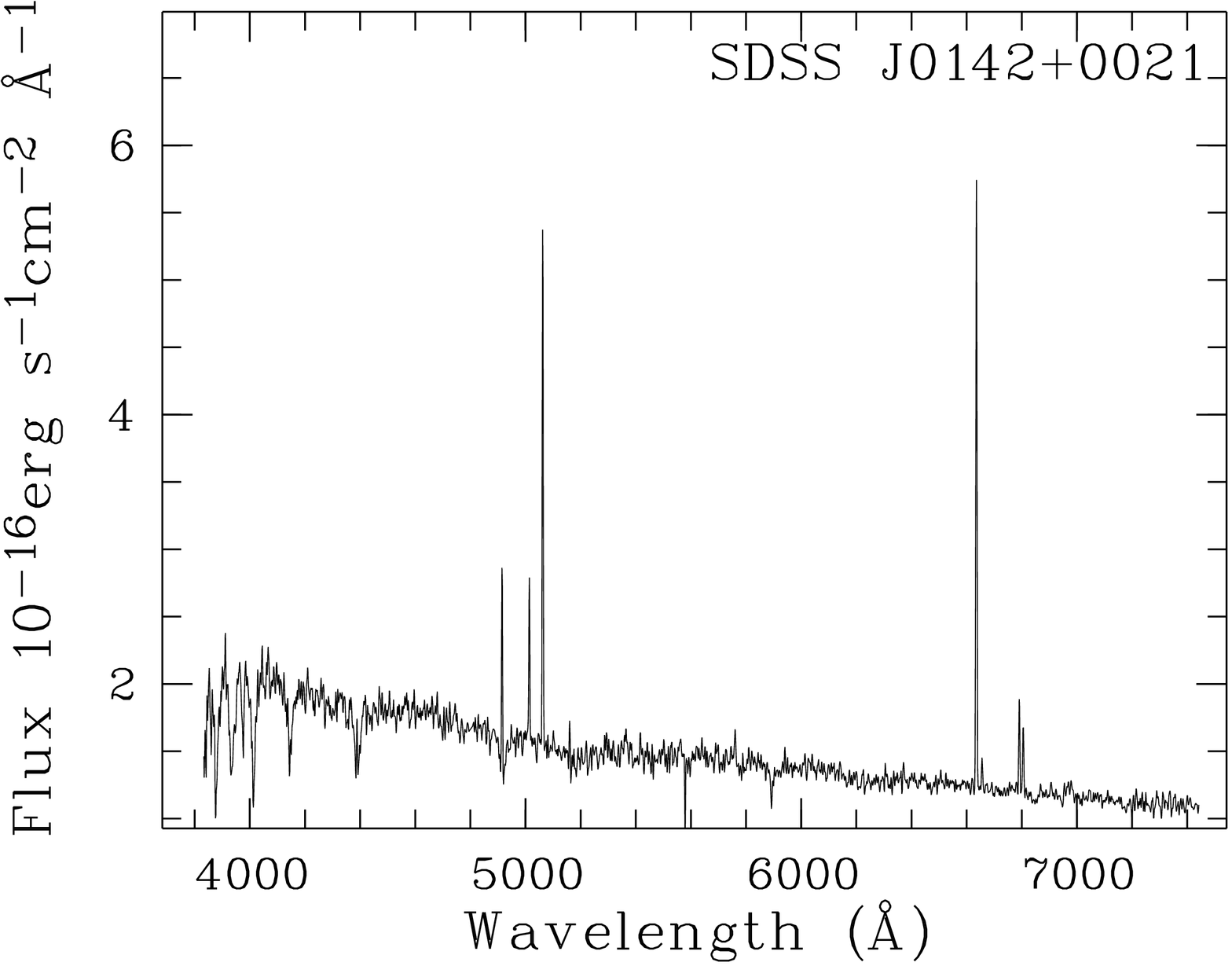}
 \includegraphics[angle=-90,width=0.4\linewidth]{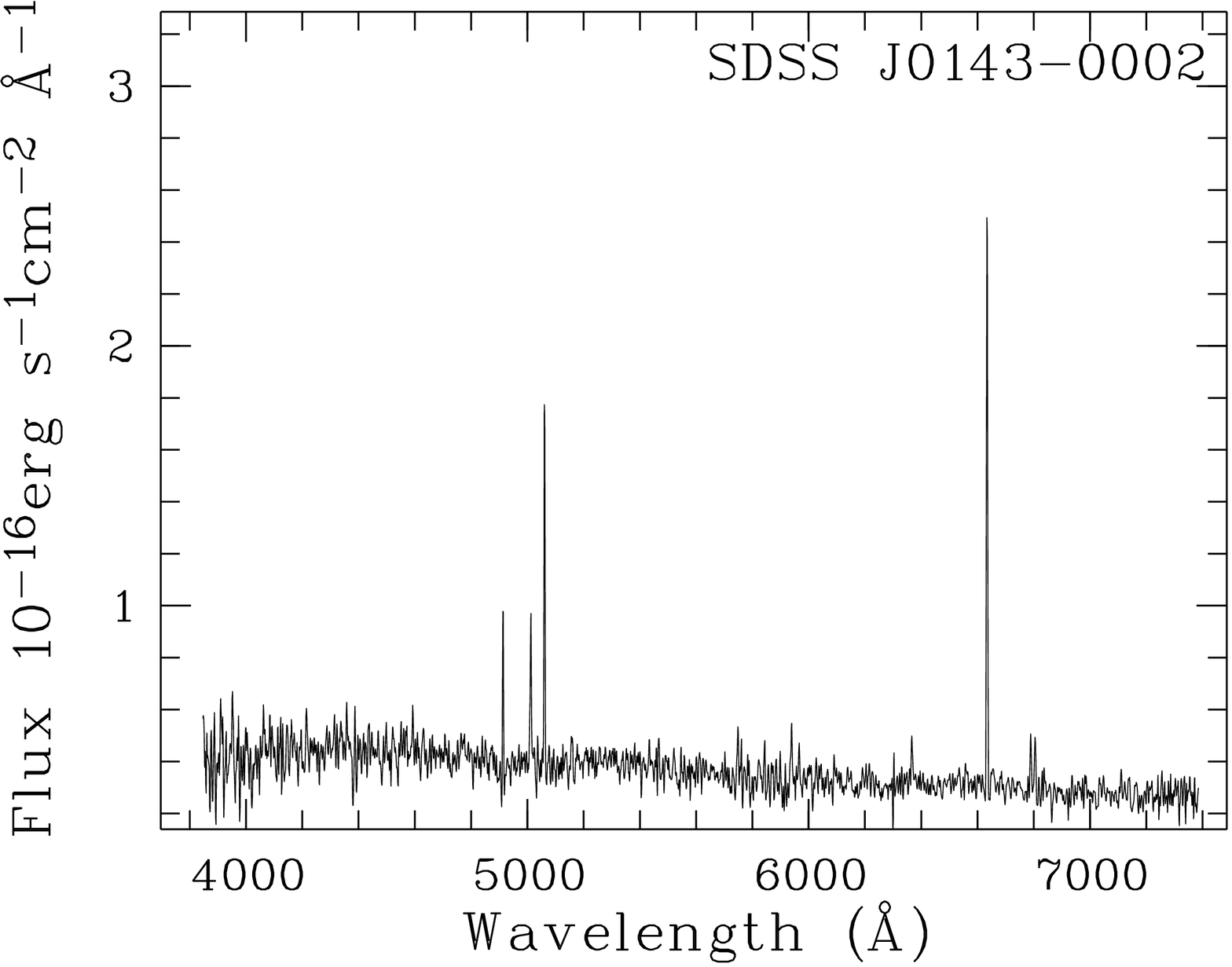}
\includegraphics[angle=-90,width=0.4\linewidth]{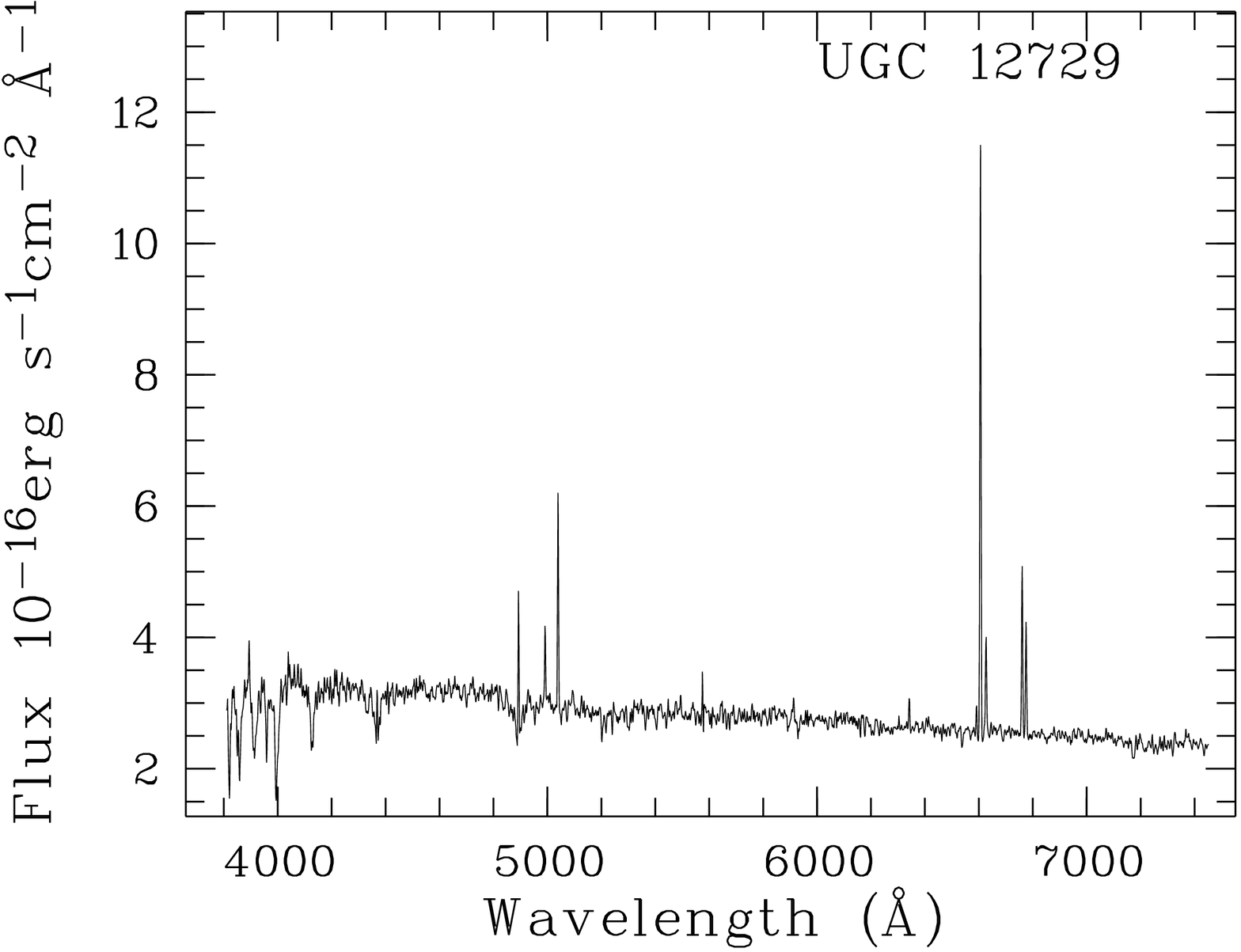}
\caption{\label{fig:SDSSspecs}
Spectra of 7 HII regions in 6 Eridanus void galaxies obtained from SDSS DR12.
}
\end{figure*}

The emission lines measurements were done in the same way described
in the Section \ref{sec:salt}.
The measured emission-line intensities for SDSS spectra with their
errors are presented in Tables \ref{t:Intens_SDSS1}-\ref{t:Intens_SDSS3}.
The measured line flux values in the $H\beta$ line are given in units
of $10^{-16}$~erg~s$^{-1}$~cm$^{-2}$. The 1D spectra of all objects studied
are available in Fig \ref{fig:SDSSspecs}.

For some of our objects we performed our own photometry on the base of
SDSS calibrated $u, g, r, i, z$ images \citep{SDSS_phot,Lupton01},
where for some of them we used the model SDSS magnitudes. In these cases the
$B$ band magnitudes presented in our Table~\ref{tab:sample},
were calculated from $g$ and $r$ magnitudes using formula from
\citet{Lupton05}.

\onecolumn
\clearpage
\newcounter{qq}
\begin{longtable}{r|l|c|r|r|c|l|c|l|c}
\caption{The current galaxy sample in the equatorial zone of Eridanus
void}\label{tab:sample}\\
\multicolumn{1}{c|}{\#}                 &
\multicolumn{1}{c|}{Name or}            &
\multicolumn{2}{c|}{Coordinates (J2000)} &
\multicolumn{1}{c|}{$V_{\rm hel}$,}     &
\multicolumn{1}{c|}{D, Mpc}             &
\multicolumn{1}{l|}{B$_{\rm tot}$,}     &
\multicolumn{1}{c|}{M$_{\rm B}$,}  &
\multicolumn{1}{c|}{Type}     &
\multicolumn{1}{r}{$D_{\rm NN}$}    \\
&
\multicolumn{1}{c|}{prefix }  &
\multicolumn{2}{c|}{ }  &
\multicolumn{1}{c|}{\kms }  &
\multicolumn{1}{c|}{ }  &
\multicolumn{1}{c|}{mag }  &
\multicolumn{1}{c|}{mag }  &
\multicolumn{1}{c|}{ }  &
\multicolumn{1}{c}{Mpc }  \\
&
\multicolumn{1}{c|}{ (1) }  &
\multicolumn{1}{c|}{ (2) }  &
\multicolumn{1}{c|}{ (3) }  &
\multicolumn{1}{c|}{ (4) }  &
\multicolumn{1}{c|}{ (5) }  &
\multicolumn{1}{l|}{ (6) }  &
\multicolumn{1}{c|}{ (7) }  &
\multicolumn{1}{c|}{ (8) }  &
\multicolumn{1}{c}{ (9) }  \\
\\[-0.2cm] \hline \\[-0.2cm]

\endhead

\qq&APM B0001+0008         &00 04 21.61&+00 25 35.3&3795 &54.4 &19.67$^{1}$ &-14.2 & BCG & 4.2 \\
\qq&SDSS J0013-0106    &00 13 40.76&-01 06 41.4&3477 &49.9 &19.37$^{1}$ &-14.3 & dI & 7.1  \\
\qq&SDSS J0015+0104    &00 15 20.68&+01 04 37.0&2035 &30.6 &18.31$^{2}$ &-14.2 & dI & 7.1  \\
\qq&SDSS J0016+0108    &00 16 28.25&+01 08 01.9&3103 &44.8 &19.27$^{1}$ &-14.1 & BCG & 5.8  \\
\qq&UM 240             &00 25 07.43&+00 18 45.7&3234 &46.5 &17.32$^{1}$ &-16.1 & Sm & 8.1  \\
\qq&UM 38              &00 27 51.52&+03 29 23.1&1371 &21.1 &15.91$^{3}$ &-15.7 & Sb & 5.7  \\
\qq&6dF J0027-0311     &00 27 55.29&-03 11 00.9&3231 &46.2 &15.84$^{1}$ &-17.6 & Sc & 7.4  \\
\qq&UM 40              &00 28 26.60&+05 00 15.9&1344 &20.8 &15.36$^{4}$ &-16.3 & Sb & 5.2  \\
\qq&KUG 0027+015       &00 29 48.02&+01 51 17.6&3436 &49.3 &16.17$^{5}$ &-17.4 & Sd & 7.9  \\
\qq&UGC 300            &00 30 04.10&+03 30 47.0&1346 &20.8 &15.86$^{6}$ &-15.8 & IAB & 5.6 \\
\qq&UGC 313            &00 31 26.11&+06 12 24.3&2085 &31.0 &14.48$^{7}$ &-18.1 & Sbc & 7.2 \\
\qq&APM B0029+0226     &00 31 45.35&+02 42 53.3&2380 &34.9 &17.90$^{8}$ &-14.9 & Sm & 7.0  \\
\qq&UGC 320            &00 32 30.93&+02 34 27.0&2374 &34.8 &15.7$^{9}$ &-17.1 & Sc & 6.9  \\
\qq&UGC 328            &00 33 22.09&-01 07 16.9&1985 &29.3 &15.49$^{10}$ &-16.9 & Sd & 5.5  \\
\qq&SDSS J0037-0040    &00 37 20.42&-00 40 43.0&2030 &29.9 &19.27$^{1}$ &-13.2 & BCG? & 6.3  \\
\qq&HIPASS J0041-01b   &00 41 39.66&-02 00 42.0&1949 &28.6 &16.89$^{11}$ &-15.5 & I & 4.8 \\
\qq&CGCG410-002        &00 44 48.42&+05 08 08.7&2894 &41.9 &15.12$^{1}$ &-18.1 & E? & 9.1  \\
\qq&NGC 259a           &00 47 17.82&-02 32 41.0&3508 &49.9 &18.30$^{12}$ &-15.3 & S & 3.8  \\
\qq&IC 52              &00 48 23.78&+04 05 30.7&1961 &29.1 &15.19$^{4}$ &-17.2 & Sb & 6.1  \\
\qq&UGC 527            &00 51 49.90&+03 06 21.0&1951 &28.8 &15.89$^{1}$ &-16.5 & Sm & 5.9  \\
\qq&UM 285             &00 51 58.73&-01 40 18.7&1908 &28.0 &17.24$^{13}$ &-15.2 & BCD & 4.8  \\
\qq&ARK 18             &00 51 59.62&-00 29 12.2&1621 &24.1 &14.76$^{13}$ &-17.2 & S0 & 3.4  \\
\qq&MCG-01-03-027      &00 52 17.23&-03 57 59.8&1412 &21.1 &15.69$^{1}$ &-16.1 & Scd & 2.9  \\
\qq&CGCG 384-019       &00 52 52.80&+01 12 50.7&1799 &26.7 &15.90$^{1}$ &-16.3 & SABd & 4.8  \\
\qq&6dF J0055-0601     &00 55 53.66&-06 01 20.0&2641 &37.8 &16.05$^{1}$ &-17.0 & E? & 4.1  \\
\qq&WINGS J0057-0124   &00 57 03.50&-01 24 42.6&3389 &48.2 &19.38$^{14}$ &-14.2 & Sd & 6.2  \\
\qq&SDSS J0057-0021    &00 57 12.60&-00 21 57.7&2866 &41.1 &19.30$^{1}$ &-13.9 & BCG & 7.5  \\
\qq&MRK 965            &00 57 28.75&-04 09 34.0&2713 &38.8 &15.81$^{1}$ &-17.3 & S0-a & 5.3  \\
\qq&SDSS J0057+0052    &00 57 56.58&+00 52 09.1&2299 &33.4 &17.43$^{1}$ &-15.3 & I & 3.4  \\
\qq&MCG-01-03-072      &01 02 22.91&-04 30 30.9&1764 &25.8 &15.69$^{11}$ &-16.5 & Im pec & 2.4  \\
\qq&6dF J0103-0302     &01 03 09.69&-03 02 55.8&1784&26.1 &15.79$^{15}$&-16.4 & dI & 3.1  \\
\qq&MRK 970            &01 03 10.43&-03 36 36.6&2587 &37.1 &14.55$^{1}$ &-18.4 & S & 3.7  \\
\qq&6dF J0105-0424     &01 05 18.59&-04 24 29.7&1801 &26.3 &16.01$^{15}$ &-16.2 & S? & 3.2  \\
\qq&6dF J0105-0251     &01 05 55.46&-02 51 36.0&1794 &26.2 &16.85$^{1}$ &-15.4 & dI & 3.2  \\
\qq&SDSS  J0106-0250   &01 06 03.06&-02 50 52.4&1750 &25.6 &18.33$^{1}$ &-13.8 & S & 2.8  \\
\qq&CGCG 384-074       &01 06 30.98&-02 11 48.0&3489 &49.5 &15.75$^{1}$ &-17.8 & Sc & 6.4  \\
\qq&UGC 695            &01 07 46.44&+01 03 49.2& 628 &10.5 &14.95$^{16}$ &-15.2 & Sd & 3.3  \\
\qq&UGC 711            &01 08 36.90&+01 38 30.0&1982 &29.0 &14.49$^{17}$ &-17.9 & SBcd & 4.8  \\
\qq&SHOC 053           &01 09 07.97&+01 07 15.5&1156 &17.7 &16.98$^{1}$ &-14.3 & I & 4.9  \\
\qq&SDSS J0110-0000    &01 10 03.68&-00 00 36.4&1136 &17.4 &19.34$^{1}$ &-12.0 & dI & 4.8  \\
\qq&PGC 135629         &01 12 50.68&+01 02 48.8&1105 &16.9 &16.10$^{8}$ &-15.1 & Sm & 4.9  \\
\qq&NGC 428            &01 12 55.71&+00 58 53.6&1152 &17.6 &12.16$^{18}$ &-19.1 & SABm & 4.6  \\
\qq&UGC 772            &01 13 39.40&+00 52 27.9&1161 &17.7 &16.28$^{19}$ &-15.0 & I & 4.6  \\
\qq&MCG +00-04-049     &01 14 20.48&+00 55 01.9&1129 &17.2 &17.04$^{11}$ &-14.3 & Sm & 4.7  \\
\qq&FGC0155            &01 21 23.80&-01 51 46.0&2217 &31.9 &16.50$^{9}$ &-16.2 & Sd & 2.3  \\
\qq&UM 323a            &01 26 16.02&-00 24 30.3&1938 &28.1 &17.40$^{1}$ &-14.9 & Sd & 3.7  \\
\qq&UGC 1014	       &01 26 23.53&+06 16 40.7&1129 &31.2 &14.79$^{4}$ &-17.8 & Sm & 2.3  \\
\qq&UM 323             &01 26 46.60&-00 38 46.1&2132 &27.8 &16.12$^{20}$ &-16.2 & iE BCD & 3.6  \\
\qq&UGC 1102	       &01 32 29.30&+04 35 54.0&1954 &28.6 &14.37$^{4}$ &-18.0 & S? & 3.6  \\
\qq&UGC 1105	       &01 32 39.90&+04 38 30.0&2027 &29.6 &16.95$^{1}$ &-15.5 & Im & 4.0  \\
\qq&SDSS J0137-0033    &01 37 08.07&-00 33 53.7&2924 &41.5 &17.23$^{1}$ &-16.0 & SABd & 5.4  \\
\qq&SDSS J0139-0027    &01 39 18.59&-00 27 30.4&3453 &48.7 &18.91$^{1}$ &-14.6 & dI & 4.6  \\
\qq&SDSS J0142+0018    &01 42 48.13&+00 18 14.6&3222 &45.5 &18.39$^{1}$ &-15.0 & dI & 4.4  \\
\qq&SDSS J0142+0021    &01 42 59.31&+00 21 36.5&3243 &45.9 &18.04$^{1}$ &-15.4 & I & 4.3  \\
\qq&SDSS J0143-0002    &01 43 13.36&-00 02 50.9&3165 &44.8 &19.00$^{1}$ &-14.4 & BCG & 4.6  \\
\qq&UGC 12729          &23 40 20.75&+01 14 45.1&1870 &28.2 &15.05$^{21}$ &-17.3 & S0/a & 4.5  \\
\qq&APM B2341-0330     &23 43 55.00&-03 13 35.0&2080 &30.8 &16.99$^{1}$ &-15.6 & Sd & 4.4  \\
\qq&B2342-0223         &23 44 48.20&-02 06 53.5&2018 &30.0 &16.77$^{1}$ &-15.7 & dI & 5.1  \\
\qq&UGC 12769          &23 45 19.22&-01 16 24.3&2078 &30.9 &16.58$^{1}$ &-16.0 & Sd & 5.3  \\
\qq&SDSS J2347+0126    &23 47 21.47&+01 26 25.4&2674 &38.9 &18.43$^{1}$ &-14.6 & dI & 2.1  \\
\qq&SDSS J2354-0005    &23 54 37.30&-00 05 01.7&2311 &34.1 &18.79$^{2}$ &-14.0 & dI & 4.6  \\
\qq&MCG-01-01-003      &23 55 26.96&-06 14 14.9&1242 &19.1 &15.87$^{22}$ &-15.6 & S0-a & 6.2  \\
\qq&SDSS J2356+0141    &23 56 26.83&+01 41 16.7&2496 &36.7 &17.61$^{1}$ &-15.3 & dI & 3.5  \\
\qq&UGC 12857          &23 56 47.62&+01 21 18.0&2477 &36.4 &14.70$^{23}$ &-18.2 & Sbc & 3.6  \\
\qq&HIPASS J2358+04    &23 58 07.15&+04 49 07.3&3035 &44.2 &17.66$^{11}$ &-15.8 & Sd & 2.0  \\
\qq&HIPASS J2359+02    &23 59 13.23&+02 43 24.3&2616 &38.4 &16.40$^{24}$ &-16.6 & Sc & 4.0  \\
\end{longtable}
$^{1}$ derived from SDSS model magnitudes in $g$ and $r$, using the formula
by \citet{Lupton05};
$^{2}$ \citet{Eridan_XMD}; 
$^{3}$ \citet{Koleva2014}; 
$^{4}$ \citet{Lu1998}; 
$^{5}$ \citet{Galaz06}; 
$^{6}$ \citet{vanZee97}; 
$^{7}$ \citet{Jansen2000}; 
$^{8}$ \citet{Impey96}; 
$^{9}$ \citet{Karachentsev99}; 
$^{10}$ calculated with $g,r$ derived in \citet{West10}; 
$^{11}$ our estimation from SDSS $g,r$ frames; 
$^{12}$ \citet{Zaritsky1993}; 
$^{13}$ \citet{GildePaz03}; 
$^{14}$ \citet{Fritz2011}; 
$^{15}$ \citet{Paturel00}; 
$^{16}$ \citet{Cook2014}; 
$^{17}$ obtained with SDSS $g,r$ in \citet{Bizyaev2014}; 
$^{18}$ \citet{Smoker99}; 
$^{19}$ \citet{Smoker96}; 
$^{20}$ \citet{Koleva2014}; 
$^{21}$ calculated with SDSS $g,r$ derived in \citet{Fernandes13}; 
$^{22}$ \citet{Paturel00b}; 
$^{23}$ \citet{Verdes05}; 
$^{24}$ \citet{Karachentsev08}
\twocolumn

\section[]{ABUNDANCE DETERMINATION AND RESULTS}
\label{sec:results}

\subsection{Methods of O/H determination}
\label{OH}


Oxygen abundances were calculated in the way described by \citet{Pap7}.
Namely, for case of well-measured faint auroral line
[O{\sc iii}]~$\lambda$4363\AA, the direct $T_{\rm e}$ method was used for
a two-zone model of \HII\ regions and procedures described by
\citet{Kniazev08} were used. The updated atomic data are used as in
\citet{Izotov06}.

To correct the measured emission-line fluxes for extinction
and the underlying young star cluster Balmer absorptions, we used an iterative
procedure from \citet{Izotov94}.
It finds the optimal combination of two physical parameters affecting the
original nebular spectrum of \HII-region: the dust extinction coefficient near
H$\beta$ line C(H$\beta$) and the equivalent width of Balmer lines of the
underlying continuum of young star cluster EW(abs). The respective correction
results in the Balmer line intensities
best approximating (within the adopted flux errors) recombination ratios.
The \citet{Whitford58} extinction law was used. As noticed in \citet{Pap7},
its use makes the negligible effect to the derived O/H with respect to the
more recent extinction law variants.

In about half of our targets, the line [O{\sc iii}]~$\lambda$4363 was either
undetected or too weak. In these cases we applied the 'semi-empirical' method
of \citet{IT07}. This method exploits the well-fitted dependence of
$T_{\rm e}$ on the total intensity of the oxygen lines
[O{\sc iii}]~$\lambda\lambda$4959,5007 and [O{\sc ii}]~$\lambda$3727 with
respect to intensity of H$\beta$. The latter dependence, in turn, was derived
from models of \citet{Stasinska03} which approximate well the relations
between equivalent width EW(H$\beta$) and the strong line intensities as found
on the large sample of extragalactic \HII\ regions representing the whole
range of observed O/H. The essence of this semi-empirical method is the
identity of the subsequent calculations of all parameters to the direct
$T_{\rm e}$ method once $T_{\rm e}$ is estimated from the strong oxygen lines.
Since the estimate of $T_{\rm e}$ via method of \citet{IT07} is
primarely based on the large sample of extragalactic \HII-regions, its
applicability to our sample of void star-forming galaxies is consistent with
an idea that in most of our galaxies we observe rather typical \HII-regions.
Due to the limited range of the method calibration,
12+$\log$(O/H)$\lesssim$7.9~dex \citep{IT07,Pap7}, we also limit its use by
this range of O/H. 

Similar to work of \citet{Pap7}, to derive a more robust O/H value, we
used besides the semi-empirical method, two additional empirical O/H
estimators from \citet{PT05} and \citet{Yin07} (PT05 and Yin07, respectively),
which show rather small internal scatter ($\sim$0.08~dex). They also use the
relative intensities of lines [O{\sc iii}]~$\lambda\lambda$4959,5007 and
[O{\sc ii}]~$\lambda$3727. While both, the semi-empirical and Yin07 methods
are applicable only for 12+$\log$(O/H) $\lesssim$ 7.9, the PT05 was used
mostly for lower branch case (12+$\log$(O/H) $\lesssim$8.1) since the absolute
majority of our targets have O/H in this regime.

\citet{PVT10} and \citet{PM11} suggested the empirical O/H calibrators based
on the strong oxygen lines and on the doublets
[N{\sc ii}]~$\lambda\lambda$6548,6584 and
[S{\sc ii}]~$\lambda\lambda$6716,6730. For a high signal-to-noise (S-to-N)
spectra, these estimators show a rather small internal scatter of O/H of
$\sim$0.08--0.10~dex. However, for a low S-to-N ratio of the line
[N{\sc ii}]~$\lambda$6584, the accuracy of these estimators decreases
substantially. Also, \citet{SanchezAlmeida16} show that $\sim$10\% of low-Z
objects have an elevated ratio of N/O and an enhanced flux of [N{\sc ii}]
lines. For such cases, one expects that the O/H with these methods should be
systematically higher than values of O/H derived with the direct $T_{\rm e}$
method. The latest version of the empirical calibrators is known
as the 'Counterpart' method, which is also based on strong and medium
intensity emission lines \citep{PGM12}. It also has the similar limitations.
For this reason, we use these methods selectively,  for additional control of
O/H estimates that do not depend on the line [N{\sc ii}]~$\lambda$6584.

For SDSS spectra of the void galaxies, the [O{\sc ii}]~$\lambda$3727
line falls outside the available spectral range.
In papers by \citet{Kniazev03}
and \citet{Kniazev04b} the modified $T_{\rm e}$ method was suggested. For
the case of measured [O{\sc iii}]~$\lambda$4363, it exploits the intensity
of the weak auroral doublet \mbox{[O{\sc ii}]~$\lambda\lambda$7320,7330} to
calculate the number of O$^+$ ions. Besides, for the cases of undetected
[O{\sc iii}]~$\lambda$4363 line, we used an iterative estimate of
I($\lambda$3727) from I($\lambda\lambda$7320,7330) as suggested in
\citet{void_OH} with the subsequent use of the semi-empirical method.

In the paper by \citet{Pap7} a comparison of several empirical and
semi-empirical methods with the direct $T_{\rm e}$ method was performed
and the formulae, accounting for small offsets of these methods with
respect of the direct $T_{\rm e}$ method, were presented. We used these
formulae to calculate the corrected values of O/H estimated with the
semi-empirical method by \citet{IT07}, and with the empirical PT05
and Yin07 methods.
For more details on different methods of O/H estimation see \citet{Pap7}.

\begin{figure*}
 \centering
 \includegraphics[angle=-90,width=0.49\linewidth]{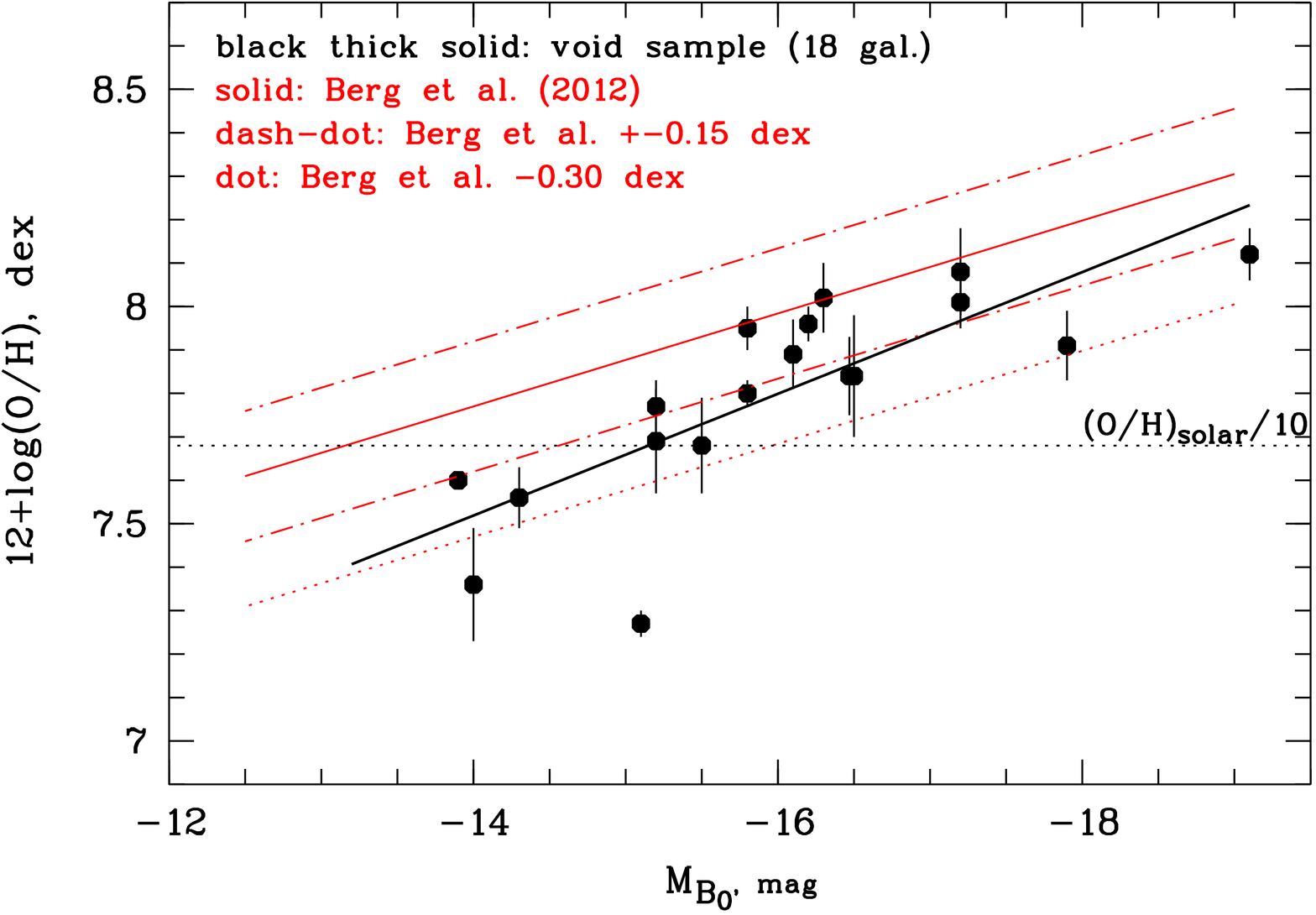}
 \includegraphics[angle=-90,width=0.49\linewidth]{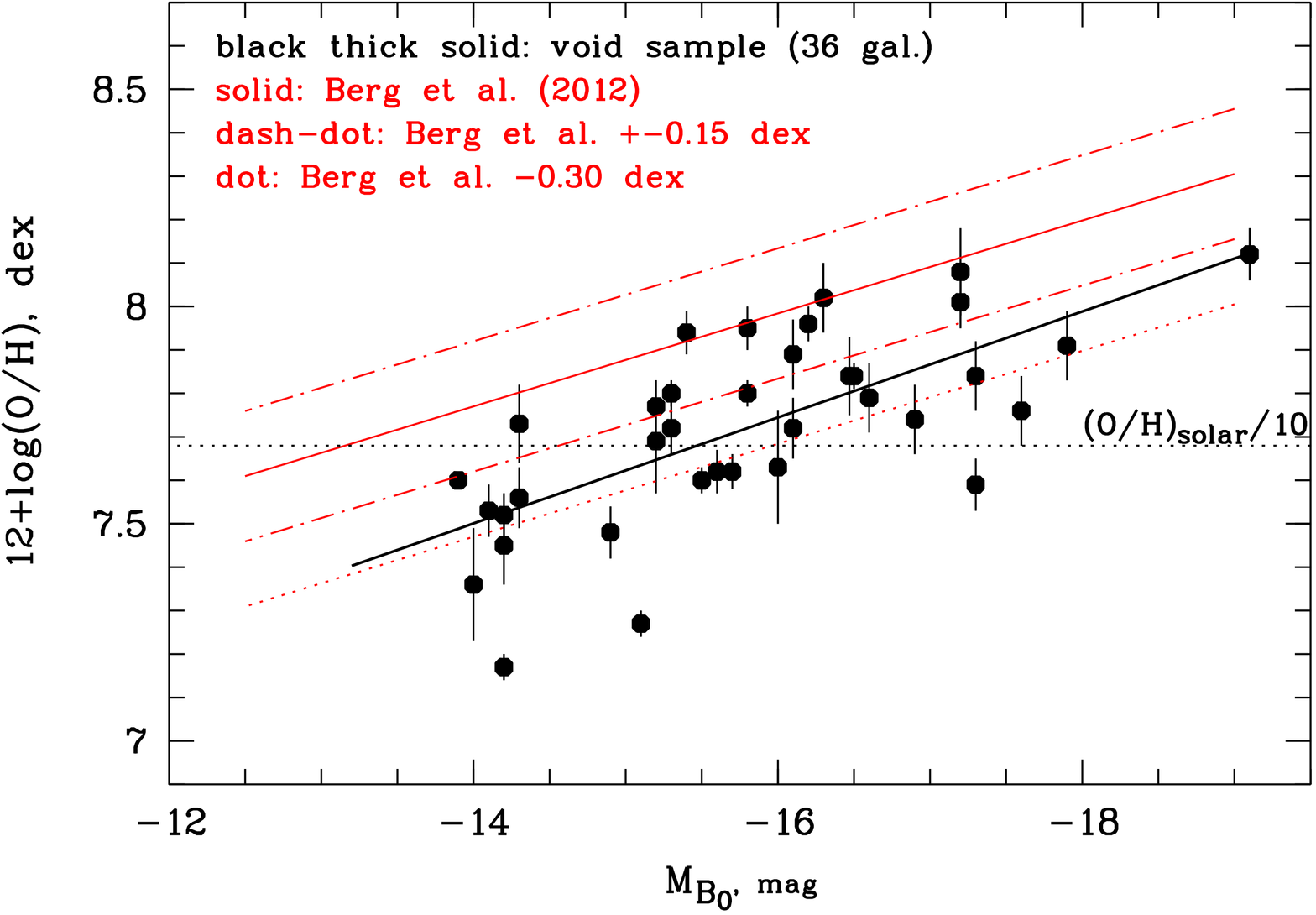}
  \caption{\label{fig:ZvsMB}
The relation between $\log$(O/H) and the absolute blue magnitude $M_{\rm B}$
for the subsample of galaxies in the Eridanus void with O/H as derived with
the direct $T_{\rm e}$ method (left panel) and for all Eridanus void galaxies
(right panel). The solid line (red) shows the linear regression
for the control sample from the Local Volume by \citet{Berg12}.
Two dot-dashed lines (red) at both sides of the reference line show the
r.m.s. deviation of their sample from the linear regression (0.15 dex).
The lowest line (dashed red, --0.30~dex) separates the region where the
most deviating metal-poor dwarfs are situated.
The horizontal dotted black line marks the value Z\sunn/10 which corresponds
to 12+$\log$(O/H)=7.69 for Z\sunn\ from \citet{AllendePrieto2001} and
\citet{Asplund09}.
}
\end{figure*}

\begin{table*}
\begin{center}
\caption{Journal of SALT (top) and BTA (bottom) spectral observations}
\label{tab:Obs}
\hoffset=-2cm
\begin{tabular}{l|l|l|l|c|c} \hline  \hline \\ [-0.2cm]
\MC{1}{c|}{Name} &
\MC{1}{c|}{Date} &
\MC{1}{c|}{Expos.}&
\MC{1}{c|}{PA} &
\MC{1}{c|}{Seeing}&
\MC{1}{c}{Air}  \\

\MC{1}{c|}{ } &
\MC{1}{c|}{ } &
\MC{1}{c|}{time, s}&
\MC{1}{c|}{ } &
\MC{1}{c|}{\arcsec} &
\MC{1}{c}{mass}\\

\MC{1}{c|}{(1)} &
\MC{1}{c|}{(2)} &
\MC{1}{c|}{(3)} &
\MC{1}{c|}{(4)} &
\MC{1}{c|}{(5)} &
\MC{1}{c}{(6)} \\
\\[-0.2cm] \hline \\[-0.2cm]
SDSS J0013-0106          & 2013.08.03 & 3$\times$800&310  & 2.0      & 1.24 \\
UM 38                    & 2012.11.12 & 3$\times$950& 75  & 1.7      & 1.29 \\
6dF J0027-0311           & 2013.07.31 & 3$\times$800& 55  & 2.0--2.6 & 1.22 \\
UM 40                    & 2013.07.30 & 3$\times$800&  8  & 1.7      & 1.27 \\
APM B0029+0226           & 2012.10.12 & 3$\times$760& 60  & 2.0      & 1.35 \\
UGC 328                  & 2013.08.10 & 3$\times$800& 342 & 2.0      & 1.27 \\
HIPASS J0041-01b         & 2012.07.24 & 3$\times$950& 90  & 1.5      & 1.21 \\
IC 52                    & 2013.07.30 & 3$\times$800& 95  & 1.5      & 1.28 \\
UM 285                   & 2012.07.24 & 2$\times$950& 0   & 1.6      & 1.16 \\
MCG -01-03-027           & 2013.09.13 & 3$\times$900& 18  & 2.5      & 1.22 \\
J0057+0052               & 2013.07.10 & 3$\times$850& 55  & 2.5      & 1.20 \\
MCG -01-03-072           & 2012.11.12 & 3$\times$950& 75  & 2.7      & 1.29 \\
6dF J0105-0251           & 2012.10.24 & 2$\times$1100& 70 & 2.1      & 1.18 \\
MCG 01-04-005            & 2013.09.10 & 2$\times$900& 147 & 1.8      & 1.27 \\
MCG 01-04-005            & 2013.09.24 & 2$\times$800& 147 & 2.5      & 1.23 \\
SDSS J0137-0033(w/o O/H) & 2013.09.07 & 1$\times$900& 95  & 2.0      & 1.25 \\
UGC 12729                & 2013.07.29 & 3$\times$800& 75  & 1.4      & 1.20 \\
APM B2341-0330           & 2013.07.11 & 3$\times$900& 60  & 2.0      & 1.20 \\
B2342-0223               & 2012.11.19 & 3$\times$870& 18  & 1.5      & 1.31 \\
UGC 12769                & 2013.09.04 & 3$\times$900& 29  & 1.6      & 1.21 \\
SDSS J2356+0141          & 2013.09.07 & 3$\times$900& 88  & 2.0--2.5 & 1.25 \\
HIPASS J2359+02          & 2013.09.06 & 1$\times$900& 12  & 1.5      & 1.31 \\
HIPASS J2359+02          & 2013.10.23 & 3$\times$850& 12  & 2.5--3.0 & 1.22 \\
\hline \\[-0.2cm]
UGC 527                  & 2013.12.28 & 3$\times$900&24.5 & 1.0      & 1.32 \\
MRK 965                  & 2013.12.29 & 2$\times$900&40.3 & 1.7      & 1.47 \\
UGC 711                  & 2013.12.28 & 2$\times$600&179.5& 1.6      & 1.41 \\
MCG +00-04-049           & 2009.01.22 & 2$\times$900& 29  & 2.2      & 1.66 \\
\hline \hline \\[-0.2cm]
\end{tabular}
\end{center}
\end{table*}

\subsection{Table with O/H determinations}
\label{OHtable}

In Table~\ref{tab:main} we present the adopted O/H values derived from
our observations and the analysis of the SDSS spectra. The table  includes
the following information.

\begin{description}
\item[$\bullet$] Column 1: Common name or SDSS prefix; 
\item[$\bullet$] Column 2: Epoch J2000 R.A.
\item[$\bullet$] Column 3: Epoch J2000 Declination
\item[$\bullet$] Column 4: 12+$\log$(O/H) derived via the direct $T_{\rm e}$
method with its 1$\sigma$ error;
\item[$\bullet$] Column 5: the value of O/H as adopted for the further
analysis.
\item[$\bullet$] Column 6: The sources of O/H in Column 5 indicated
as follows:
\subitem (1) O/H from the BTA, SALT and SDSS data evaluated with the direct 
$T_{\rm e}$ method (for IC52, values in Col.4,5 are the weighted means of
O/H derived via the direct $T_{\rm e}$ method for spectra a) and c)); for
SDSS spectra [O{\sc ii}]$\lambda\lambda$7320,7330 line fluxes were used to
determine the O/H via the modified direct $T_{\rm e}$ method
\citep{Kniazev03,Kniazev04b};

\subitem (2) the most robust O/H estimate from the SALT, BTA, or SDSS
spectra, based on the weighted mean from the semi-empirical method
\citep{IT07} and the empirical methods from \citet{PT05} and \citet{Yin07};
for all three methods, O/H estimates were slightly corrected to reduce small
systematics with respect to O/H derived via the direct $T_{\rm e}$ method
(see the end of subsection \ref{OH});

\subitem (3) the same as in (2), but including in the weighted mean the
values of O/H derived with the direct $T_{\rm e}$ method;

\subitem (4) the same as (3), but including in the weighted mean the
values of O/H derived with the empirical estimators from \citet{PM11}
and \citet{PVT10}, \citet{PGM12} which incorporate, in addition to the
strong oxygen lines, the intensities of [N{\sc ii}]$\lambda$6584 and
[S{\sc ii}] $\lambda\lambda$6716,6730. Due to the mentioned above known
cases of the enhanced intensities of [N{\sc ii}]$\lambda\lambda$6548,6584,
we used these estimators only where they did not show the systematic offset
with respect to O/H derived only with the strong oxygen lines;

\subitem (5) the value of 12+$\log$(O/H) was adopted from the literature
and re-evaluated to the new scale (\citet{Izotov06}), if necessary. Letters
after
(5) refer to the source in the footnote of the table. For galaxies with more
than one estimation of O/H the adopted 12+$\log$(O/H) is the weighted mean
value (for APMB0001+0008 and UGC772).
For cases where the calculated error of the weighted mean was less than 0.03
dex, we adopted the minimal value of 0.03 dex.
\end{description}

\subsection{Comments on individual galaxies}
\label{commentsOH}

\textit{MCG-01-04-005}. The galaxy was initially included in the void sample,
 and its spectrum was obtained with SALT (see Table \ref{t:Intens_MCG}).
However, the galaxy was later excluded from the sample since it seems to
belong to the NGC450 group ($D_{\rm proj}= $0.7 Mpc to NGC448, with
$\delta V = $33~\kms). We include its spectrum
analysis here, but do not use it in the void galaxy sample study.

\textit{UM285=J0051-0140, MCG+00-04-049=J0114+0055}. We adopted their
$12+\log$(O/H)$_{\rm Te}$ as the final value. We notice, however, that
they well match $12+\log$(O/H) derived by the other methods.


\textit{MCG-01-03-027=J0052-0357, SDSS J0057+0052, MCG-01-04-005}.
Due to the noisy signal of the line [O{\sc iii}]$\lambda$4959, we
adopted its flux according to the theoretical ratio as
I([O{\sc iii}]$\lambda$5007)/2.98.

\textit{SDSS J0013-0106}. While a SDSS spectrum of this galaxy exists,
we use its SALT spectrum in our analysis. The O/H value derived from the SDSS data agree with that from the SALT spectrum, but only under certain
assumptions on I([O{\sc ii}]3727) and I([O{\sc iii}]5007,5949). Therefore
we do not show the SDSS spectrum and do not use O/H derived from it.

\textit{UGC12729=J2340+0114}. We used both the SALT and SDSS spectra. The
value of I([O{\sc ii}]3727) for the SDSS spectrum was adopted to be the
same as for the SALT spectrum. Both values of O/H are in a good agreement.

\textit{IC52=J0048+0405}. The adopted value of O/H is the weighted mean
of the two values derived with the direct $T_{\rm e}$ method for spectra
from regions a) and c).

\textit{Ark18=J0051-0029.} There are two spectra available in SDSS for
this galaxy. We present values of $12+\log$(O/H) for both regions separately
 in Table~\ref{tab:main}. The O/H derived for the central part of Ark18 via
the semi-empirical method is significantly lower than O/H for the non-central
 part. We adopt for the whole galaxy the latter value obtained via the
modified $T_{\rm e}$ method.

\textit{NGC428}. In the paper by \citet{PGK14} it is shown that, based
of SDSS data, this relatively bright spiral galaxy has flat metallicity
gradient. Their estimations of 12+$\log$(O/H) (8.20$\pm$0.06) are consistent
 with our value 8.12$\pm$0.06 within uncertainties. We therefore adopt
12+$\log$(O/H) from the direct $T_{\rm e}$ method for one of \HII\
regions in the galaxy.

\textit{UGC12769=J2345-0116} is a spiral with the possible metallicity
gradient. We give estimations of 12+$\log$(O/H) for two regions in this
 galaxy: 7.87$\pm$0.06 near the center and 7.55$\pm$0.05 at
$\sim$5$\arcsec$\ from the center. Usually for such systems
with a visible gradient of O/H, the characteristic value of
O/H is adopted as measured at $R = R_{\mathrm 25}/2$. Both
our regions with the measured 12+$\log$(O/H) are located inside
this radius. Therefore the adopted value (the weighted mean,
7.63$\pm$0.13) should be treated as the upper limit for this galaxy.

\textit{Mrk965=J0057-0409}. On both BPT diagnostic diagrams
\citep{BPT}, I([O{\sc iii}])/I(H$\beta$) vs I([N{\sc ii}])/I(H$\alpha$) and
I([O{\sc iii}])/I(H$\beta$) vs I([S{\sc ii}])/I(H$\alpha$),
all three emission-line knots for which the 1d spectra were
subtracted fall into regions of photo-ionization. We therefore
use the standard methods applicable for \HII\ regions to derive
the estimates of their O/H. It is worth noticing though, that the
standard BPT diagrams are constructed for solar metallicities. For
the lower metallicities, the contribution of shock waves could be
significant even in the part occupied by \HII\ regions on the classical
diagrams for solar metallicities \citep{Allen08}. That could be the case
for Mrk965 with its very disturbed morphology, high
I([O{\sc ii}]3727/H$\beta$) and I([N{\sc ii}]6584/H$\beta$)
and low I([O{\sc iii}]4959,5007/H$\beta$). The estimated
values of 12+$\log$(O/H) should be used with caution.

\textit{SHOC053=J0109+0107}. In the paper by \citet{Kniazev04b}
the value of 12+$\log$(O/H)=7.54$\pm$0.15 was obtained via the
modified $T_{\rm e}$ method based on the SDSS spectrum. \citet{Guseva09}
measured its 12+$\log$(O/H)=7.98$\pm$0.04 via the direct $T_{\rm e}$
method on the spectrum obtained with the 3.6m ESO telescope. However,
from the  visual inspection of their spectrum, it looks like the
accuracy of [O {\sc iii}]~$\lambda$4363 line measurement might be
overestimated.
Therefore, their value of O/H should be used with caution. The value
of 12+$\log$(O/H) for SHOC053 estimated via the semi-empirical method
with the use of line fluxes provided by \citet{Guseva09} is 7.73$\pm$0.09.
This value is in a better agreement with that from \citet{Kniazev04b} and
corresponds much better to the galaxy's absolute magnitude
($M_{\rm B} = -14.3$). We thus use this estimation as the
adopted value for SHOC053.

\newcounter{cc}
\begin{table*}
\begin{center}
\caption{\label{tab:main}
Main parameters of Eridanus void galaxies with new and updated O/H data}
\vspace{0.3cm}
\footnotesize{
\begin{tabular}{r|l|c|r|l|l|r} \hline \\[-0.2cm]
\multicolumn{1}{c|}{\#}                 &
\multicolumn{1}{c|}{Name or}            &
\multicolumn{2}{c|}{Coordinates (J2000)} &
\multicolumn{1}{r}{12+$\log$(O/H),}	&
\multicolumn{1}{r}{12+$\log$(O/H),}     &
\multicolumn{1}{r}{Source}       \\
&
\multicolumn{1}{c|}{prefix }  &
\multicolumn{2}{c|}{of objects}  &
\multicolumn{1}{c}{direct}  &
\multicolumn{1}{c|}{adopted}  &
\multicolumn{1}{c|}{}  \\
&
\multicolumn{1}{c|}{ (1) }  &
\multicolumn{1}{c|}{ (2) }  &
\multicolumn{1}{c|}{ (3) }  &
\multicolumn{1}{c|}{ (4) }  &
\multicolumn{1}{c|}{ (5) }  &
\multicolumn{1}{c|}{ (6) }    \\

\\[-0.2cm] \hline \\[-0.2cm]

\cc&SDSS J0013-0106         &00 13 40.8& -01 06 41 & --            &7.52$\pm$0.05 & 2, SALT\\
\cc&UM~38              &00 28 51.5& +03 29 23 & 7.95$\pm$0.05 &7.95$\pm$0.05 & 1, SALT\\
\cc&6dF J0027-0311     &00 27 55.3& $-$03 11 01 & --            &7.76$\pm$0.08 & 2, SALT\\
\cc&UM~40              &00 28 26.6& +05 00 16 & 7.97$\pm$0.10 &7.97$\pm$0.10 & 1, SALT\\
\cc&APM~B0029+0226	   &00 31 45.3& +02 42 53 & --            &7.48$\pm$0.06 & 2, SALT\\	
\cc&UGC~328            &00 33 22.1& $-$01 07 17 & --            &7.74$\pm$0.08 & 2, SALT\\
\cc&HIPASS J0041-01b   &00 41 43.1& $-$01 59 18 & 7.68$\pm$0.11 &7.60$\pm$0.03 & 3, SALT\\
\cc&IC~52              &00 48 23.8& +04 05 31 & 8.01$\pm$0.06 &8.01$\pm$0.06 & 1, SALT\\
\cc&UGC 527            &00 51 49.9& +03 06 21 & 7.84$\pm$0.14 &7.84$\pm$0.03 & 4, BTA\\
\cc&UM 285             &00 51 58.7& $-$01 40 19 & 7.77$\pm$0.06 &7.77$\pm$0.06 & 1, SALT\\
\cc&ARK 18 (a)         &00 51 59.7& $-$00 29 21 & 8.08$\pm$0.10 &8.08$\pm$0.10 & 1, SDSS\\
\cc&ARK 18 (b,nuc)     &00 51 59.6& $-$00 29 12 & --            &7.55$\pm$0.04 & 2, SDSS\\
\cc&MCG-01-03-027      &00 52 17.2& $-$03 58 00 & --            &7.72$\pm$0.07 & 2, SALT\\
\cc&MRK 965            &00 57 28.7& $-$04 09 34 & --            &7.59$\pm$0.06 & 2, BTA\\
\cc&SDSS J0057+0052    &00 57 56.6& +00 52 09 & --            &7.80$\pm$0.03 & 2, SALT\\
\cc&MCG-01-03-072      &01 02 22.9& -04 30 31 & 7.84$\pm$0.09 &7.84$\pm$0.09 & 1, SALT\\
\cc&6dF J0105-0251     &01 05 55.4& -02 51 36 & --            &7.94$\pm$0.05 & 2, SALT\\
\cc&UGC 711            &01 08 36.9& +01 38 30 & 7.91$\pm$0.08 &7.91$\pm$0.08 & 1, BTA\\
\cc&NGC 428            &01 12 55.7& +00 58 54 & 8.12$\pm$0.06 &8.12$\pm$0.06 & 1,5a, SDSS\\
\cc&MCG +00-04-049     &01 14 20.5& +00 55 02 & 7.56$\pm$0.07 &7.56$\pm$0.07 & 1, BTA\\
\cc&UGC 12729          &23 40 20.7& +01 14 45 & --            &7.84$\pm$0.08 & 2, SALT,SDSS\\
\cc&APM B2341-0330     &23 43 55.0& -03 13 35 & --            &7.62$\pm$0.05 & 2, SALT\\
\cc&B2342-0223         &23 44 48.2& -02 06 54 & --            &7.62$\pm$0.04 & 2, SALT\\
\cc&UGC 12769          &23 45 19.2& -01 16 24 & --            &7.63$\pm$0.13 & 2, SALT\\
\cc&SDSS J2356+0141         &23 56 26.8& +01 41 17 & --            &7.72$\pm$0.06 & 2, SALT\\
\cc&HIPASS J2359+02    &23 59 13.2& +02 43 26 & --            &7.79$\pm$0.08 & 2, SALT\\
\\[-0.25cm] \hline \\[-0.2cm]
\cc&APM B0001+0008         &00 04 21.6& +00 25 35 &      --       &7.45$\pm$0.09& 5b \\
\cc&SDSS J0015+0104    &00 15 20.7& +01 04 37 &      --       &7.17$\pm$0.03& 5c \\
\cc&SDSS J0016+0108    &00 16 28.2& +01 08 02 &      --       &7.53$\pm$0.06& 5b \\
\cc&UM 240             &00 25 07.4& +00 18 46 & 7.89$\pm$0.08 &7.89$\pm$0.08& 5d \\
\cc&UGC 300            &00 30 04.1& +03 30 47 & 7.80$\pm$0.03 &7.80$\pm$0.03& 5e \\
\cc&SDSS J0057-0021    &00 57 12.6& -00 21 58 & 7.60$\pm$0.01 &7.60$\pm$0.01& 5b \\
\cc&UGC 695            &01 07 46.4& +01 03 49 & 7.69$\pm$0.12 &7.69$\pm$0.12& 5f \\
\cc&SHOC 053           &01 09 08.0& +01 07 15 &      --       &7.73$\pm$0.09& 5b \\
\cc&UGC 772            &01 13 39.4& +00 52 28 & 7.27$\pm$0.03 &7.27$\pm$0.03& 5g \\
\cc&UM 323              &01 26 46.5& -00 38 45 & 7.96$\pm$0.04 &7.96$\pm$0.04& 5h \\
\cc&SDSS J2354-0005    &23 54 37.3& -00 05 02 & 7.36$\pm$0.13 &7.36$\pm$0.13& 5b \\
\end{tabular}
}
\end{center}
a) \citet{PGK14}
b) \citet{Guseva09}
c) \citet{Guseva17} 
d) \citet{Kniazev04b}
e) \citet{vanZee97}
f) \citet{Berg12}
g) \citet{Izotov12} - weighted mean for several knots
h) \citet{Izotov06} 
\end{table*}

\section[]{DISCUSSION AND SUMMARY}
\label{sec:dis}

\subsection{Eridanus void galaxy sample in context}
\label{ssec:context}

The sample of 66 galaxies presented here that reside in the equatorial
part of the Eridanus void is only the second one of its kind with a large
number of galaxies covering a broad range of luminosities: $M_{\rm B}$
between --12.0$^m$ and --19.5$^m$. It is useful to compare this void
galaxy sample with the first sample of its kind, the 108 galaxies in
the Lynx-Cancer void.

The main differences in the sample properties are related to their
median distances, which, for the Eridanus void, is about a factor of
1.7 larger. This translates to the difference in the median
$M_{\rm B}$ $\sim$ -15.8 for the Eridanus sample versus $\sim$ -14.45 for
the Lynx-Cancer void sample \citep{Pap6}. Due to the same limit in apparent
magnitudes for the large galaxy redshift surveys ($B \sim 18.5^m$), this
difference in distance also results in a much reduced relative number of
low luminosity dwarfs. There are only four galaxies in the Eridanus void
with $M_{\rm B} > -14^{m}$ in comparison to 40 such objects in the updated
Lynx-Cancer void sample \citep{Pap6,Pap7}.

It is worth mentioning that a substantial part of the lowest luminosity
dwarfs, among which the most unusual void objects were discovered
\citep{CP2013,void_LSBD,Pap7,CPE2017}, were found with the dedicated
search for faint void galaxies, in particular as possible counterparts
of already known void dwarfs. In this context, the SDSS Stripe82
\citep{Stripe82}, which covers the central part of the Eridanus void
equatorial region ($\sim 18$\% of the whole sky area for the void sample
in the question) and goes about 2 magnitudes deeper than the SDSS itself,
can serve as a good basis to search for the fainter galaxy population in
this void.

\subsection{Relationship between O/H and galaxy luminosity} 
\label{ssec:OHvsLum}

The main issue we wish to address with this new Eridanus
void galaxy sample is their evolutionary status
in comparison to similar galaxies in a denser
environment. One of the goals is to extend our mass study of galaxy
evolution in the nearby Lynx-Cancer void \citep{PaperI,void_OH,Pap6,Pap7}
to another void sample, selected with the same criteria and
studied with the same method, and to examine how the main
results obtained for the Lynx-Cancer void sample are similar or
different to those derived on a completely independent sample. 
        
We adopt, as a comparison galaxy sample in a denser environment, the sample
of \citet{Berg12}, representing the Local Volume galaxies with well defined,
velocity independent distances and luminosities
and the gas O/H values derived via the direct $T_{\rm e}$ method. The same
sample was used for comparison in the Lynx-Cancer galaxy void study
\citep{Pap7}.

In Fig.~\ref{fig:ZvsMB} we plot $\log$(O/H) versus $M_{\rm B}$ of our
void galaxies along with the data of \citet{Berg12} 'control' sample.
 Our galaxies are shown with filled 'symbols' with error bars in O/H.
The control sample is presented by the solid line of their sample's linear
regression, while the two dashed lines parallel to the solid one show
the control sample's scatter around the linear regression with
$\sigma$=0.15 dex in $\log$(O/H). The dotted line parallel to the solid
one, displaced by --0.30 dex, shows the deviation of --2$\sigma$. The
galaxies below this dotted line are treated as the most extreme outliers.
The horizontal dotted black line marks the value Z\sunn/10 which corresponds
to 12+$\log$(O/H)=7.69 for Z\sunn\ from \citet{AllendePrieto2001} and
\citet{Asplund09}.

Similar to the analysis in \citet{Pap7}, we plot in the left panel of
Fig.~\ref{fig:ZvsMB} separately the subsample of galaxies with O/H derived
with the direct $T_{\rm e}$ method. In the right panel, we show the O/H
values of the whole Eridanus galaxy sample. As one can see, similar to
the results for the Lynx-Cancer void sample, there is no bias of the
directly derived O/H with respect to the whole Eridanus sample.
The black solid line in both panels represents the linear regression
on the whole Eridanus sample and it lies significantly below the
linear regression for the control sample.

\begin{figure*}
 \centering
 \includegraphics[angle=-90,width=0.8\linewidth]{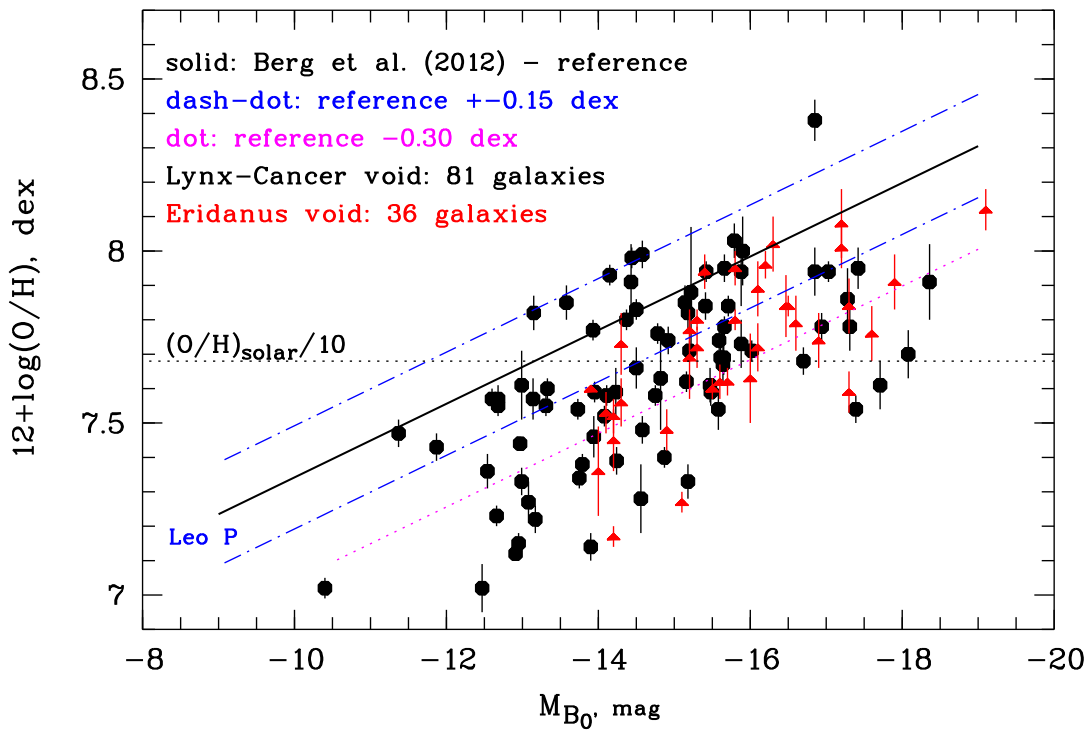}
  \caption{\label{fig:OHvsMB_both}
The relation between $\log$(O/H) and the absolute blue magnitude $M_{\rm B}$
for 36 Eridanus void galaxies (red triangles) and 81 Lynx-Cancer void
galaxies (black circles). The solid line shows the linear regression
for the control sample from the Local Volume by \citet{Berg12}.
Two dash-dotted lines on the both sides of this reference line show the
r.m.s. deviation of their sample from the linear regression (0.15 dex). The
dotted line (displaced at --0.30~dex from the reference one) separates the
region where the most deviating metal-poor dwarfs are situated.
}
\end{figure*}

This result is well consistent with the  conclusion drawn by \citet{Pap7}
from the gas metallicity study for 81 Lynx-Cancer void galaxies. To increase
the statistical significance of finding lower O/H values in void galaxies,
we combine in Fig.~\ref{fig:OHvsMB_both} the data for galaxies from both
voids, with a total of 117 objects, thus increasing the number of analyzed
void galaxies by a factor of 1.5. This plot clearly demonstrates
systematically lower O/H for the full range of void galaxy luminosities.
This fact is naturally treated as evidence of the slower evolution of void
galaxies with respect to similar objects residing in denser environments.

Besides, for the least luminous dwarfs we see a small subgroup of the
lowest O/H objects with a highly elevated mass fraction of gas. Due to the
observational selection effects, most of these unusual dwarfs come from the
Lynx-Cancer void sample \citep{Pap7}. 
For the Eridanus void, the only known representative of this group is SDSS
J0015+0104 \citep{Eridan_XMD}, with the lowest 12+$\log$(O/H)=7.17
\citep{Guseva17} among this void galaxies published to date. Another galaxy
SDSS J0110-0000, the faintest dwarf in this void sample ($M_{\rm B}$=--12),
was found very recently to have similar properties and even the lower O/H
(Pustilnik et al. 2018, in preparation). This emphasizes the bias due to the
observational selection for finding such unusual galaxies in voids at
$D > 25$ Mpc.

\subsection{Notes on individual interesting void galaxies}
\label{ssec:notes}

\textit{Ark18 = J0051--0029}. The SDSS Stripe82 images of this galaxy
reveal an underlying LSB disk. The metallicity in the center may be lower
than in the bright non-central \HII-region due to an accretion event of
fresh gas to the galaxy's center.
The object is in the list of "Local Orphan Galaxies"\ (LOG)
by \citet{Karachentsev11} which includes the most isolated nearby galaxies.

\textit{UM40 = J0028+0500; 6dF J0055--0601.} The LSB structures around the
brighter parts of both galaxies are well visible on deep DECaLS (The Dark
Energy Camera Legacy Survey, \citet{Blum2016}) images. These could be the
underlying LSB disks or the result of recent interaction.
UM40 also is in the list of LOG.

\textit{J0142+0018, J0142+0021, J0143-0002.} Galaxies J0142+0018 with
$V_{\rm hel} =$ 3222~\kms\ and J0142+0021 with $V_{\rm hel} =$~3243~\kms\ --
comprise a pair with a projected separation of 50 kpc. Galaxy J0143--0002
with $V_{\rm hel} =$ 3165 \kms\ may belong to the same large scale structure
(filament).

\subsection{Relation to previous works and models}
\label{ssec:previous_works}



Despite to rather little studies known to date on the relation
of galaxy evolution to their global environment, it is worth to compare
our results with previous ones. As already mentioned, this work is
motivated by the results on gas metallicity for the galaxy sample in the
nearby Lynx-Cancer void \citep{Pap7}.
In Sec.~\ref{ssec:OHvsLum} we demonstrate that our results and conclusions
for the Eridanus void galaxy sample are completely consistent with their.
A similar study, but limited by only seven galaxies from the Void Galaxy
Survey (VGS) was presented by \citet{Kreckel15}. They did not find from their
data an indication for the reduced metallicity in void galaxies, in opposite
to our results. However, as discussed in \citet{Pap7}, if we take into
account only three their objects with reliable O/H, derived via the direct
T$_{\rm e}$ method, they appear consistent with our results.

A recent mass study of extremely metal-poor (XMP) galaxies
(namely with 12+$\log$(O/H)$\leq$7.69), based on the SDSS spectra,
was presented by \citet{SanchezAlmeida16}. It was supplemented with analysis
of their large-scale environment based on constrained N-body cosmological
simulations of the local Universe \citep{Nuza2014}. With a reservation on
the different definition of voids in that work, the pronounced increase of
probability to find XMP galaxies in voids with respect of sheets, and
especially of 'knots' (groups) and filaments, is striking. This finding is
in full agreement with our results on generally reduced metallicity of void
late-type galaxies for the same luminosity in comparison to similar galaxies
in denser environments.

One more evidence for DM halo assembly bias as a function of environment
was presented recently by \citet{Tojeiro17}. They classify (using a tidal
tensor method) 'Galaxy and Mass Assembly' (GAMA) galaxies to reside in voids,
sheets, filaments or knots, and find that low-mass haloes residing in knots
are older than haloes of the same mass in voids. This result is consistent
with our conclusion on the less advanced phase of void galaxy evolution
with respect of their counterparts in denser environments.

As concerned to comparison with model works, they, as far as we are aware, up
to date give too little concrete predictions on galaxy properties in the
void-type environment. Especially this relates to the baryonic mass range
($M_{\rm bar} \sim 10^{7}$ to 3$\times 10^{8}$ M\sunn) in which we find
the most unusual objects mimicing the youngest local population.
We mention paper by \citet{Hahn07} indicating the general delay in time
of DM haloes formation in voids with respect of that in denser structures,
but their mass resolution is not suitable to assess the majority of void
population.

In a more targeted study, \citet{Kreckel11b} present modelling
of void galaxy stellar and gas properties in the standard CDM cosmology.
Despite their CDM/gas mass resolution (1.5$\times$10$^{8}$/2$\times$10$^{7}$
M\sunn) is somewhat coarse for the least massive void galaxies, they are able
to conclude the following. The most luminous galaxies with M$_{\rm r} < -18$
do not show trends in their properties for various density of environment.
However, dwarfs in voids with M$_{\rm r} > -16$ appear bluer, with higher
SFR and lower mean stellar ages than galaxies in the average density
environments. These predictions seem find a support from our results on the
slower evolution of void galaxies and the presence of a fraction of unusual,
looking very young, dwarfs.

In the other related paper, \citet{Aragon2013} explore dynamics of void
structure. Since the interior of void expands with respect of the surrounding
region of the Universe, voids
mimic the time machine, in the sense that the develompent of structures in
voids is significantly delayed with respect of that in denser surroundings.
From this, one can suggest that at least part of void DM halos (and related
galaxies) formed at later epochs. Due to much reduced rate of galaxy
interactions in voids, they
could keep their unevolved state till the current epochs and thus show up
as rather unusual objects. Probably such late galaxies are one of the options
for the most metal-poor and gas-rich void dwarfs.

\subsection{Conclusions}
\label{ssec:conclusions}

Summarizing the results and discussion above, we draw the following
conclusions:

\begin{enumerate}
\item
The galaxy sample residing in the equatorial part of the large `Eridanus' void
is compiled based on the criterion of sufficiently large distance
($D_{\rm NN} >$2.0 Mpc) to the luminous \textbf{($M_{\rm B} < M_{\rm B}^{*} +1.0 = -19.5^{m}$)} galaxies delineating the void region. The current number of
identified void galaxies within the selected part of the void is 66, with the
range of $M_{\rm B}$ = [--12.0,--19.1] and a median value of $-15.8^{m}$.
\item
The analysis of the new spectral observations for 23 void galaxies with SALT
and BTA, complemented with 3 available SDSS spectra, allowed us to determine
the gas-phase oxygen abundance for 25 void objects. Compiling these data with
O/H estimates for another 11 void galaxies available in the literature, we
analyse "$\log$(O/H) vs $M_{\rm B}$" relation for 36 objects.
The comparison of this relation with that for the control sample of similar
galaxies residing in the denser environment of the Local Volume
\citep{Berg12} results in the conclusion that the Eridanus void galaxies
show systematically lower values of O/H, that is, the direct indication of their less advanced phase of evolution, and hence, either of the
slower chemical evolution, or later formation, or both.
\item
The combined results of the previous study of gas O/H in 81 galaxies in the
nearby Lynx-Cancer void with the current data for 36 galaxies of the Eridanus
void sample give unambiguous evidence for substantially reduced O/H
(on average by $\sim$40\%) for a given galaxy luminosity in the void
environment. Besides, a small fraction of void galaxies appear to be
unusually gas-rich and extremely metal-poor with properties of a recent
(one - a few Gyr) turn-on of star formation. This finding is consistent
with theoretical expectations on the 'special' unusual properties of part
of the void galaxy population.  Therefore, the dedicated search for such
objects among void sample galaxies can be one of the most efficient means
of their detection.
\item
The unusual void galaxies mentioned above appear to be intrinsically faint
and are mostly LSB dwarfs, with $M_{\rm B} \gtrsim -14$. This points out
the importance of the severe observational selection effect in search for
this void population. If one uses the existing spectral surveys with the
typical depth of $B \sim 18.5^m-19^m$, only the nearby voids, with
$D \lesssim 25$ Mpc, can be studied. In more distant voids, such objects
can be searched via dedicated spectroscopy or \HI\ observations of potential
fainter LSB companions of the known, more luminous void dwarfs.
\end{enumerate}

\section*{Acknowledgements}

AYK, ESE and SAP acknowledge the partial support of the initial part of
this project (assembling the void sample, observations, data reduction)
through the RFBR grant No.~14-02-00520. The analysis of results and
preparation of publication was conducted with support
 to SAP and ESE from Russian Scientific Fund Program RSCF~14-12-00965.
AYK acknowledges the support from the National Research Foundation (NRF)
of South Africa.
We thank the anonymous reviewer for questions, comments and suggestions
which helped us to improve the paper quality. We acknowledge the usage of the HyperLeda database (http://leda.univ-lyon1.fr).
This research has made use of the NASA/IPAC Extragalactic
Database (NED), which is operated by the Jet Propulsion Laboratory,
California Institute of Technology, under contract with the National
Aeronautics and Space Administration.
The authors acknowledge the spectral and photometric data and the related
information available in the SDSS database used for this study.
Funding for SDSS-III has been provided by the Alfred P. Sloan Foundation,
the Participating Institutions, the National Science Foundation, and the
U.S. Department of Energy Office of Science. The SDSS-III web site is
http://www.sdss3.org/.
SDSS-III is managed by the Astrophysical Research Consortium for the
Participating Institutions of the SDSS-III Collaboration including
the University of Arizona, the Brazilian Participation Group, Brookhaven
National Laboratory, Carnegie Mellon University, University of Florida,
the French Participation Group, the German Participation Group, Harvard
University, the Instituto de Astrofisica de Canarias, the Michigan
State/Notre Dame/JINA Participation Group, Johns Hopkins University,
Lawrence Berkeley National Laboratory, Max Planck Institute for Astrophysics,
Max Planck Institute for Extraterrestrial Physics, New Mexico State
University, New York University, Ohio State University, Pennsylvania
State University, University of Portsmouth, Princeton University, the
Spanish Participation Group, University of Tokyo, University of Utah,
Vanderbilt University, University of Virginia, University of
Washington, and Yale University.


\section*{APPENDIX}
\label{sec:append}

\newcounter{qqq}
\begin{table*}
\begin{center}
 \caption{\label{tab:vel}
New galaxies with SALT and BTA measured velocities}
\vspace{0.3cm}
\footnotesize{
\begin{tabular}{r|l|c|r|c|r|r|r|r|r} \hline \\[-0.2cm]
\multicolumn{1}{c|}{\#}                 &
\multicolumn{1}{c|}{Name or}            &
\multicolumn{2}{c|}{Coordinates (J2000)} &
\multicolumn{1}{c|}{g}            &
\multicolumn{1}{c|}{Date of}            &
\multicolumn{1}{c|}{$V_{\rm hel}$,}     &
\multicolumn{1}{c|}{Telescope}\\
&
\multicolumn{1}{c|}{prefix }  &
\multicolumn{2}{c|}{ }  &
\multicolumn{1}{c|}{ }  &
\multicolumn{1}{c|}{observations}            &
\multicolumn{1}{r|}{\kms }  \\
&
\multicolumn{1}{c|}{ (1) }  &
\multicolumn{1}{c|}{ (2) }  &
\multicolumn{1}{c|}{ (3) }  &
\multicolumn{1}{c|}{ (4) }  &
\multicolumn{1}{c|}{ (5) }  &
\multicolumn{1}{c|}{ (6) }  &
\multicolumn{1}{c|}{ (7) }\\
\\[-0.2cm] \hline \\[-0.2cm]
\qqq&SDSS J0001+1350 & 00 01 41.3 & +13 50 33.2 & 17.4 & 2012.11.16 & 6332$\pm$19 & BTA \\
\qqq&SDSS J0005-0035 & 00 05 10.2 & -00 35 14.3 & 19.1 & 2012.11.16 & 9937$\pm$3 & SALT  \\
\qqq&SDSS J0006-0027 & 00 06 51.4 & -00 27 53.8 & 19.4 & 2012.12.03 & 9875$\pm$2 & SALT  \\
\qqq&SDSS J0008+1611 & 00 08 37.1 & +16 11 32.7 & 18.7 & 2012.11.17 & 20904$\pm$10 & BTA \\
\qqq&SDSS J0013+1523 & 00 13 31.3 & +15 23 25.2 & 18.5 & 2012.11.17 & 10951$\pm$10 & BTA \\
\qqq&SDSS J0014+1351 & 00 14 06.7 & +13 51 43.3 & 17.8 & 2012.11.17 & 1783$\pm$8 & BTA  \\
\qqq&SDSS J0019+1515 & 00 19 12.4 & +15 15 01.3 & 18.8 & 2012.11.17 & 7534$\pm$11 & BTA \\
\qqq&SDSS J0038+0106 & 00 38 23.8 & +01 06 22.3 & 19.0 & 2012.09.22 & 5849$\pm$5 & SALT  \\
\qqq&SDSS J0107+0110 & 01 07 50.9 & +01 10 20.3 & 18.5 & 2012.11.15 & 15786$\pm$8 & SALT  \\
\qqq&PGC 135629 & 01 12 50.7 & +01 02 48.5 & 18.7 & 2012.11.17 & 1103$\pm$8 & BTA \\
\qqq&SDSS J2340+1607 & 23 40 52.9 & +16 07 21.4 & 16.6 & 2012.11.17 & 4025$\pm$7 & BTA \\
\qqq&SDSS J2340+1355 & 23 40 58.7 & +13 55 00.8 & 18.9 & 2012.11.17 & 18342$\pm$3 & BTA \\
\qqq&SDSS J2341+1354 & 23 40 59.5 & +13 54 42.9 & 19.0 & 2012.11.17 & 18271$\pm$3 & BTA \\
\qqq&SDSS J2341+1515 & 23 41 43.8 & +15 15 20.9 & 19.4 & 2012.11.17 & 7656$\pm$8 & BTA \\
\qqq&SDSS J2344-0106 & 23 44 27.1 & -01 06 31.6 & 20.1 & 2012.09.08 & 6706$\pm$5 & SALT  \\
\qqq&SDSS J2344-0041 & 23 44 37.9 & -00 41 23.4 & 19.3 & 2012.11.17 & 6979$\pm$2 & SALT  \\
\qqq&SDSS J2347+0126 & 23 47 21.5 & +01 26 25.4 & 18.3 & 2012.07.28 & 2674$\pm$10 & SALT   \\
\qqq&SDSS J2355-0126 & 23 55 39.2 & -01 26 26.8 & 17.6 & 2012.07.10 & 9391$\pm$13 & SALT   \\
\qqq&SDSS J2355-0140 & 23 55 48.9 & -01 40 27.6 & 17.9 & 2012.07.25 & 8177$\pm$4 & SALT   \\
\qqq&SDSS J2356+0141 & 23 56 26.8 & +01 41 17.4 & 17.2 & 2012.07.25 & 2496$\pm$13 & SALT   \\
\qqq&SDSS J2359-0052 & 23 59 32.7 & -00 52 50.6 & 19.3 & 2012.12.04 & 10075$\pm$5 & SALT    \\

\\[-0.2cm] \hline \\[-0.2cm]
\end{tabular}
\begin{tablenotes}

\item The radial velocity of PGC~135629 was known from \HI-mapping with VLA
\citep{Smoker96} to be $V_{\rm hel} =$ 1105~\kms. We attempted to find
\HII-regions in this LSB dwarf suitable for O/H determination. The only
faint region of H$\alpha$ emission was visible on the acquisition image with
the medium width filter SED665. The H$\alpha$ line of this region in our
spectrum is also rather weak.

\end{tablenotes}
}
\end{center}
\end{table*}



\title[Study of galaxies in the Eridanus void. I. Sample and Abundances ]
{Study of galaxies in the Eridanus void. I. Sample and element abundances}
\author[A.Y.~Kniazev, E.S.~Egorova, S.A.~Pustilnik]
{A.Y. Kniazev,$^{1,2,3}$ E.S.~Egorova,$^{3}$ S.A. Pustilnik$^{4}$ \\
\rule{-4pt}{20pt}
$^1$ South African Astronomical Observatory, PO Box 9, 7935 Observatory,
   Cape Town, South Africa\\
$^2$ Southern African Large Telescope Foundation, PO Box 9, 7935 Observatory,
   Cape Town, South Africa \\
$^{3}$ Sternberg Astronomical Institute, Lomonosov Moscow State University,
Moscow, Russia \\
$^4$ Special Astrophysical Observatory of RAS, Nizhnij Arkhyz,
  Karachai-Circassia 369167, Russia
}


\begin{table*}
\centering{
\caption{Measured and corrected line intensities, and derived oxygen abundances (SALT data)}
\begin{threeparttable}
\label{t:Intens_SALT1}
\begin{tabular}{l|c|c|c|c|c|c} \hline
\rule{0pt}{10pt}
& \MC{2}{c|}{J0013-0106}
& \MC{2}{c|}{UM 38}
& \MC{2}{c}{6dF J0027-0311}     \\ \hline
\rule{0pt}{10pt}
$\lambda_{0}$(\AA) Ion    & F($\lambda$)/F(H$\beta$)&I($\lambda$)/I(H$\beta$) & F($\lambda$)/F(H$\beta$)&I($\lambda$)/I(H$\beta$)  & F($\lambda$)/F(H$\beta$)&I($\lambda$)/I(H$\beta$) \\ \hline
3727\ [O\ {\sc ii}]\                & 1.965$\pm$0.885 & 1.486$\pm$0.900 & 2.673$\pm$0.066 & 2.549$\pm$0.074 & 2.941$\pm$0.300 & 3.728$\pm$0.402 \\
3868\ [Ne\ {\sc iii}]\              &       ...       &       ...       & 0.292$\pm$0.018 & 0.277$\pm$0.019 & 0.184$\pm$0.091 & 0.224$\pm$0.115 \\
3889\ He\ {\sc i}\ +\ H8\           &       ...       &       ...       & 0.097$\pm$0.015 & 0.203$\pm$0.040 & 0.177$\pm$0.102 & 0.249$\pm$0.173 \\
3967\ [Ne\ {\sc iii}]\ +\ H7\       &       ...       &       ...       & 0.126$\pm$0.008 & 0.223$\pm$0.019 &       ...       &       ...       \\
4101\ H$\delta$\                    &       ...       &       ...       & 0.178$\pm$0.010 & 0.273$\pm$0.020 & 0.220$\pm$0.080 & 0.288$\pm$0.139 \\
4340\ H$\gamma$\                    & 0.319$\pm$0.172 & 0.454$\pm$0.357 & 0.398$\pm$0.011 & 0.465$\pm$0.017 & 0.392$\pm$0.054 & 0.457$\pm$0.100 \\
\end{tabular}
\begin{tablenotes}
\item For J0013-0106, 6dF J0027-0311 the adopted O/H in Table~\ref{tab:main} is the weighted mean of all 3 available methods.

For UM 38 the adopted O/H in Table~\ref{tab:main} is $O/H_{\rm Te}$.
\end{tablenotes}

\end{threeparttable}
}
\end{table*}

\begin{table*}
\centering{
\caption{
Measured and corrected line intensities, and derived oxygen abundances (SALT data)}
\label{t:Intens_SALT2}
\begin{threeparttable}
\begin{tabular}{l|c|c|c|c|c|c} \hline
\rule{0pt}{10pt}
& \MC{2}{c|}{UM 40}
& \MC{2}{c|}{APM B0029+0226}
& \MC{2}{c}{UGC 328}     \\ \hline
\rule{0pt}{10pt}
$\lambda_{0}$(\AA) Ion    & F($\lambda$)/F(H$\beta$)&I($\lambda$)/I(H$\beta$) & F($\lambda$)/F(H$\beta$)&I($\lambda$)/I(H$\beta$)  & F($\lambda$)/F(H$\beta$)&I($\lambda$)/I(H$\beta$) \\ \hline
3727\ [O\ {\sc ii}]\                & 2.171$\pm$0.075 & 1.880$\pm$0.081 & 2.476$\pm$0.458 & 1.868$\pm$0.485 & 3.241$\pm$0.382 & 3.183$\pm$0.389 \\
3868\ [Ne\ {\sc iii}]\              & 0.507$\pm$0.068 & 0.438$\pm$0.070 &       ...       &       ...       &       ...       &       ...       \\
3889\ He\ {\sc i}\ +\ H8\           & 0.039$\pm$0.048 & 0.208$\pm$0.399 &       ...       &       ...       &       ...       &       ...       \\
3967\ [Ne\ {\sc iii}]\ +\ H7\       & 0.065$\pm$0.020 & 0.212$\pm$0.098 &       ...       &       ...       &       ...       &       ...       \\
4101\ H$\delta$\                    & 0.090$\pm$0.015 & 0.241$\pm$0.060 &       ...       &       ...       & 0.196$\pm$0.043 & 0.213$\pm$0.057 \\
4340\ H$\gamma$\                    & 0.389$\pm$0.021 & 0.478$\pm$0.034 & 0.261$\pm$0.129 & 0.474$\pm$0.374 & 0.438$\pm$0.077 & 0.447$\pm$0.084 \\
\end{tabular}
\begin{tablenotes}
\item For UM 40 the adopted O/H in Table~\ref{tab:main} is $O/H_{\rm Te}$

For APM B0029+0226, UGC 328 the adopted O/H in Table~\ref{tab:main} is the weighted mean of all 3 available methods.
\end{tablenotes}

\end{threeparttable}
}
\end{table*}

\begin{table*}
\centering{
\caption{
Measured and corrected line intensities, and derived oxygen abundances (SALT data)}
\begin{threeparttable}
\label{t:Intens_SALT3}
\vspace{0.3cm}
\begin{tabular}{l|c|c|c|c|c|c} \hline
\rule{0pt}{10pt}
& \MC{2}{c|}{HIPASS J0041-01b}
& \MC{2}{c|}{IC 52 (a)}
& \MC{2}{c}{IC 52 (b)}     \\ \hline
\rule{0pt}{10pt}
$\lambda_{0}$(\AA) Ion    & F($\lambda$)/F(H$\beta$)&I($\lambda$)/I(H$\beta$) & F($\lambda$)/F(H$\beta$)&I($\lambda$)/I(H$\beta$)  & F($\lambda$)/F(H$\beta$)&I($\lambda$)/I(H$\beta$) \\ \hline
3727\ [O\ {\sc ii}]\                & 1.514$\pm$0.064 & 1.502$\pm$0.067 & 2.538$\pm$0.068 & 2.395$\pm$0.074 & 3.890$\pm$0.183 & 3.635$\pm$0.238 \\
3868\ [Ne\ {\sc iii}]\              & 0.176$\pm$0.018 & 0.175$\pm$0.018 & 0.200$\pm$0.012 & 0.188$\pm$0.013 &       ...       &       ...       \\
3889\ He\ {\sc i}\ +\ H8\           & 0.124$\pm$0.019 & 0.134$\pm$0.026 & 0.113$\pm$0.010 & 0.215$\pm$0.026 &       ...       &       ...       \\
3967\ [Ne\ {\sc iii}]\ +\ H7\       & 0.154$\pm$0.021 & 0.162$\pm$0.027 & 0.139$\pm$0.008 & 0.228$\pm$0.019 &       ...       &       ...       \\
4101\ H$\delta$\                    & 0.204$\pm$0.018 & 0.212$\pm$0.025 & 0.184$\pm$0.010 & 0.264$\pm$0.019 & 0.042$\pm$0.021 & 0.282$\pm$0.252 \\
4340\ H$\gamma$\                    & 0.401$\pm$0.035 & 0.411$\pm$0.030 & 0.412$\pm$0.013 & 0.466$\pm$0.018 & 0.264$\pm$0.039 & 0.462$\pm$0.103 \\
\end{tabular}
\begin{tablenotes}
\item For HIPASS J0041-01b the adopted O/H in Table~\ref{tab:main} is the weighted mean of all 4 available methods.

For IC52 the adopted O/H in Table~\ref{tab:main} is the weighted mean of $O/H_{\rm Te}$ for spectra a) and c)\textbf{.
}\end{tablenotes}
\end{threeparttable}
}
\end{table*}

\begin{table*}
\centering{
\caption{
Measured and corrected line intensities, and derived oxygen abundances (SALT data)}
\begin{threeparttable}
\label{t:Intens_SALT4}
\vspace{0.3cm}
\begin{tabular}{l|c|c|c|c|c|c} \hline
\rule{0pt}{10pt}
& \MC{2}{c|}{IC 52 (c)}
& \MC{2}{c|}{UM 285}
& \MC{2}{c}{MCG-01-03-027}     \\ \hline
\rule{0pt}{10pt}
$\lambda_{0}$(\AA) Ion    & F($\lambda$)/F(H$\beta$)&I($\lambda$)/I(H$\beta$) & F($\lambda$)/F(H$\beta$)&I($\lambda$)/I(H$\beta$)  & F($\lambda$)/F(H$\beta$)&I($\lambda$)/I(H$\beta$) \\ \hline
3727\ [O\ {\sc ii}]\                & 2.785$\pm$0.081 & 2.790$\pm$0.097 & 1.588$\pm$0.064 & 1.648$\pm$0.074 & 3.173$\pm$0.471 & 2.923$\pm$0.571 \\
3868\ [Ne\ {\sc iii}]\              & 0.173$\pm$0.032 & 0.170$\pm$0.036 & 0.292$\pm$0.019 & 0.299$\pm$0.021 &       ...       &       ...       \\
3889\ He\ {\sc i}\ +\ H8\           & 0.077$\pm$0.025 & 0.215$\pm$0.099 & 0.127$\pm$0.017 & 0.190$\pm$0.030 &       ...       &       ...       \\
3967\ [Ne\ {\sc iii}]\ +\ H7\       & 0.110$\pm$0.020 & 0.226$\pm$0.056 & 0.181$\pm$0.013 & 0.243$\pm$0.022 &       ...       &       ...       \\
4101\ H$\delta$\                    & 0.150$\pm$0.020 & 0.261$\pm$0.048 & 0.182$\pm$0.012 & 0.241$\pm$0.021 &       ...       &       ...       \\
4340\ H$\gamma$\                    & 0.392$\pm$0.026 & 0.475$\pm$0.032 & 0.447$\pm$0.020 & 0.487$\pm$0.025 &       ...       &       ...       \\
\end{tabular}

 \begin{tablenotes}
\item For UM 285 the adopted O/H in Table~\ref{tab:main} is $O/H_{\rm Te}$.

For MCG-01-03-027 the adopted O/H in Table~\ref{tab:main} is the weighted mean of all 3 available methods.
\end{tablenotes}
 \end{threeparttable}
  }
\end{table*}

\begin{table*}
\centering{
\caption{
Measured and corrected line intensities, and derived oxygen abundances (SALT data)}
\begin{threeparttable}
\label{t:Intens_SALT5}
\vspace{0.3cm}
\begin{tabular}{l|c|c|c|c|c|c} \hline
\rule{0pt}{10pt}
& \MC{2}{c|}{J0057+0052}
& \MC{2}{c|}{MCG-01-03-072 (a)}
& \MC{2}{c}{MCG-01-03-072 (b)}     \\ \hline
\rule{0pt}{10pt}
$\lambda_{0}$(\AA) Ion    & F($\lambda$)/F(H$\beta$)&I($\lambda$)/I(H$\beta$) & F($\lambda$)/F(H$\beta$)&I($\lambda$)/I(H$\beta$)  & F($\lambda$)/F(H$\beta$)&I($\lambda$)/I(H$\beta$) \\ \hline
3727\ [O\ {\sc ii}]\                & 5.421$\pm$1.588 & 5.121$\pm$2.170 & 1.970$\pm$0.080 & 2.244$\pm$0.096 & 2.855$\pm$0.204 & 2.850$\pm$0.223 \\
3868\ [Ne\ {\sc iii}]\              &       ...       &       ...       & 0.281$\pm$0.043 & 0.315$\pm$0.049 & 0.227$\pm$0.059 & 0.225$\pm$0.062 \\
3889\ He\ {\sc i}\ +\ H8\           &       ...       &       ...       & 0.156$\pm$0.040 & 0.175$\pm$0.051 & 0.138$\pm$0.055 & 0.178$\pm$0.080 \\
3967\ [Ne\ {\sc iii}]\ +\ H7\       &       ...       &       ...       & 0.215$\pm$0.021 & 0.239$\pm$0.030 & 0.264$\pm$0.056 & 0.301$\pm$0.072 \\
4101\ H$\delta$\                    &       ...       &       ...       & 0.245$\pm$0.017 & 0.267$\pm$0.026 & 0.184$\pm$0.032 & 0.232$\pm$0.053 \\
4340\ H$\gamma$\                    &       ...       &       ...       & 0.459$\pm$0.020 & 0.486$\pm$0.026 & 0.455$\pm$0.032 & 0.490$\pm$0.045 \\
\end{tabular}
\end{threeparttable}
\begin{tablenotes}
\item For J0057+0052 the adopted O/H in Table~\ref{tab:main} is the weighted mean of all 3 available methods.

For MCG-01-03-072 the adopted O/H in Table~\ref{tab:main} is $O/H_{\rm Te}$ for spectra a).
\end{tablenotes}
}
\end{table*}

\begin{table*}
\centering{
\caption{
Measured and corrected line intensities, and derived oxygen abundances (SALT data)}
\begin{threeparttable}
\label{t:Intens_SALT6}
\vspace{0.3cm}
\begin{tabular}{l|c|c|c|c|c|c} \hline
\rule{0pt}{10pt}
& \MC{2}{c|}{6dF J0105-0251}
& \MC{2}{c|}{UGC 12729}
& \MC{2}{c}{APM B2341-0330}     \\ \hline
\rule{0pt}{10pt}
$\lambda_{0}$(\AA) Ion    & F($\lambda$)/F(H$\beta$)&I($\lambda$)/I(H$\beta$) & F($\lambda$)/F(H$\beta$)&I($\lambda$)/I(H$\beta$)  & F($\lambda$)/F(H$\beta$)&I($\lambda$)/I(H$\beta$) \\ \hline
3727\ [O\ {\sc ii}]\                & 3.864$\pm$0.272 & 4.884$\pm$0.398 & 4.433$\pm$0.501 & 4.762$\pm$0.823 & 3.377$\pm$0.442 & 3.162$\pm$0.478 \\
4340\ H$\gamma$\                    & 0.356$\pm$0.049 & 0.472$\pm$0.087 & 0.168$\pm$0.098 & 0.482$\pm$0.502 &       ...       &       ...       \\
4861\ H$\beta$\                     & 1.000$\pm$0.075 & 1.000$\pm$0.092 & 1.000$\pm$0.130 & 1.000$\pm$0.201 & 1.000$\pm$0.135 & 1.000$\pm$0.177 \\
4959\ [O\ {\sc iii}]\               & 1.140$\pm$0.090 & 1.019$\pm$0.089 & 0.795$\pm$0.152 & 0.541$\pm$0.148 & 0.525$\pm$0.076 & 0.475$\pm$0.075 \\
5007\ [O\ {\sc iii}]\               & 3.160$\pm$0.191 & 2.792$\pm$0.186 & 2.486$\pm$0.264 & 1.666$\pm$0.255 & 1.645$\pm$0.175 & 1.487$\pm$0.175 \\
5876\ He\ {\sc i}\                  & 0.095$\pm$0.024 & 0.070$\pm$0.019 & 0.259$\pm$0.048 & 0.135$\pm$0.037 &       ...       &       ...       \\
\end{tabular}
 \end{threeparttable}
 \begin{tablenotes}
 \item For 6dF J0105-0251, UGC 12729, APM B2341-0330 the adopted O/H in Table~\ref{tab:main} is the weighted mean of all 3 available methods.
 \end{tablenotes}
 }
\end{table*}

\begin{table*}
\centering{
\caption{
Measured and corrected line intensities, and derived oxygen abundances (SALT data)}
\begin{threeparttable}
\label{t:Intens_SALT7}
\vspace{0.3cm}
\begin{tabular}{l|c|c|c|c|c|c} \hline
\rule{0pt}{10pt}
& \MC{2}{c|}{B2342-0223}
& \MC{2}{c|}{UGC 12769 (a)}
& \MC{2}{c}{UGC 12769 (b)}     \\ \hline
\rule{0pt}{10pt}
$\lambda_{0}$(\AA) Ion    & F($\lambda$)/F(H$\beta$)&I($\lambda$)/I(H$\beta$) & F($\lambda$)/F(H$\beta$)&I($\lambda$)/I(H$\beta$)  & F($\lambda$)/F(H$\beta$)&I($\lambda$)/I(H$\beta$) \\ \hline
3727\ [O\ {\sc ii}]\                & 4.450$\pm$0.477 & 3.688$\pm$0.533 & 2.701$\pm$0.480 & 2.240$\pm$0.497 & 4.253$\pm$0.797 & 5.290$\pm$1.041 \\
4340\ H$\gamma$\                    & 0.282$\pm$0.059 & 0.476$\pm$0.147 & 0.205$\pm$0.092 & 0.346$\pm$0.211 & 0.411$\pm$0.093 & 0.466$\pm$0.123 \\
4861\ H$\beta$\                     & 1.000$\pm$0.109 & 1.000$\pm$0.146 & 1.000$\pm$0.158 & 1.000$\pm$0.193 & 1.000$\pm$0.136 & 1.000$\pm$0.150 \\
4959\ [O\ {\sc iii}]\               & 0.497$\pm$0.077 & 0.385$\pm$0.077 & 0.552$\pm$0.111 & 0.458$\pm$0.111 & 0.553$\pm$0.137 & 0.533$\pm$0.136 \\
5007\ [O\ {\sc iii}]\               & 1.545$\pm$0.149 & 1.193$\pm$0.149 & 2.261$\pm$0.276 & 1.875$\pm$0.276 & 1.719$\pm$0.215 & 1.642$\pm$0.211 \\
6363\ [O\ {\sc i}]\                 & 0.092$\pm$0.030 & 0.067$\pm$0.029 &       ...       &       ...       &       ...       &       ...       \\
\end{tabular}
 \end{threeparttable}
 \begin{tablenotes}
 \item For B2342-0223 the adopted O/H in Table~\ref{tab:main} is the weighted mean of all 3 available methods.
 
 For UGC12769 the adopted O/H in Table~\ref{tab:main} is the weighted mean of all 3 available methods for both regions.
 \end{tablenotes}
 }
\end{table*}

\begin{table*}
\centering{
\caption{
Measured and corrected line intensities, and derived oxygen abundances (SALT data)}
\begin{threeparttable}
\label{t:Intens_SALT8}
\vspace{0.3cm}
\begin{tabular}{l|c|c|c|c|c|c} \hline
\rule{0pt}{10pt}
& \MC{2}{c|}{J2356+0141 (a)}
& \MC{2}{c|}{J2356+0141 (b)}
& \MC{2}{c}{HIPASS J2359+02}     \\ \hline
\rule{0pt}{10pt}
$\lambda_{0}$(\AA) Ion    & F($\lambda$)/F(H$\beta$)&I($\lambda$)/I(H$\beta$) & F($\lambda$)/F(H$\beta$)&I($\lambda$)/I(H$\beta$)  & F($\lambda$)/F(H$\beta$)&I($\lambda$)/I(H$\beta$) \\ \hline
3727\ [O\ {\sc ii}]\                & 3.903$\pm$0.550 & 4.077$\pm$0.677 & 3.468$\pm$0.356 & 2.684$\pm$0.398 & 3.653$\pm$0.400 & 2.801$\pm$0.424 \\
4340\ H$\gamma$\                    & 0.344$\pm$0.058 & 0.472$\pm$0.103 & 0.233$\pm$0.061 & 0.475$\pm$0.214 & 0.214$\pm$0.065 & 0.403$\pm$0.190 \\
4861\ H$\beta$\                     & 1.000$\pm$0.081 & 1.000$\pm$0.101 & 1.000$\pm$0.072 & 1.000$\pm$0.111 & 1.000$\pm$0.125 & 1.000$\pm$0.172 \\
4959\ [O\ {\sc iii}]\               & 0.723$\pm$0.085 & 0.617$\pm$0.084 & 0.968$\pm$0.075 & 0.684$\pm$0.075 & 1.353$\pm$0.146 & 1.033$\pm$0.146 \\
5007\ [O\ {\sc iii}]\               & 2.672$\pm$0.184 & 2.265$\pm$0.181 & 3.110$\pm$0.177 & 2.192$\pm$0.175 & 3.934$\pm$0.365 & 3.005$\pm$0.365 \\
5876\ He\ {\sc i}\                  &       ...       &       ...       & 0.099$\pm$0.048 & 0.066$\pm$0.045 & 0.127$\pm$0.044 & 0.097$\pm$0.044 \\
\end{tabular}
\end{threeparttable}
\begin{tablenotes}
\item For J2356+0141 the adopted O/H in Table~\ref{tab:main} is the weighted mean of all 3 available methods for both regions.

For HIPASS J2359+02 the adopted O/H in Table~\ref{tab:main} is the weighted mean of all 3 available methods.
\end{tablenotes}
}
\end{table*}

\begin{table*}
\centering{
\caption{
Measured and corrected line intensities, and derived oxygen abundances (SALT data)}
\begin{threeparttable}
\label{t:Intens_MCG}
\vspace{0.3cm}
\begin{tabular}{l|c|c} \hline
\rule{0pt}{10pt}
& \MC{2}{c|}{MCG-01-04-005}  \\ \hline
\rule{0pt}{10pt}
$\lambda_{0}$(\AA) Ion    & F($\lambda$)/F(H$\beta$)&I($\lambda$)/I(H$\beta$)  \\ \hline
3727\ [O\ {\sc ii}]\                & 4.994$\pm$0.804 & 4.074$\pm$1.083   \\
4861\ H$\beta$\                     & 1.000$\pm$0.169 & 1.000$\pm$0.297   \\
4959\ [O\ {\sc iii}]\               & 0.651$\pm$0.162 & 0.408$\pm$0.160   \\
5007\ [O\ {\sc iii}]\               & 1.942$\pm$0.286 & 1.208$\pm$0.280   \\
6548\ [N\ {\sc ii}]\                & 0.171$\pm$0.088 & 0.084$\pm$0.068   \\
6563\ H$\alpha$\                    & 5.280$\pm$0.653 & 2.807$\pm$0.597   \\
\end{tabular}
\end{threeparttable}
\begin{tablenotes}
\item The galaxy was initially included in the void sample, and its spectrum was obtained with SALT.However, later the galaxy was excluded from the sample since it seems to belong to NGC450 group. We include its spectrum analysis here, but do not use it in the void galaxy sample study.
\end{tablenotes}
 }
\end{table*}

\begin{table*}
\centering{
\caption{
Measured and corrected line intensities, and derived oxygen abundances (BTA data)}
\begin{threeparttable}
\label{t:Intens_BTA1}
\vspace{0.3cm}
\begin{tabular}{l|c|c|c|c|c|c} \hline
\rule{0pt}{10pt}
& \MC{2}{c|}{UGC 527 (a)}
& \MC{2}{c|}{UGC 527 (b)}
& \MC{2}{c}{Mrk 965 (a)}     \\ \hline
\rule{0pt}{10pt}
$\lambda_{0}$(\AA) Ion    & F($\lambda$)/F(H$\beta$)&I($\lambda$)/I(H$\beta$) & F($\lambda$)/F(H$\beta$)&I($\lambda$)/I(H$\beta$)  & F($\lambda$)/F(H$\beta$)&I($\lambda$)/I(H$\beta$) \\ \hline
3727\ [O\ {\sc ii}]\                &  1.642$\pm$0.302  &  1.730$\pm$0.329 & 4.141$\pm$0.219 & 4.141$\pm$0.227 & 4.878$\pm$0.236 & 4.635$\pm$0.332 \\
3967\ [Ne\ {\sc iii}]\ +\ H7\      &  ...            & ...                 & 0.136$\pm$0.027 & 0.136$\pm$0.052 & ...             & ...    \\
4101\ H$\delta$\                    &  0.162$\pm$0.021  & 0.217$\pm$0.036  & 0.276$\pm$0.036 & 0.276$\pm$0.051 & ...             & ... \\
4340\ H$\gamma$\                    &  0.463$\pm$0.042  & 0.504$\pm$0.050  & 0.583$\pm$0.062 & 0.583$\pm$0.070 & 0.197$\pm$0.030 & 0.468$\pm$0.122 \\
4363\ [O\ {\sc iii}]\               &  0.109$\pm$0.039  & 0.109$\pm$0.041  &  ...            & ...             & ...             & ... \\
4861\ H$\beta$\                     &  1.000$\pm$0.042  & 1.000$\pm$0.045  & 1.000$\pm$0.043 & 1.000$\pm$0.051 & 1.000$\pm$0.056 & 1.000$\pm$0.084 \\
\end{tabular}
\end{threeparttable}
\begin{tablenotes}
\item For UGC 527 the adopted O/H in Table~\ref{tab:main} is the weighted mean of all available methods for both regions.

For Mrk 965 the adopted O/H in Table~\ref{tab:main} is the weighted mean of all available methods for three regions.
\end{tablenotes}
}
\end{table*}

\begin{table*}
\centering{
\caption{
Measured and corrected line intensities, and derived oxygen abundances (BTA data)}
\begin{threeparttable}
\label{t:Intens_BTA2}
\vspace{0.3cm}
\begin{tabular}{l|c|c|c|c|c|c} \hline
\rule{0pt}{10pt}
& \MC{2}{c|}{Mrk 965 (b)}
& \MC{2}{c|}{Mrk 965 (c)}
& \MC{2}{c}{UGC 711}     \\ \hline
\rule{0pt}{10pt}
$\lambda_{0}$(\AA) Ion    & F($\lambda$)/F(H$\beta$)&I($\lambda$)/I(H$\beta$) & F($\lambda$)/F(H$\beta$)&I($\lambda$)/I(H$\beta$)  & F($\lambda$)/F(H$\beta$)&I($\lambda$)/I(H$\beta$) \\ \hline
3727\ [O\ {\sc ii}]\                & 4.313$\pm$0.214 & 4.889$\pm$0.358 & 2.934$\pm$0.136 & 3.176$\pm$0.177 & 3.039$\pm$0.116 & 3.082$\pm$0.126 \\
3868\ [Ne\ {\sc iii}]\              &  ...            & ...             & ...             & ...             & 0.347$\pm$0.025 & 0.351$\pm$0.026   \\
3889\ He\ {\sc i}\ +\ H8\           &  ...            & ...             & ...             & ...             & 0.111$\pm$0.017 & 0.119$\pm$0.029    \\
3967\ [Ne\ {\sc iii}]\ +\ H7\       &  ...            & ...             & ...             & ...             & 0.212$\pm$0.021 & 0.220$\pm$0.030   \\
4101\ H$\delta$\                    &  ...            & ...             & ...             & ...             & 0.241$\pm$0.021 & 0.249$\pm$0.030 \\
4340\ H$\gamma$\                    & 0.177$\pm$0.017 & 0.469$\pm$0.078 & 0.360$\pm$0.031 & 0.477$\pm$0.058 & 0.492$\pm$0.022 & 0.498$\pm$0.028 \\
\end{tabular}
\end{threeparttable}
}
\begin{tablenotes}
\item For Mrk 965 the adopted O/H in Table~\ref{tab:main} is the weighted mean of all available methods for three regions.

For UGC 711 the adopted O/H in Table~\ref{tab:main} is $O/H_{\rm Te}$.
\end{tablenotes}
\end{table*}

\begin{table*}
\centering{
\caption{
Measured and corrected line intensities, and derived oxygen abundances (BTA data)}
\begin{threeparttable}
\label{t:Intens_BTA3}
\vspace{0.3cm}
\begin{tabular}{l|c|c} \hline
\rule{0pt}{10pt}
& \MC{2}{c|}{MCG +00-04-049} \\ \hline
\rule{0pt}{10pt}
$\lambda_{0}$(\AA) Ion     & F($\lambda$)/F(H$\beta$)&I($\lambda$)/I(H$\beta$)   \\ \hline
3727\ [O\ {\sc ii}]\               & 1.770$\pm$0.205 & 2.071$\pm$0.244 \\
4340\ H$\gamma$\                    & 0.444$\pm$0.030 & 0.477$\pm$0.040  \\
4363\ [O\ {\sc iii}]\               & 0.067$\pm$0.017 & 0.071$\pm$0.018  \\
4861\ H$\beta$\                     & 1.000$\pm$0.062 & 1.000$\pm$0.067  \\
4959\ [O\ {\sc iii}]\               & 0.878$\pm$0.046 & 0.866$\pm$0.045 \\
5007\ [O\ {\sc iii}]\               & 2.529$\pm$0.123 & 2.479$\pm$0.122  \\
\end{tabular}
\end{threeparttable}

\begin{tablenotes}
\item For MCG +00-04-049 the adopted O/H in Table~\ref{tab:main} is $O/H_{\rm Te}$.
\end{tablenotes}
}
\end{table*}

\begin{table*}
\centering{
\caption{
Measured and corrected line intensities, and derived oxygen abundances (SDSS data)}
\begin{threeparttable}
\label{t:Intens_SDSS1}
\vspace{0.3cm}
\begin{tabular}{l|c|c|c|c|c|c} \hline
\rule{0pt}{10pt}
& \MC{2}{c|}{Ark18 (a)}
& \MC{2}{c|}{Ark18 (b)}
& \MC{2}{c}{NCG428}     \\ \hline
\rule{0pt}{10pt}
$\lambda_{0}$(\AA) Ion    & F($\lambda$)/F(H$\beta$)&I($\lambda$)/I(H$\beta$) & F($\lambda$)/F(H$\beta$)&I($\lambda$)/I(H$\beta$)  & F($\lambda$)/F(H$\beta$)&I($\lambda$)/I(H$\beta$) \\ \hline
(3727\ [O\ {\sc ii}])\             &       ...       &       ...       & 1.583$\pm$0.333 & 1.600$\pm$0.392 & 2.809$\pm$0.111 & 3.413$\pm$0.138\\
3835\ H9\ \                        & 0.060$\pm$0.007 & 0.075$\pm$0.013 &  ...            &  ...            & 0.057$\pm$0.004 & 0.067$\pm$0.007\\
3868\ [Ne\ {\sc iii}]\             & 0.264$\pm$0.007 & 0.302$\pm$0.009 &  ...            &  ...            & 0.234$\pm$0.008 & 0.277$\pm$0.010\\
3889\ He\ {\sc i}\ +\ H8\          & 0.177$\pm$0.007 & 0.207$\pm$0.012 &  ...            &  ...            & 0.184$\pm$0.006 & 0.217$\pm$0.009 \\
3967\ [Ne\ {\sc iii}]\ +\ H7\      & 0.172$\pm$0.010 & 0.199$\pm$0.015 &  ...            &  ...            & 0.210$\pm$0.010 & 0.243$\pm$0.013  \\
4026\ He\ {\sc i}\                 & 0.019$\pm$0.005 & 0.022$\pm$0.006 &  ...            &  ...            & 0.009$\pm$0.004 & 0.011$\pm$0.005 \\
\end{tabular}
\end{threeparttable}
\begin{tablenotes}
\item Ark 18: Values for both spectra are presented in Table~\ref{tab:main} separately. For spectra a) the adopted O/H in Table~\ref{tab:main} is $O/H_{\rm Te}$, for spectra b) the adopted O/H in Table~\ref{tab:main} is the weighted mean of all 3 available methods.

NGC 428: $T_{\rm e}$ method was used as adopted O/H in Table~\ref{tab:main}.
\end{tablenotes}
 }
\end{table*}

\begin{table*}
\centering{
\caption{
Measured and corrected line intensities, and derived oxygen abundances (SDSS data)}
\begin{threeparttable}
\label{t:Intens_SDSS2}
\vspace{0.3cm}
\begin{tabular}{l|c|c|c|c|c|c} \hline
\rule{0pt}{10pt}
& \MC{2}{c|}{J0142+0018}
& \MC{2}{c|}{J0142+0021}
& \MC{2}{c}{J0143-0002}     \\ \hline
\rule{0pt}{10pt}
$\lambda_{0}$(\AA) Ion    & F($\lambda$)/F(H$\beta$)&I($\lambda$)/I(H$\beta$) & F($\lambda$)/F(H$\beta$)&I($\lambda$)/I(H$\beta$)  & F($\lambda$)/F(H$\beta$)&I($\lambda$)/I(H$\beta$) \\ \hline
%
4340\ H$\gamma$\                    & 0.284$\pm$0.057 & 0.431$\pm$0.126 & 0.170$\pm$0.047& 0.476$\pm$0.265  & 0.315$\pm$0.053 & 0.469$\pm$0.107 \\
4861\ H$\beta$\                     & 1.000$\pm$0.072 & 1.000$\pm$0.099 & 1.000$\pm$0.072 & 1.000$\pm$0.124 & 1.000$\pm$0.073 & 1.000$\pm$0.089 \\
4959\ [O\ {\sc iii}]\               & 0.576$\pm$0.066 & 0.479$\pm$0.066 & 0.843$\pm$0.071 & 0.543$\pm$0.071 & 0.487$\pm$0.067 & 0.431$\pm$0.066 \\
5007\ [O\ {\sc iii}]\               & 1.522$\pm$0.090 & 1.264$\pm$0.090 & 2.634$\pm$0.144 & 1.699$\pm$0.144 & 1.464$\pm$0.091 & 1.285$\pm$0.089 \\
6548\ [N\ {\sc ii}]\                & 0.076$\pm$0.078 & 0.063$\pm$0.078 & 0.015$\pm$0.061 & 0.009$\pm$0.060 & 0.024$\pm$0.072 &0.017$\pm$0.055\\
6563\ H$\alpha$\                    & 3.187$\pm$0.179 & 2.746$\pm$0.205 & 3.870$\pm$0.211 & 2.752$\pm$0.255 & 3.997$\pm$0.263 & 2.806$\pm$0.225 \\
6584\ [N\ {\sc ii}]\                & 0.277$\pm$0.113 & 0.230$\pm$0.113 & 0.197$\pm$0.086 & 0.125$\pm$0.085 & 0.084$\pm$0.090 & 0.057$\pm$0.069 \\
6717\ [S\ {\sc ii}]\                & 0.247$\pm$0.039 & 0.205$\pm$0.039 & 0.506$\pm$0.045 & 0.321$\pm$0.046 & 0.280$\pm$0.088 & 0.187$\pm$0.066 \\
6731\ [S\ {\sc ii}]\                & 0.312$\pm$0.046 & 0.259$\pm$0.046 & 0.370$\pm$0.041 & 0.235$\pm$0.042 & 0.306$\pm$0.079 & 0.204$\pm$0.059 \\
				                    & \MC {2}{c|}{~}                    & \MC {2}{c|}{~}                    & \MC {2}{c}{~}             \\ \hline
C(H$\beta$)\ dex                    & \MC {2}{c|}{0.000$\pm$0.073}      & \MC {2}{c|}{0.010$\pm$0.071}      & \MC {2}{c}{0.365$\pm$0.086}         \\
EW(abs)\ \AA\                       & \MC {2}{c|}{1.600$\pm$0.337}      & \MC {2}{c|}{2.550$\pm$0.201}      & \MC {2}{c}{1.450$\pm$0.471}         \\
F(H$\beta$)\                        & \MC {2}{c|}{4.58$\pm$0.23}       & \MC {2}{c|}{6.24$\pm$0.31}       & \MC {2}{c}{3.34$\pm$0.17}         \\
EW(H$\beta$)\ \AA\                  & \MC {2}{c|}{7.35$\pm$0.38}        & \MC {2}{c|}{4.56$\pm$0.23}        & \MC {2}{c}{13.48$\pm$0.70}         \\
Rad. vel.\ \kms\                    & \MC {2}{c|}{3303}                 & \MC {2}{c|}{3324$\pm$9}           & \MC {2}{c}{3234$\pm$12}         \\
				                    & \MC {2}{c|}{~}                    & \MC {2}{c|}{~}                    & \MC {2}{c}{~}            \\ \hline
12+log(O/H)(PM10)\                  & \MC {2}{c|}{8.22$\pm$0.14}        & \MC {2}{c|}{7.69$\pm$0.21}        & \MC {2}{c}{7.36$\pm$0.35}        \\
   \hline
\MC{3}{l}{~~} \\
\end{tabular}
\end{threeparttable}
\begin{tablenotes}
\item For J0142+0018, J0142+0021, J0143-0002 due to poor quality of spectra the values of 12+log(O/H) should be considered only as indicative. They were not used in our analysis. In the case of J0142+0018 the value 12+log(O/H) estimated via PM method could be overestimated (see subsection \ref{OH}) so it should be considered as upper limit.
\end{tablenotes}
 }
\end{table*}

\begin{table*}
\centering{
\caption{
Measured and corrected line intensities, and derived oxygen abundances (SDSS data)}
\begin{threeparttable}
\label{t:Intens_SDSS3}
\vspace{0.3cm}
\begin{tabular}{l|c|c} \hline
\rule{0pt}{10pt}
& \MC{2}{c|}{UGC 12729} \\ \hline
\rule{0pt}{10pt}
$\lambda_{0}$(\AA) Ion     & F($\lambda$)/F(H$\beta$)&I($\lambda$)/I(H$\beta$)   \\ \hline
(3727\ [O\ {\sc ii}])\               & 4.459$\pm$0.597 & 6.059$\pm$1.016 \\
4340\ H$\gamma$\                    & 0.262$\pm$0.042 & 0.474$\pm$0.109  \\
4861\ H$\beta$\                    & 1.000$\pm$0.105 & 1.000$\pm$0.135   \\
4959\ [O\ {\sc iii}]\               & 0.660$\pm$0.121 & 0.528$\pm$0.118 \\
5007\ [O\ {\sc iii}]\               & 1.795$\pm$0.166 & 1.411$\pm$0.159  \\
6548\ [N\ {\sc ii}]\                & 0.241$\pm$0.128 & 0.118$\pm$0.076  \\
6563\ H$\alpha$\                    & 5.591$\pm$0.442 & 2.821$\pm$0.295   \\
6584\ [N\ {\sc ii}]\                & 1.048$\pm$0.245 & 0.509$\pm$0.146  \\
6717\ [S\ {\sc ii}]\                & 1.528$\pm$0.125 & 0.717$\pm$0.078   \\
6731\ [S\ {\sc ii}]\                & 1.009$\pm$0.091 & 0.472$\pm$0.055  \\
				                   & \MC {2}{c}{~}             \\ \hline
C(H$\beta$)\ dex                                 & \MC {2}{c|}{0.695$\pm$0.103}                        \\
EW(abs)\ \AA\                                    & \MC {2}{c|}{0.650$\pm$0.118}                      \\
F(H$\beta$)\                                     & \MC {2}{c|}{8.06$\pm$0.60}                        \\
EW(H$\beta$)\ \AA\                               & \MC {2}{c|}{3.18$\pm$0.24}                        \\
Rad. vel.\ \kms\                                 & \MC {2}{c|}{1971$\pm$3}                        \\
				                          & \MC {2}{c}{~}            \\ \hline
$T_{\rm e}$(OIII)(K)\                            & \MC {2}{c|}{13899$\pm$1388}                      \\
$T_{\rm e}$(OII)(K)\                             & \MC {2}{c|}{13272$\pm$891}                      \\
O$^{+}$/H$^{+}$($\times$10$^5$)\                 & \MC {2}{c|}{8.175$\pm$2.322}                     \\
O$^{++}$/H$^{+}$($\times$10$^5$)\                & \MC {2}{c|}{1.949$\pm$0.556}                      \\
O/H($\times$10$^5$)\                             & \MC {2}{c|}{10.120$\pm$2.388}                       \\
12+log(O/H)(Te)\                                 & \MC {2}{c|}{...}                       \\
12+log(O/H)(IT07$_{corr}$)\                               & \MC {2}{c|}{7.98$\pm$0.12}                       \\
   \hline
\MC{3}{l}{~~} \\
\end{tabular}
\end{threeparttable}
\begin{tablenotes}
\item For UGC12729 the adopted O/H was estimated from SALT spectra. The value obtained with SDSS spectra is in good agreement with it.
\end{tablenotes}
 }
\end{table*}

\label{lastpage}
\end{document}